\documentclass{article}     % onecolumn (second format)

\usepackage{amsmath, amsfonts,amsthm,amssymb}
\usepackage{dsfont}

\usepackage[english]{babel}

\usepackage[left=1.75cm, top=2.7cm, bottom=2.7cm,right=1.75cm]{geometry}

\usepackage{verbatim}
\usepackage{mathrsfs}
\usepackage{bm}
\usepackage{color}
\usepackage{graphicx}

\usepackage{url}

\usepackage{bm}

\usepackage{colortbl}

 \newtheorem{thm}{Theorem}[section]
 
 \newtheorem{lem}[thm]{Lemma}
 
 \newtheorem{prop}[thm]{Proposition}
 \theoremstyle{definition}

 \newtheorem{rem}[thm]{Remark}
 \numberwithin{equation}{section}

\newcommand{\be}{\begin{equation}}
\newcommand{\ee}{\end{equation}}
\newcommand{\bq}{\begin{eqnarray}}
\newcommand{\eq}{\end{eqnarray}}

\newcommand{\nn}{\nonumber}

    \def\ed{{\,\stackrel{\frak {D}}{=}\,}}

    \def\calF{{\mathcal F}}

    \def\calN{{\mathcal N}}

    \def\bbr{{\mathbb R}}
    \def\bbe{{\mathbb E}}
    \def\bbp{{\mathbb P}}

    \def\ed{\;{\stackrel{\frak {D}}{=}}\;}

    \def\c{{c}}

    \def\C{{C}}

  \definecolor{Red}{rgb}{1.00, 0.00, 0.00}
    
    \definecolor{DRed}{rgb}{0.7, 0.3, 0.00}
    
    \definecolor{Green}{rgb}{0.2, 0.5, 0.2}%{0.5, 0.00, 0.00}
    
    \definecolor{Blue}{rgb}{0.00, 0.00, 1.00}%{0.00, 0.00, 1.00}
    
    \definecolor{PaleGrey}{rgb}{.6, .6, .6}

\title{High-order short-time expansions for ATM option prices under the CGMY model}
\author{Jos\'e E. Figueroa-L\'{o}pez\thanks{Department of Statistics, Purdue University, West Lafayette, IN, 47907,  USA ({\tt figueroa@purdue.edu}). {Research supported in part by the {NSF Grants: DMS-0906919, DMS-1149692.}}}
 \and Ruoting Gong\thanks{School of Mathematics, Georgia Institute of Technology, Atlanta, GA, 30332, USA ({\tt rgong@math.gatech.edu}).}
 \and Christian Houdr\'e\thanks{School of Mathematics, Georgia Institute of Technology, Atlanta, GA, 30332, USA ({\tt houdre@math.gatech.edu}).}
}

\date{
}

\begin{document}
\maketitle
\begin{abstract}
The short-time asymptotic behavior {of} option prices for a variety of models with jumps has received much attention in recent years. In the present work, a novel second-order approximation for ATM option prices under the CGMY L\'evy model is derived and, then, extended to a model with an additional independent Brownian component. Our method of proof is based on an integral representation of the option price involving the tail probability of the log-return process under the share measure and a suitable change of probability measure under which the process becomes stable. This approach is sufficiently efficient to produce the third-order asymptotic behavior of the option prices and, moreover, is expected to apply to many other popular classes of L\'evy processes which satisfy the fundamental property of being stable under a suitable change of probability measure. Our results shed new light on the connection between both the volatility of the continuous component and the jump parameters and the behavior of ATM option prices near expiration. In the case of an additional Brownian component, the second-order term, in time-$t$, is of the form $ d_{2}\, t^{(3-Y)/2}$, with the coefficient $d_{2}$ depending only on the overall jump intensity parameter $C$ and the tail-heaviness parameter $Y$. This extends the known result that the leading term is $(\sigma/\sqrt{2\pi})t^{1/2}$, where $\sigma$ is the volatility of the continuous component. In contrast, under a pure-jump CGMY model, the dependence on the two parameters $C$ and $Y$ is already reflected in the leading term, which is of the form $d_{1} t^{1/Y}$. Information on the relative frequency of negative and positive jumps appears only in the second-order term, which is shown to be of the form $d_{2} t$ and whose order of decay turns out to be independent of $Y$. The third-order asymptotic behavior of the option prices as well as the asymptotic behavior of the corresponding Black-Scholes implied volatilities are also addressed. Our numerical results show that first-order term typically exhibits extremely poor performance and that the second-order term significantly improves the approximation's accuracy.

\vspace{0.3 cm}


\vspace{0.3 cm}


\end{abstract}

\section{Introduction}

It is generally recognized that the standard option pricing model of Black-Scholes is inconsistent with options data, while remaining a widely used model in practice because of its simplicity. Exponential L\'{e}vy models generalize the classical Black-Scholes setup by allowing jumps in stock prices while preserving the independence and stationarity of returns. There are several reasons for introducing jumps in financial modeling. First of all, asset prices do jump, and some risks simply cannot be handled within continuous-paths models. Second, historical asset prices exhibit distributions with so-called stylized features, such as heavy tails, high kurtosis, volatility clustering and leverage effects, which are hard to replicate within purely-continuous frameworks. Finally, market prices of vanilla options exhibit skewed implied volatilities (relative to changes in the strikes), in contrast to the classical Black-Scholes model which predicts a flat implied volatility smile. Moreover, the fact that the implied volatiliy smile and skewness phenomenon becomes much more pronounced for short maturities is a clear indication of the presence of jumps.

One of the first applications of jump processes in financial modeling is due to Mandelbrot \cite{Mandelbrot}, who suggested a pure-jump stable L\'evy process $Z$ to model power-like tails and self-similar behavior in cotton price returns.
Merton \cite{Merton:76} and Press \cite{Press} subsequently considered option pricing and hedging problems under an exponential compound Poisson process with Gaussian jumps and an additive independent non-zero Brownian component. A similar exponential compound Poisson jump-diffusion model was more recently studied in Kou~\cite{Kou:2002}, where the jump sizes are distributed according to an asymmetric Laplace law. For infinite activity exponential L\'{e}vy models, Barndorff-Nielsen~\cite{BNS:1998} introduced the normal inverse Gaussian (NIG) model, while the extension to the generalized hyperbolic class was studied by Eberlein, Keller and Prause~\cite{EKP:1998}. Madan and Seneta~\cite{MadSen:1990} introduced the symmetric variance gamma (VG) model while its asymmetric extension was later studied by Madan and Milne~\cite{MadMil:1991} and Madan, Carr and Chang~\cite{MCC:1998}. Both models are built on Brownian subordination; the main difference being that the log-return process in the NIG model is an infinite variation process with stable like ($\alpha=1$) behavior of small jumps, while in the VG model, the log-price is of finite variation with infinite but relatively low activity of small jumps. The class of ``tempered stable" processes was first introduced by Koponen~\cite{Koponen:1995} and further developed by Carr, Geman, Madan and Yor~\cite{CGMY:2002}, who introduced the terminology CGMY. The CGMY model is a particular case of the more general KoBoL class of  \cite{BL02} and was also previously proposed for financial modeling in \cite{Cont:1997} and \cite{Matacz}. Nowadays, the CGMY model is considered to be a prototype of the general class of models with jumps and enjoys widespread applicability.

Stemming in part from its importance for model calibration and testing, small-time asymptotics of option prices have received a lot of attention in recent years (see, e.g., \cite{BBF:2002}, \cite{BBF2:2004}, \cite{FFF:2010}, \cite{FFK:2012}, \cite{FigForde:2012}, \cite{FordeJac:2009}, \cite{FordeJac:2010}, \cite{FordeJacLee:2010}, \cite{Gatheral:2009}, \cite{Henry:2009}, \cite{Paulot:2009}, \cite{Rop10}, \cite{Tankov}). We shall review here only the studies most closely related to ours,  focusing in particular on the at-the-money (ATM) case. Carr and Wu~\cite{CW03} first analyzed, partially via heuristic arguments, the first order asymptotic behavior of an It\^{o} semimartingale with jumps. Concretely, ATM option prices of pure-jump models of bounded variation decrease at the rate $O(t)$, while they are just $O(\sqrt{t})$ under the presence of a Brownian component. By considering a stable pure-jump component,~\cite{CW03} also showed that, in general, the rate could be $O(t^{\beta})$, for some $\beta\in(0,1)$. Muhle-Karbe and Nutz~\cite{MuhNut:2009} formally showed that, under the presence of a continuous-time component, the leading term of ATM option prices is of order $\sqrt{t}$, for a relatively general class of It\^{o} models, while for a more general type of It\^{o} processes with $\alpha$-stable-like small jumps, the leading term is $O(t^{1/{}\alpha})$ (see also~\cite[Proposition 4.2]{FigForde:2012}, \cite[Theorem 3.7]{FigGH:2011}, and~\cite[Proposition 5]{Tankov} for related results in exponential L\'{e}vy models). However, none of the these papers obtained second or higher order asymptotics for the ATM option prices, which are arguably more relevant for calibration purposes, given that the most liquid options are of this type.

In the present paper, we study the small-time behavior for at-the-money (ATM) call (or equivalently, put) option prices
\begin{equation}\label{CallPriceDfn}
	\bbe\left(S_{t}-S_{0}\right)^{+}=S_{0}\bbe\left(e^{X_{t}}-1\right)^{+},
\end{equation}
under the exponential L\'evy model
\begin{equation}\label{ExpLvMdl}
	S_{t}:=S_{0}e^{X_{t}},
\end{equation}
where $X$ is the superposition of a CGMY L\'evy process $(L_{t})_{t\geq{}0}$  and of an independent Brownian motion ${(\sigma W_{t})_{t\geq{}0}}$; i.e.,
\begin{equation}\label{CGMY1stDfn}
	X_{t} = L_{t}+\sigma W_{t},
\end{equation}
where $(W_{t})_{t\geq{}0}$ is a standard Brownian motion independent of $L$.
Here, as usual, $x^{+}$ is the positive part of $x$. The first order {asymptotic behavior of (\ref{CallPriceDfn}) in short-time under the model (\ref{CGMY1stDfn}) takes the form:
\begin{equation}\label{1stOAp}
	\lim_{t\to{}0}t^{-1/Y}\mathbb{E}(S_t-S_0)^{+}= S_{0}\mathbb{E}(Z^{+}),
\end{equation}
where $Z$ is a symmetric {stable} random variable {with $\alpha=Y$} under $\mathbb{P}$. When $\sigma\neq{}0$, $Z\sim \calN(0,\sigma^{2})$ ($\alpha=2$) and, thus, $\mathbb{E}(Z^{+})=\sigma/\sqrt{2\pi}$ (see \cite{Tankov} and \cite{Rop10}).
 When $\sigma=0$ {and $\alpha=Y$,} the characteristic function of $Z$ is explicitly given (see \cite{FigForde:2012} and \cite{Tankov}) by
\[
	\bbe e^{iuZ}=e^{-2C\Gamma(-Y)|\cos(\frac{1}{2}Y\pi)|\,|u|^{Y}}.
\]
{In} that case, (see (25.6) in \cite{Sato:1999}),
\begin{equation}\label{1stOrdFA}
	d_{1}:=\mathbb{E}(Z^{+})=\frac{1}{\pi}\Gamma\left(1-\frac{1}{Y}\right)\left(2 C \Gamma(-Y)\left|\cos\left(\frac{\pi Y}{2}\right)\right|\right)^{1/Y}.
\end{equation}
Interestingly enough, under the presence of a continuous component, the first-order asymptotic term {only reflects} information on the continuous-time volatility, in sharp contrast with the pure-jump case where the leading term depends on the overall jumps-intensity parameter $C$ and the index $Y$, which in turn controls the tail-heaviness of the distributions.

Below, we also obtain a second order correction term for the approximation (\ref{1stOAp}).
The derivation of the second-order results builds on two facts. First, as in \cite{FigForde:2012}, we make use of the following representation of Carr and Madan \cite{CM09}:
\be\label{CMR}
	{\frac{1}{S_0}\mathbb{E}(S_t-S_{0})^{+}=\mathbb{P}^*(X_t > E)  =\int_{0}^{\infty} e^{-x} \mathbb{P}^*(X_t > x) dx}\,,\ee
where $\mathbb{P}^*$ is the martingale probability measure obtained when one takes the stock as the num\'{e}raire (i.e., $\mathbb{P}^{*}(A):=\mathbb{E}\left( S_{t} 1_{A}\right)$) and $E$ is an independent mean-one exponential random variable under $\mathbb{P}^*$. The measure $\mathbb{P}^{*}$ is sometimes called the share measure (see \cite{CM09}). {Notice that under $\bbp^{*}$, $(X_{t})_{t\geq 0}$ also admits a decomposition similar to (\ref{CGMY1stDfn}),
\begin{align}\label{CGMY1stDfnshare}
X_{t}=L_{t}^{*}+\sigma W_{t}^{*},\quad t\geq 0,
\end{align}
where $W^{*}:=(W_{t}^{*})_{t\geq 0}$ is a Wiener process and $L^{*}:=(L_{t}^{*})_{t\geq{}0}$ is also a CGMY process, independent of $W^{*}$.} Second, we change probability measures from $\bbp^{*}$ to a probability measure $\widetilde{\bbp}$, under which {$(L_{t}^{*})_{t\geq 0}$} is a stable L\'evy process and {$(W_{t}^{*})_{t\geq{}0}$} is {still} a standard Brownian motion independent of {$L^{*}$}. We show that the second-order asymptotic behavior of the ATM call option price (\ref{CallPriceDfn}) in short-time is then of the form
	\[
		\frac{1}{S_{0}}\mathbb{E}(S_t-S_{0})^{+}= d_{1} t^{\frac{1}{Y}}+d_{2} t +o(t), \qquad (t\to{}0),
	\]
in the pure-jump CGMY case ($\sigma=0$), while in the case of a non-zero independent Brownian component ($\sigma\neq{}0$),	
	\[
		\frac{1}{S_{0}}\mathbb{E}(S_t-S_{0})^{+}= d_{1} t^{\frac{1}{2}}+d_{2} t^{\frac{3-Y}{2}} +{o\left(t^{\frac{3-Y}{2}}\right)}, \qquad (t\to{}0),
	\]
for different constants $d_{1}$ and $d_{2}$ that we will determine explicitly. To wit, we found that, under the presence of a nonzero Gaussian component, the second-order term depends only on the overall jump intensity parameter $C$ and the tail-{heaviness} parameter $Y$. The parameters $G$ and $M$ (which control the relative frequency of negative and positive jumps) do not appear until the next order term.  However, for a pure-jump case, the parameters $G$ and $M$ are already present in the second-order term. The above asymptotic behaviors should also be compared to the corresponding behavior under the standard Black-Scholes model, where it is known that (see, e.g., {\cite[Corollary 3.4]{FordeJacLee:2010}})
	\[
		\mathbb{E}(e^{\sigma W_{t}-\frac{\sigma^{2}}{2}t}-1)^{+}= \frac{\sigma}{\sqrt{2\pi}} t^{\frac{1}{2}}-\frac{\sigma^{3}}{24 \sqrt{2\pi}}t^{\frac{3}{2}}+O(t^{\frac{5}{2}}).
	\]	
Our method of proof is sharp enough to produce the third-order asymptotic behavior of the option prices (see Remark \ref{ThirdOrderAppPJ} and \ref{ThirdOrderAppGen} below) and, moreover, is expected to apply to other popular classes of L\'evy processes, which satisfy the fundamental property of being stable under a suitable change of probability measure such as tempered stable processes in the sense of Rosi\'nski \cite{Rosinski:2007} (this will be presented elsewhere). Finally, the asymptotic behavior of the corresponding Black-Scholes implied volatilities are also addressed. 	

The present paper is organized as follows. Section 2 contains preliminary results on the CGMY model, some probability measure transformations, and asymptotic results for stable L\'{e}vy processes which will be needed throughout the paper. Section 3 establishes the second-order asymptotics of the call option price under the pure-jump CGMY model ($\sigma=0$). Section 4 establishes the second-order asymptotics of the call-option price under the CGMY model with an additional independent non-zero Brownian component ($\sigma\neq 0$). In Section \ref{Numerics}, we assess the performance of our asymptotic expansions through a detailed numerical analysis. The proofs of our main results are deferred to the Appendices.

\section{Setup and preliminary results}\label{Sec:Setup}

\subsection{The CGMY model}
Throughout, $(L_{t})_{t\geq{}0}$ stands for a CGMY L\'evy process defined on a complete {filtered} probability space {$(\Omega,\calF,(\calF_{t})_{t\geq{}0},\bbp)$} with corresponding parameters $C, G, M>0$ and $Y\in(1,2)$. That is, $L$ is a pure-jump L\'evy process with characteristic function
\begin{equation}\label{PJCGMY}
	\bbe\left(e^{iuL_{t}}\right)=\exp\left(t\left[i  \c u+C \Gamma(-Y)\left((M-iu)^Y+(G+iu)^Y-M^Y-G^Y\right)\right]\right).
\end{equation}
Let ${X_{t}=\sigma W_{t}+L_{t}}$, $t\geq 0$, where $(W_{t})_{t\geq{}0}$ is a standard Brownian motion, independent of {$(L_{t})_{t\geq 0}$, defined on {$(\Omega,\calF,(\calF_{t})_{t\geq{}0},\bbp)$}. We call the process $X$ the (generalized) CGMY model.

We assume zero interest rate and that $\bbp$ is a martingale measure for the exponential L\'evy model $S_{t}=S_{0}e^{X_{t}}$. In particular, $M>1$ and the characteristic function $\varphi_t$ of {$X_{t}$ is} given by
\begin{equation}\label{CFVCGMY}
	\varphi_t(u)=\bbe\left(e^{iuX_{t}}\right)=\exp\left(t\left[i\c u-\frac{\sigma^{2}u^{2}}{2}+C \Gamma(-Y)\left((M-iu)^Y+(G+iu)^Y-M^Y-G^Y\right)\right]\right),
\end{equation}
with
\begin{align}
	 c &= -C\Gamma(-Y)\left((M-1)^{Y}+(G+1)^{Y}-M^{Y}-G^{Y}\right)-\frac{\sigma^{2}}{2};\label{cVValMartM}
\end{align}
see, e.g., Proposition 4.2 in \cite{Tankov}. In particular, we note that $\gamma :=\bbe X_{1}=\bbe L_{1}$ is given by
\begin{equation}
     \gamma =c-C Y\Gamma(-Y) (M^{Y-1}-G^{Y-1}).\label{VMeanMartM}
\end{equation}
The L\'evy triplet of $(X_{t})_{t\geq 0}$ (relative to the truncation function $x{\bf 1}_{\{|x|\leq{}1\}}$) is denoted by $(b,\sigma^{2},\nu)$. Thus,  $\nu$ and $b$ are given by
\begin{align}
    \label{LevyMeasCGMY}
	\nu(dx)&=\left(\frac{C e^{-Mx}}{x^{1+Y}}\, {\bf 1}_{\{x> 0\}} +\frac{C e^{Gx}}{|x|^{1+Y}}\,{\bf 1}_{\{x<0\}}\right)dx,\\
	b&=\c -\int_{|x|>1} x\nu(dx)-C Y\Gamma(-Y) (M^{Y-1}-G^{Y-1}).
	\label{bTripletCGMY}
\end{align}
Without loss of generality, we also assume throughout that $(X_{t})_{t\geq 0}$ is the canonical process $X_{t}(\omega)=\omega(t)$ defined on the canonical space $\Omega=\mathbb{D}([0,\infty),\mathbb{R})$ (the space of c\`{a}dl\`{a}g functions $\omega:[0,\infty)\to\mathbb{R}$) equipped with the $\sigma$-field $\mathcal{F}=\sigma(X_{s}:s\geq{}0)$ and the right-continuous filtration $\mathcal{F}_{t}:=\cap_{s>t}\sigma(X_{u}:u\leq{}s)$.

\subsection{Probability measure transformations}
Following a density transformation construction as given in Sato \cite{Sato:1999} (see Definition 33.4 and Example 33.4 there) and using the martingale condition $\bbe e^{X_{t}}=1$, we define a probability measure $\mathbb{P}^{*}$ on $(\Omega,\mathcal{F})$ such that
\begin{equation}\label{DSM}
	\frac{d {\bbp}^{*}|_{\calF_{t}}}{d \bbp |_{\calF_{t}}}=e^{X_{t}}, \qquad (t\geq{}0);
\end{equation}
i.e., $\mathbb{P}^{*}(B)=\mathbb{E}\left(e^{X_{t}} {{\bf 1}_{B}} \right)$, for any $B\in\mathcal{F}_{t}$ and $t\geq 0$.  The measure $\mathbb{P^*}$ can be interpreted as the martingale measure when using the stock price as the num\'{e}raire. Under $\bbp^{*}$, $(X_{t})_{t\geq{}0}$ is also a L\'{e}vy process and its characteristic function is given by
\begin{equation}\label{CFCGMYShare}
	\mathbb{E}^*(e^{iu X_t})=\exp\left(t\left[i  \c^*u-\frac{\sigma^{2}u^{2}}{2}+ C \Gamma(-Y)\left((M^*-iu)^Y+(G^*+iu)^Y-{M^*}^Y-{G^*}^Y\right)\right]\right),
\end{equation}
with (see Appendix \ref{AddtnProofs})
\[
	M^{*}=M-1, \quad G^{*}=G+1, \quad c^{*}=c+\sigma^{2}.
\]
It is clear from (\ref{CFCGMYShare}) that, under $\bbp^{*}$, {$(X_{t})_{t\geq 0}$ can {also} be decomposed as in (\ref{CGMY1stDfnshare}), where $(W_{t}^{*})_{t\geq 0}$ is again a Wiener process while $(L_{t}^{*})_{t\geq{}0}$ is still a CGMY process, independent of $W^{*}$, but with parameters $C$,  $Y$, $M=M^{*}$, and $G=G^{*}$.} Hereafter, we denote the L\'evy triplet of $(X_{t})_{t\geq 0}$ under $\bbp^{*}$ by $(b^{*},(\sigma^{*})^{2},\nu^{*})$, where $\sigma^{*}=\sigma$, $\nu^{*}(dx)=e^{x}\nu(dx)$, and
\begin{equation}
	b^*:={\c^{*}} -\int_{|x|>1} x\nu^*(dx)-C Y\Gamma(-Y) ((M^*)^{Y-1}-(G^*)^{Y-1}).
	\label{b*TripletCGMY}
\end{equation}

As explained in the introduction, an important tool in the sequel is to change the probability measures from $\bbp^{*}$ to a probability measure $\widetilde{\bbp}$, under which {$(L_{t}^{*})_{t\geq 0}$} is a stable L\'evy process and {$(W_{t}^{*})_{t\geq{}0}$ is still} a Wiener process independent of {$L^{*}$}.
 Concretely, let
\[
	\tilde{\nu}(dx):= C |x|^{-Y-1}dx,\qquad
	{\tilde{b}=b^{*}+\int_{|x|\leq{}1} x(\tilde{\nu}-\nu^{*})(dx).}
\]
Note that $\tilde{\nu}$ is the L\'evy measure of a symmetric stable L\'evy process and, also,
\[
	\tilde{\nu}(dx)=e^{\varphi(x)} \nu^{*}(dx),
\]
with
\[
	\varphi(x):=M^{*} x \, {\bf 1}_{\{x> 0\}} -G^{*} x\,{\bf 1}_{\{x<0\}}.
\]
Hence, by virtue of Theorem 33.1 in \cite{Sato:1999}, there exists a probability measure $\widetilde{\bbp}$ locally equivalent\footnote{Equivalently, there exists a process $(U_{t})_{t}$ such that $\widetilde{\bbp}(B)=\bbe^{*}(e^{U_{t}}{\bf 1}_{B})$, for $t\geq{}0$ and $B\in\mathcal{F}_{t}$.} to $\bbp^{*}$ such that $(X_{t})_{t\geq 0}$ is a L\'evy process with L\'evy triplet $(\tilde{b},\sigma^{2},\tilde\nu)$ under $\widetilde{\bbp}$. Throughout, $\widetilde{\bbe}$ denotes the expectation under $\widetilde{\bbp}$.

In light of (\ref{b*TripletCGMY}) and since $\widetilde{\bbe} X_{1}=\widetilde{\bbe} {L_{1}^{*}}=\tilde{b}+\int_{\{|x|>{}1\}} x\tilde{\nu}(dx)$,
it can be shown (see Appendix \ref{AddtnProofs}) that
\begin{equation}\label{Cent}
	\tilde{\gamma}:=\widetilde{\bbe} X_{1}=-C\Gamma(-Y) \left((M-1)^{Y}+(G+1)^{Y}-M^{Y}-G^{Y}\right)+\frac{\sigma^{2}}{2}.
\end{equation}
Next, {we} recall that the centered process $(Z_{t})_{t\geq 0}$,  given by
\begin{equation}\label{SSSP}
	Z_{t}:={L_{t}^{*}}-t\tilde{\gamma},
\end{equation}
is symmetric and strictly $Y$-stable\footnote{Concretely, its scale, skewness, and location parameters are $2C\Gamma(-Y)|\cos(\pi Y/2)|$, $0$, and $0$, respectively.} {under $\widetilde{\bbp}$} and, thus, is self-similar; i.e.,
\begin{equation}\label{SSCN}
	(t^{-1/Y} Z_{u t})_{u\geq{}0}\ed (Z_{u})_{u\geq{}0},
\end{equation}
for any $t>0$.
We also need the following representation of the density process (see Theorem 33.2 in \cite{Sato:1999}):
\begin{equation}\label{EMMN}
	\frac{d \widetilde{\bbp}\big|_{\calF_{t}}}{d \bbp^{*}\big|_{\calF_{t}}}=e^{U_{t}},
\end{equation}
with
\[
	U_{t}:=\lim_{\epsilon\to 0}\left(\sum_{s\leq{}t:|\Delta X_{s}|>\varepsilon} \varphi(\Delta X_{s})-t \int_{|x|>\varepsilon} (e^{\varphi(x)}-1)\nu^{*}(dx) \right).
\]

The process $(U_{t})_{t\geq 0}$ can be expressed in terms of the jump-measure $N(dt,dx):=\#\{(s,\Delta X_{s})\in dt\times dx\}$ of the process $(X_{t})_{t\geq 0}$ and its compensator $\bar{N}(dt,dx):=N(dt,dx)-\tilde{\nu}(dx)dt$ (under $\widetilde{\bbp}$); namely,
\begin{equation}\label{DcmLL}
	U_{t}=M^{*}\bar{U}_{t}^{+}-G^{*}\bar{U}_{t}^{-}+\eta t,
\end{equation}
where
\begin{align}
\label{Uplusminus}\bar{U}^{+}_{t}&:=\int_{0}^{t}\int_{(0,\infty)}x\bar{N}({ds},dx),\quad\bar{U}^{-}_{t}:=\int_{0}^{t}\int_{(-\infty,0)}x\bar{N}({ds},dx),\\
\eta&:=C\int_{0^{+}}^{\infty}(e^{-M^{*}x}-1+M^{*}x)x^{-Y-1}dx+C\int_{-\infty}^{0^{-}}(e^{G^{*}x}-1-G^{*}x)|x|^{-Y-1}dx\nonumber\\
&=C\Gamma(-Y)\left((M^{*})^{Y}+(G^{*})^{Y}\right),\label{eta}
\end{align}
where {in the last equality} we used the analytic continuation {presented in \cite{Sato:1999} (see (14.19) therein)}.
Finally, let us also note the following decomposition of the process $X$ in terms of the previously defined processes:
\begin{equation}\label{RX}
	X_{t}= Z_{t}+t\tilde\gamma+{\sigma W_{t}^{*}}=\bar{U}^{+}_{t}+\bar{U}^{-}_{t}+t\tilde{\gamma}+{\sigma W_{t}^{*}}.
\end{equation}
The following table summarizes the different probability measures used in this paper:
\begin{center}
{\small
\begin{tabular}{|c|c|c|}
\hline
& & \\
Prob. Measure & $(L_{t})_{t}$ distribution & Density wrt $\bbp$ \\
& & \\
\hline
& &\\
${\bbp}$  & ${\rm CGMY}(C,G,M,Y)$  & 1\\
& &\\
\hline
& & \\
${\bbp}^{*}$ & ${\rm CGMY}(C,G^*,M^*,Y)$ & $e^{X_{t}}$\\
& &\\
\hline
& & \\
$\widetilde{\bbp}$ & ${\rm Stable}(\beta=0;\alpha=Y)$   & $e^{X_{t}+U_{t}}$\\
& & \\
\hline
\end{tabular}
}
\end{center}

\subsection{Some {needed} properties of stable L\'evy processes}\hfill
Let us now collect some well-known results on stable L\'evy processes needed in the sequel. First, from (\ref{Uplusminus}), it is clear that $(\bar{U}_{t}^{+})_{t\geq 0}$ and $(-\bar{U}_{t}^{-})_{t\geq 0}$ are independent and identically distributed one-sided $Y$-stable processes\footnote{Concretely, its scale, skewness, and location parameters are $C|\cos(\pi Y/2)|\Gamma(-Y)$, $1$, and $0$, respectively.} under $\widetilde{\bbp}$. Hence, the common transition density of $\bar{U}_{t}^{+}$ and $-\bar{U}_{t}^{-}$, denoted by {$p(t,x)$}, exists (cf. \cite[Proposition 2.5]{Sato:1999}). Moreover, the following result for the asymptotic behavior of the transition density is known (see, e.g., \cite{Ruschendorf} and \cite{FigHou:2008}):
\[
	\lim_{t\to{}0}\frac{1}{t} p(t,u)= s(u), \qquad (u\neq{}0),
\]
where $s$ is the L\'evy density of the L\'evy process $(\bar{U}_{t}^{+})_{t\geq 0}$. In particular, since by construction the L\'evy measure of $(\bar{U}_{t}^{+})_{t\geq 0}$ is {$\tilde{\nu}_{+}(du) = Cu^{-Y-1}{\bf 1}_{\{u>0\}}du$}, the L\'evy density $s$ is just {$Cu^{-Y-1}{\bf 1}_{\{u>0\}}$, $u\neq 0$}, and we get:
\begin{align}\label{EFFOAS}
	\lim_{t\to{}0}\frac{1}{t} p(t,u)= {Cu^{-Y-1}{\bf 1}_{\{u>0\}},\quad u\neq 0.}
\end{align}
Equivalently, by the self-similarity of $(\bar{U}_{t}^{+})_{t\geq 0}$, we have
\[
	p(t,u)=t^{-1/Y} p(1,t^{-1/Y}u),
\]
and  (\ref{EFFOAS}) can be casted as follows by setting $x=t^{-1/Y} u$:
%}
%
%{$p(1,x)$} admits the following asymptotic behavior (see, e.g., (14.37) in \cite{Sato:1999}), as $x\rightarrow\infty$,
\begin{align}\label{Asystbden}
p(1,x)\sim  {C} %\frac{1}{\pi}\,\Gamma(Y+1)\sin(\pi Y)
x^{-Y-1}, \qquad (x\to{}\infty).
\end{align}
As a consequence,
\begin{align}\label{Utail}
\widetilde{\bbp}\Big(\bar{U}_{1}^{+}\geq x\Big)=\widetilde{\bbp}\Big(-\bar{U}_{1}^{-}\geq x\Big)\sim {\frac{C}{Y}}
%\frac{1}{\pi}\,\Gamma(Y)\sin(\pi Y)
x^{-Y},
\end{align}
as $x\to\infty$. Equivalently, plugging $x=t^{-1/Y}v$ and using the self-similarity of $\bar{U}^{+}$ and $\bar{U}^{-}$, we recover the well-known result:
\begin{align}\label{UtailPlus}
\lim_{t\to{}0}\frac{1}{t}\widetilde{\bbp}\Big(\pm\,\bar{U}_{t}^{\pm}\geq v\Big)=\tilde{\nu}([v,\infty))=\frac{C}{Y} v^{-Y}.
\end{align}
In particular, there {exists} $N>0$, such that for all $0<t\leq 1$ and $v>0$ satisfying $t^{-1/Y}v>N$,
\begin{align}\label{UIn1}
\widetilde{\bbp}(\bar{U}_{1}^{+}\geq t^{-1/Y}v)\leq{\frac{2C}{Y}} t v^{-Y},\quad
\quad \widetilde{\bbp}(-\bar{U}_{1}^{-}\geq t^{-1/Y}v)\leq{\frac{2C}{Y}} t v^{-Y}.
\end{align}
The following result sharpens (\ref{UIn1}). Its proof is presented in Appendix \ref{AddtnProofs}.
\begin{lem}\label{Bnd1TailSt}
	There exists {a} constant {$0<\kappa<\infty$} such that for \emph{any} $0<t\leq{}1$ and $v>0$,
\begin{align*}
\widetilde{\mathbb{P}}(\bar{U}_{1}^{+}\geq t^{-1/Y}v)\leq\kappa tv^{-Y},\quad
\quad\widetilde{\mathbb{P}}(-\bar{U}_{1}^{-}\geq t^{-1/Y}v)\leq\kappa tv^{-Y}.
\end{align*}
\end{lem}
{Therefore}, since $Z_{1}= \bar{U}^{+}_{1}-\bar{U}^{-}_{1}$,
\begin{equation}\label{KIn}
\widetilde{\bbp}(Z_{1}\geq t^{-1/Y}v)=\widetilde{\bbp}(Z_{t}\geq{}v)\leq{2^{Y+1}\kappa{}t{}v^{-Y}\leq{}8\kappa{}tv^{-Y}},
\end{equation}
for any $0<t\leq 1$ and $v>0$.  Note also that
\begin{equation}\label{AsytailZ100}
	\widetilde{\bbp}(Z_{t}\geq{}v) = \widetilde{\bbp}(Z_{1}\geq{} t^{-1/Y} v) \sim t \tilde{\nu}([v,\infty))=t\, \frac{C}{Y} v^{-Y},\quad (t\to{}0),
\end{equation}
and that the probability density $p_{Z}$ of $Z_{1}$ is such that
\begin{equation}\label{Asydenpz00}
	{p_{Z}(v)\sim C v^{-Y-1}, \qquad (v\to\infty).}
\end{equation}
The following identity for $\widetilde{U}_{t}:=M^{*}\bar{U}_{t}^{+}-G^{*}\bar{U}_{t}^{-}$ will also be needed in sequel:
\begin{align}
\widetilde{\bbe}\Big(e^{-t^{1/Y}\widetilde{U}_{1}}\Big)&=\bbe^{*}\Big(e^{-t^{1/Y}M^* \bar{U}^+_{1}}\Big)\bbe^{*}\Big(e^{t^{1/Y}G^* \bar{U}^-_{1}}\Big)=\exp(\eta t). \label{expUt}
\end{align}
The {relation above} follows from the representations (\ref{Uplusminus}), the independence of $\bar{U}^+$ and $\bar{U}^-$,  and the form of the characteristic function of a Poisson integral.

\section{The pure-jump CGMY model}\label{Sec:PureJump}

In this section, we find the second-order asymptotic behavior for the at-the-money call option prices (\ref{CallPriceDfn}) in the pure-jump CGMY model. {The proofs of all results in the section are deferred to the Appendix \ref{proofA}. Throughout} this section, $(X_{t})_{t\geq 0}$ is a L\'evy process with triplet $(b,0,\nu)$ as introduced in Section \ref{Sec:Setup}. As explained in the introduction, the first order asymptotic behavior is given by (\ref{1stOAp}){.  Before} stating our first result, we need to rewrite the call option price (\ref{CallPriceDfn}) in a suitable form.
\begin{lem}\label{Lm:NRATMPJ}
	In terms of the probability measure $\widetilde{\bbp}$ defined in (\ref{EMMN}) and the parameter $\tilde{\gamma}$ defined in (\ref{Cent}), we have that
	\begin{align}\label{LOP}
	t^{-\frac{1}{Y}}{\frac{1}{S_{0}}}\mathbb{E}(S_t-S_{0})^{+}&=e^{-(\tilde{\gamma}+\eta)t}\int_{-\tilde{\gamma} t^{1-1/Y}}^{\infty} e^{-t^{1/Y}v} \,
	\widetilde{\bbe}\left(e^{-t^{1/Y}\widetilde{U}_{1}}{\bf 1}_{\{Z_{1}\geq{}v\}}\right)dv, %\label{LOP3}
\end{align}
where $\widetilde{U}_{t}:=M^{*}\bar{U}_{t}^{+}-G^{*}\bar{U}_{t}^{-}$, {and $(\bar{U}^{+}_{t})_{t\geq{}0}$ and $(\bar{U}_{t}^{-})_{t\geq{}0}$ are defined as in (\ref{Uplusminus}).}
\end{lem}
The following result gives the second-order asymptotic behavior of at-the-money call option prices under the pure-jump CGMY model. 
\begin{thm}\label{2ndASY}
Under the exponential CGMY model (\ref{ExpLvMdl}) without Brownian component,
	\begin{equation}\label{CL2}
		 \lim_{t\to{}0}t^{\frac{1}{Y}-1}\left(t^{-\frac{1}{Y}}{\frac{1}{S_{0}}}{\mathbb{E}(S_t-S_{0})^{+}}-\widetilde{\bbe}(Z_{1}^{+})\right)={\,\vartheta+\eta+\frac{\tilde\gamma}{2}},%-\int_{0}^{\infty} \left(1-e^{-u}\right)u^{-Y}du
%		\int_{0}^{\infty}\widetilde{\bbe}\left(\widetilde{U}_{1}{\bf 1}_{\{Z_{1}\geq{}v\}}\right)dv.
	\end{equation}
	where {$\eta$ and $\tilde{\gamma}$ are respectively given as in (\ref{eta}) and (\ref{Cent}), and}
\begin{align}
	\label{vartheta}
	{\vartheta := -} C \Gamma(-Y) %{\frac{C \Gamma(1-Y)}{Y}}
%	\pi^{-1}\Gamma(Y)\sin(\pi Y)
	\Big((M^{*}+1)^{Y}+(G^{*})^{Y}\Big).
\end{align}
\end{thm}

\begin{rem}\label{Rem:ApproxPJ}
	Using (\ref{vartheta}), (\ref{eta}), and (\ref{Cent}), it turns out that
	\[
		d_{2}:=\vartheta+\eta+\frac{\tilde\gamma}{2}=\frac{C\Gamma(-Y)}{2} \left((M-1)^{Y}-M^{Y}-(G+1)^{Y}+G^{Y}\right).
	\]
	Hence, the second-order asymptotic behavior of the ATM call option price (\ref{CallPriceDfn}) in short-time is
	\begin{equation}\label{ExpAsymBehCGMY}
		\frac{1}{S_{0}}\mathbb{E}(S_t-S_{0})^{+}= d_{1} t^{{\frac{1}{Y}}}+d_{2} t +o(t), \qquad (t\to{}0),
	\end{equation}
	with $d_{1}=\widetilde{\bbe}(Z_{1}^{+})$ given as in (\ref{1stOrdFA}):
	\[
		d_{1}=\frac{1}{\pi}\Gamma\left(1-\frac{1}{Y}\right)\left(2 C \Gamma(-Y)\left|\cos\left(\frac{\pi Y}{2}\right)\right|\right)^{{\frac{1}{Y}}}.
	\]
Broadly speaking, the first-order term synthesizes only the information on the tail-heaviness index $Y$ and the overall jump-intensity parameter $C$, while the second-order term incorporates also the information on the relative intensities of negative and positive jumps (controlled by the parameters $G$ and $M$). Note also that $d_{2}< -C \Gamma(-Y)\leq -2 C$.
\end{rem}
\begin{rem}\label{ThirdOrderAppPJ}
The proof of Theorem \ref{2ndASY} (see Appendix \ref{proofA}) also provides the higher order asymptotics of the ATM call option price under the pure jump CGMY model. Indeed, it is clear that the term $D_{2}$ defined in (\ref{DecomD}) is such that ${D_{2}(t)\sim -\eta \widetilde{\bbe}(Z_{1}^{+})t}$. Moreover, the second-order term of $D_{1}(t)$ therein can be shown to be $O(t^{2-\frac{1}{Y}})$, while $D_{3}(t)=o(D_{2}(t))$, as $t\to{}0$. Therefore, as $t\to{}0$, {and since $1<Y<2$},
\begin{align}
\frac{1}{S_{0}}\mathbb{E}(S_t-S_{0})^{+}= d_{1} t^{{\frac{1}{Y}}}+d_{2} t-\eta {\widetilde{\bbe}(Z_{1}^{+}) }t^{1+\frac{1}{Y}}+o( t^{1+\frac{1}{Y}}).\label{HighAsyCGMY}
\end{align}
\end{rem}

Let $\hat{\sigma}(t)$ denote the ATM Black-Scholes implied volatility at maturity $t$ with zero interest rates. The following result gives the asymptotic behavior of $\hat{\sigma}(t)$ as $t\rightarrow 0$.
\begin{prop}\label{AsyIVPCGMY}
Under the exponential CGMY model (\ref{ExpLvMdl}) without Brownian component, the implied volatility $\hat{\sigma}(t)$ has the following small-time behavior:
\begin{align}\label{AsyIVPureCGMY}
\hat{\sigma}(t)=\sigma_{1}t^{\frac{1}{Y}-\frac{1}{2}}+\sigma_{2}t^{{\frac{1}{2}}}+o(t^{{\frac{1}{2}}}),\qquad t\rightarrow 0,
\end{align}
where
\begin{align}
\label{1stCoefIVPureCGMY}\sigma_{1}&:=\sqrt{2\pi}\,\widetilde{\bbe}(Z_{1}^{+}),\\
\label{2ndCoefIVPureCGMY}\sigma_{2}&:=\sqrt{\frac{\pi}{2}}C\Gamma(-Y)\left((M-1)^{Y}-M^{Y}-(G+1)^{Y}+G^{Y}\right).
\end{align}
\end{prop}

\section{The CGMY model with Brownian component}\label{Sect:NonZeroBrwn}

In this part, we consider the CGMY model with non-zero Brownian component. Concretely, throughout, $(X_{t})_{t\geq{}0}$ is a L\'evy process with triplet $(b,\sigma^{2},\nu)$ as introduced in Section \ref{Sec:Setup} and $\sigma\neq{}0$.
In that case, it follows from (\ref{CFVCGMY}) that
\[
	\lim_{t\rightarrow 0}\bbe^{*}\left(\exp(iu X_{t}/\sqrt{t})\right)=\exp{\left(-\frac{1}{2}\sigma^{2}u^{2}\right)},
\]
and, thus, $(X_{t}/{\sqrt{t}})$ converges weakly to the centered Gaussian distribution with variance $\sigma^{2}$. Equivalently, recalling that under $\bbp^{*}$, {$(W_{t}^{*})_{t\geq{}0}$ is} a standard Brownian motion, it follows that
\begin{align}\label{wcVsqrt}
\lim_{t\rightarrow 0}\bbp^{*}(X_{t}/{\sqrt{t}}\geq x)=\bbp^{*}({\sigma W_{1}^{*}}\geq x).
\end{align}
The first order asymptotic behavior for the ATM European call options in this mixed model was obtained in \cite{Tankov} using Fourier methods. We present, in Appendix \ref{ProofsSectGenCse}, a probabilistic proof based on (\ref{wcVsqrt}) and following an approach similar to that in \cite{FigForde:2012}.
\begin{prop}\label{1stOAsyCGMYB}
In the setting of Section \ref{Sec:Setup}, the at-the-money European call option price has the following asymptotic behavior:
\begin{align}\label{1stOApCGMYB}
\lim_{t\rightarrow 0}t^{-1/2}\bbe(S_{t}-S_{0})_{+}=S_{0}\sigma\bbe^{*}{(W_{1}^{*})_{+}}.
\end{align}
\end{prop}

Next, we give the second-order correction term for the at-the-money European call option price. As before, we change the probability measure $\bbp^{*}$ to $\widetilde{\bbp}$ so that $X_{t}=Z_{t}+t\tilde{\gamma}+{\sigma W_{t}^{*}}$, with $(Z_{t})_{t\geq{}0}$ a symmetric strictly $Y$-stable L\'evy process under $\widetilde{\bbp}$ (see (\ref{SSSP})) and $\tilde\gamma$ defined as in (\ref{Cent}). Recall also that, under both ${\bbp}^{*}$ and $\widetilde{\bbp}$, {$W^{*}$} is still a standard Brownian motion. We will also make use of the decompositions (\ref{DcmLL})-(\ref{RX}). The proof of the following result is presented in Appendix \ref{ProofsSectGenCse}.

\begin{thm}\label{2ndOAsyCGMYB}
In the setting of Section \ref{Sec:Setup}, the at-the-money European call option price is such that:
\begin{align}\label{2ndAsyCGMYB}
\lim_{t\rightarrow{}0}t^{\frac{Y}{2}-1}\Big( t^{-\frac{1}{2}}\frac{1}{S_{0}}\,\bbe(S_{t}-S_{0})_{+}-\sigma\bbe^{*}{(W_{1}^{*})_{+}}\Big)=\frac{C\sigma^{1-Y}}{Y(Y-1)}
\bbe^{*}\left({|W_{1}^{*}|^{1-Y}}\right).
\end{align}
\end{thm}

\begin{rem}\label{Rem:ApproxGen}
As well known, the $(1-Y)$-centered moment of a standard normal distribution is given by (see, e.g., (25.6) in \cite{Sato:1999}):
	\[
		{\bbe^{*}\left({|W_{1}^{*}|^{1-Y}}\right)=%\frac{2^{1-Y}\Gamma\left(\frac{1+1-Y}{2}\right)\Gamma\left(1-\frac{1-Y}{2}\right)}{\sqrt{\pi} \Gamma\left(1-\frac{1-Y}{2}\right)}
		\frac{2^{1-Y}}{\sqrt{\pi}}\Gamma\left(1-\frac{Y}{2}\right).}
	\]
Thus, the second-order asymptotic behavior of the ATM call option price (\ref{CallPriceDfn}) in short-time takes the form
	\begin{equation}\label{ExpAsymBehCGMYBM}
		\frac{1}{S_{0}}\,{\mathbb{E}(S_t-S_{0})^{+}= d_{1} t^{{\frac{1}{2}}}+d_{2} t^{\frac{3-Y}{2}} +o\left(t^{\frac{3-Y}{2}}\right), \qquad (t\to{}0),}
	\end{equation}
	with
	\begin{equation}\label{PrcDefn2ndTrm}
		d_{1}=\frac{\sigma}{\sqrt{2\pi}}, \qquad d_{2}=\frac{2^{1-Y}}{\sqrt{\pi}}\Gamma\left(1-\frac{Y}{2}\right)\frac{C\sigma^{1-Y}}{Y(Y-1)}.
	\end{equation}
	Intuitively, the first-order term only synthesizes the information about the continuous volatility parameter $\sigma$, while the second-order term incorporates also the information on the tail index parameter $Y$ and the overall jump-intensity parameter $C$. However, these two-terms do not reflect the relative intensities of negative or positive jumps (controlled by the parameters $G$ and $M$). This fact suggests that it could be necessary to develop a third-order approximation as described below.
\end{rem}

\begin{rem}\label{ThirdOrderAppGen}
The proof of Theorem \ref{2ndOAsyCGMYB} (See Appendix \ref{ProofsSectGenCse}) also provides higher order asymptotics of the ATM call option price under the {generalized} CGMY model. In fact, as mentioned in the proof, the second integral in the decomposition of $B_{t}$ is asymptotically equivalent to $(\tilde{\gamma}/2)\sqrt{t}$, while the last integral is clearly asymptotically equivalent to $(-\sigma^{2}/4)\sqrt{t}$. Then, it remains to analyze the first integral $A_{t}$ in (\ref{DecomAt}), which in the proof of Theorem \ref{2ndOAsyCGMYB} is decomposed into $I_{1}(t)$, $I_{2}(t)$ and $I_{3}(t)$. For $I_{1}(t)$, it can be shown that the second order term of $J_{12}(t,y)$ is $O(t^{2-Y})$ while the first order of $J_{11}(t,y)$ is $O(\sqrt{t})$. For $I_{2}(t)$, the first term in the decomposition (\ref{DecomI2}) is $o(\sqrt{t})$, while the second order term is $O(t^{2-Y})$. Finally, for $I_{3}(t)$, the second order of $J_{31}^{(2)}(t,y)$ and $J_{32}^{(2)}(t,y)$ {is} $O(t^{2-Y})$, while all the other terms in the decomposition of $I_{3}(t)$ are of order $o(\sqrt{t})$. Therefore, as $t\rightarrow 0$,
\begin{align}
\frac{1}{S_{0}}\,\mathbb{E}(S_t-S_{0})^{+}&= d_{1}t^{{\frac{1}{2}}}+d_{2}t^{\frac{3-Y}{2}}+d_{3}t+o(t),\qquad1<Y\leq\frac{3}{2},\label{HighAsyCGMYB1}\\
\frac{1}{S_{0}}\,\mathbb{E}(S_t-S_{0})^{+}&= d_{1}t^{{\frac{1}{2}}}+d_{2}t^{\frac{3-Y}{2}}+ d_{3}t^{\frac{5}{2}-Y}+o(t^{\frac{5}{2}-Y}),\qquad\frac{3}{2}<Y<2,\label{HighAsyCGMYB2}
\end{align}
where $d_{3}$ can be explicitly derived.
\end{rem}

The next proposition gives the small-time asymptotic behavior for the ATM Black-Scholes implied volatility, again denoted by $\hat\sigma(t)$, under the {generalized} CGMY model. Unlike the pure-jump case, we can only derive the first order asymptotics using Theorem \ref{2ndOAsyCGMYB}. In fact, the first order term of the ATM call option price under the {generalized} CGMY model is the same as the one under the Black-Scholes model. The second order term of $\hat{\sigma}_{t}$ requires higher order asymptotics of the ATM call option price. The proof is deferred to Appendix \ref{ProofsSectGenCse}.
\begin{prop}\label{AsyIVCGMYB}
Under the exponential CGMY model (\ref{ExpLvMdl}) with non-zero Brownian component, the implied volatility $\hat{\sigma}$ is such that:
\begin{align}\label{AsyIVGerCGMY}
{\hat{\sigma}(t)=\sigma+\frac{C2^{\frac{3}{2}-Y}\sigma^{1-Y}}{Y(Y-1)}\Gamma\left(1-\frac{Y}{2}\right)t^{1-\frac{Y}{2}}+o\left(t^{1-\frac{Y}{2}}\right),\quad t\rightarrow 0.}
\end{align}
\end{prop}

\section{Numerical examples}\label{Numerics}

In this part, we assess the performance of the previous approximations through a detailed numerical analysis.

\subsection{The numerical methods}

Let us first select a suitable numerical method to compute the ATM option prices by considering two methods: Inverse Fourier Transform (IFT) and Monte Carlo (MC).

Before introducing the IFT method, let us set some notations. The characteristic function corresponding to the Black-Scholes model with volatility $\Sigma$ is given by
\[
	\varphi_{t}^{BS,\Sigma}(u)=\exp\left(-\frac{\Sigma^{2} t}{2}\left(v^{2}+iv\right)\right).
\]
The corresponding call option price at the log-moneyness {$k=\log(S_{0}/K)$} under the Black-Scholes model with volatility $\Sigma$ is denoted by {$C_{BS}^{\Sigma}(k)$} ; that is,
\[
	{{C_{BS}^{\Sigma}(k)}=S_{0}e^{-r t} \mathbb{E}(e^{(r-\Sigma^{2}/2)t+\Sigma W_{t}}-e^{k})_{+}}.
\]
Let us also recall that the characteristic function under the {generalized} CGMY model is denoted by $\varphi_{t}$ (see (\ref{CFVCGMY})) and let us denote the corresponding call option price at log-moneyness {$k$ by $C(k)$}.
The IFT method is based on the following inversion formula (see Section 11.1.3 in \cite{CT04}):
\begin{equation}\label{InvFormOptP}
	{z_{_{T}}(k):=C(k) - C_{BS}^{\Sigma}(k)=\frac{1}{2\pi} \int_{-\infty}^{\infty}e^{-i v k} \zeta_{_{T}}(v) dv},
\end{equation}
where
\begin{align}\label{DfnZeta}
	\zeta_{_{T}}(v)&:=e^{ivr} \frac{\varphi_{_{T}}(v-i )- \varphi_{_{T}}^{BS,\Sigma}(v-i)}{iv(1+iv)}.
\end{align}
In our case, we fix $r=0$ and, since we are only interested in ATM option prices, we set {$k=0$}. In order to compute numerically the integral in (\ref{InvFormOptP}), we use the Simpson's rule:
\[
	z_{_{T}}(0):=\frac{1}{2\pi} \int_{-\infty}^{\infty} \zeta_{_{T}}(v) dv=\Delta \sum_{m=0}^{{P-1}} w_{m} \zeta_{_{T}}(v_{m}),
\]
with {$\Delta=Q/(P-1)$, $v_{m}=-Q/2+m\Delta$}, and $w_{0}=1/2$, {$w_{2 \ell-1}=4/3$, and $w_{2 \ell}=2/3$, for {$\ell=1,\dots,P/2$}}.

We also consider a Monte Carlo method based on the risk-neutral option price representation under the probability measure $\widetilde{\bbp}$. Under this probability measure and using the notation (\ref{Uplusminus}) as well as the relations (\ref{DcmLL}) and (\ref{RX}), we have:
\begin{align*}
	\bbe(e^{X_{T}}- 1)_{+}&={\bbe}^{*}\left(e^{-X_{T}}\left(e^{X_{T}}-1\right)_{+}\right)= \widetilde{\bbe}\left(e^{-U_{T}}\left(1-e^{-X_{T}}\right)_{+}\right)\\
	&=\widetilde{\bbe}\left(e^{-M^{*}\bar{U}^{+}_{T}+G^* \bar{U}^{-}_{T}-\eta T}\left(1-e^{-\bar{U}^{+}_{T}-\bar{U}^{-}_{T}-T \tilde\gamma -{\sigma W_{T}^{*}}}\right)_{+}\right),
\end{align*}
which can be easily computed by Monte Carlo method using the fact that, under $\widetilde{\bbp}$, the variables $\bar{U}^{+}_{T}$ and $-\bar{U}^{-}_{T}$ are independent $Y$-stable random variables with  scale, skewness, and location parameters $T C|\cos(\pi Y/2)|\Gamma(-Y)$, $1$, and $0$, respectively. Standard simulation methods are available to generate stable random variables.
We consider the following set of parameters for the CGMY component:
\[
	{C=0.5,\quad G=2,\quad M=3.6,\quad Y=1.5}.
\]
Figure \ref{Figure1} compares the first- and second-order approximations as given in Remarks \ref{Rem:ApproxPJ} and \ref{Rem:ApproxGen} to the prices based on the Inverse Fourier Transform (IFT-based price) and the Monte Carlo method (MC-based price) under both the pure-jump case and the {generalized} CGMY case with $\sigma=0.4$. For the MC-based price, we use $100,000$ simulations, while for the IFT-based method, we use {$P=2^{14}$} and {$Q=800$}. As it can be seen, it is not easy to integrate numerically the characteristic function (\ref{DfnZeta}) since in this case $T$ is quite small and, therefore, the characteristic functions $\varphi_{T}$ and $\varphi_{T}^{BS,\Sigma}$ are quite flat. The Monte Carlo method turns out to be much more accurate and faster.
\begin{figure}[htp]
    {\par \centering
    \includegraphics[width=8.0cm,height=9.0cm]{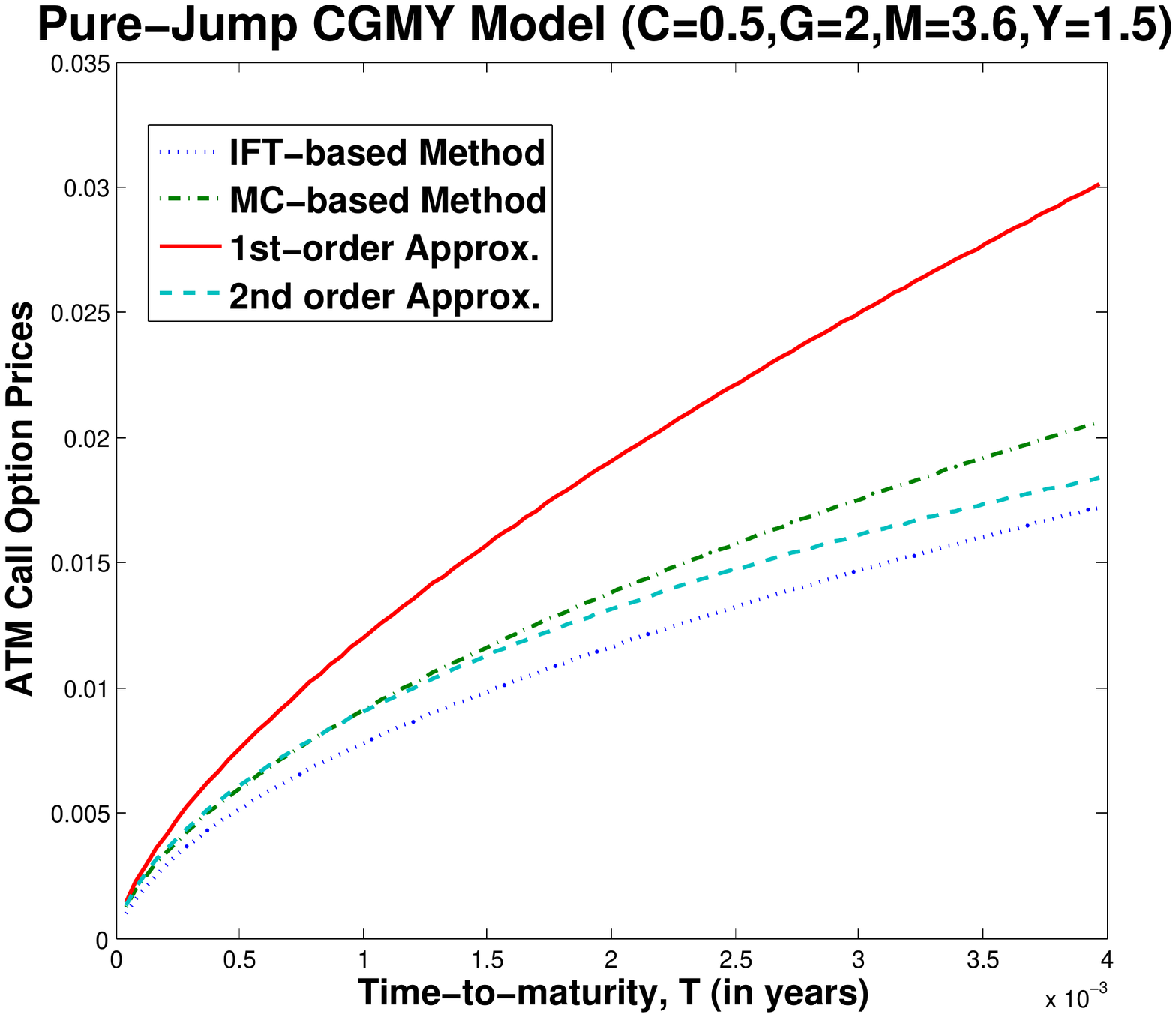}
    \includegraphics[width=8.0cm,height=9.0cm]{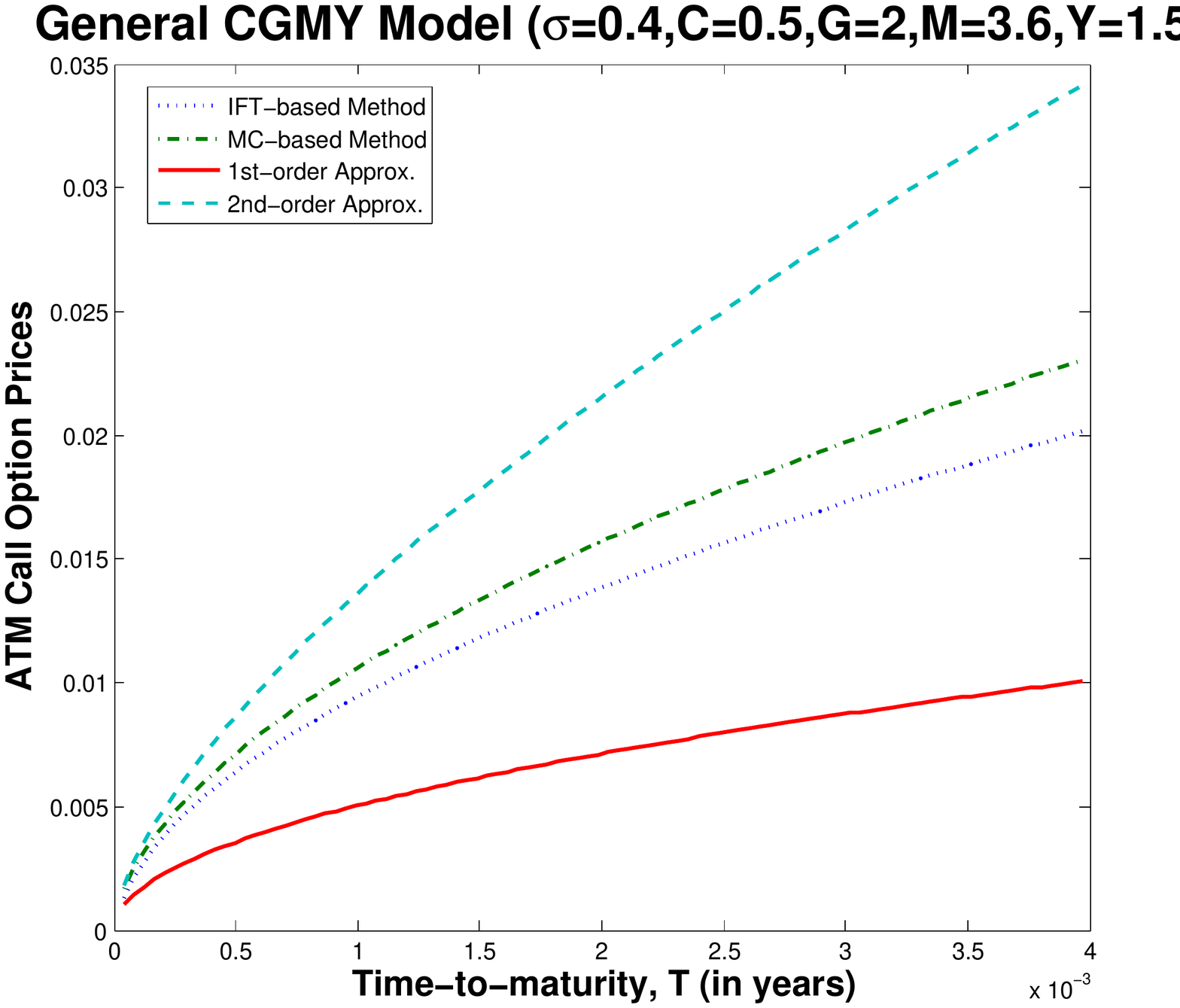}
    \par}\vspace{-1.2 cm}
    \caption{
    Comparisons of ATM call option prices for two methods (Inverse Fourier Transform and Monte-Carlo method) with the first- and second-order approximations. MC-based price is based on $100,000$ simulations while the IFT-based method is based on the parameter values {$P=2^{14}$} and {$Q=800$}. The parameter $\sigma$ in the {generalized} CGMY model is set to be $0.1$.}\label{Figure1}
\end{figure}

\subsection{Results for different parameter settings}

Here, we investigate the performance of the approximations for different settings of parameters:
\begin{enumerate}
	\item Figure \ref{Figure2} compares the 1st- and 2nd-order approximations with the MC prices for different values of $C$, fixing the values of all the other parameters.  In the pure-jump case, the 2nd order approximation is significantly better for moderately small values of $C$, but for larger values of $C$, this is not the case unless $T$ is extremely small. For a nonzero continuous component, the 1st order approximation is extremely bad as it only takes into account the parameter $\sigma$.

	\item Figure \ref{Figure3} compares the 1st- and 2nd-order approximations with the MC prices for different values of $Y$, fixing the values of all the other parameters.  In both cases, the 2nd order approximation is significantly better for values of $Y$ around $1.5$, which is consistent with the observation that $|d_{2}|\to\infty$ as $Y\to{}1$ or $Y\to{}2$. For a nonzero continuous component, the 1st order approximation is again extremely bad as compared to the 2nd order approximation.
	
	\item In the left panel of Figure \ref{Figure4}, we analyze the effect of the relative intensities of negative jumps compared to positive jumps in the pure-jump CGMY case. That is, we fix the values $M$ to be $4$ and consider different values for $G$.  As expected, since the first order approximation does not take into account this information, the 2nd-order approximation performs significantly better.
	
	\item In the right panel of Figure \ref{Figure4}, we analyze the effect of the volatility of the continuous component in the {generalized} CGMY case. The 2nd order approximation is, in general, much better than the 1st-order approximation and, interestingly enough, the quality of the 2nd order approximations improves as the values of $\sigma$ increases. In fact, it seems that the 2nd-order approximation and the MC prices collapse to a steady curve as $\sigma$ increases.
	\end{enumerate}
\begin{figure}[htp]
    {\par \centering
    \includegraphics[width=8.0cm,height=9.0cm]{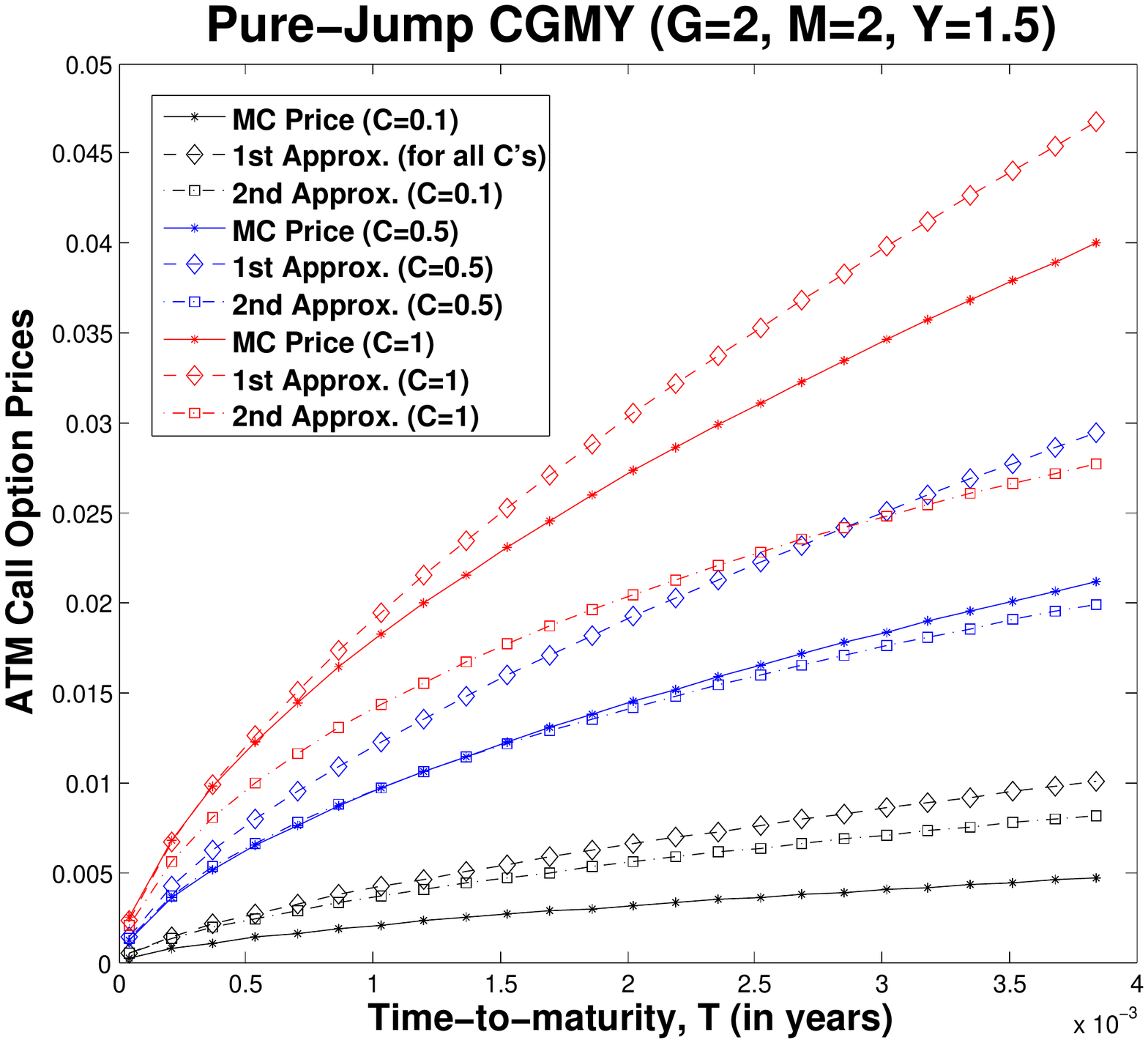}
    \includegraphics[width=8.0cm,height=9.0cm]{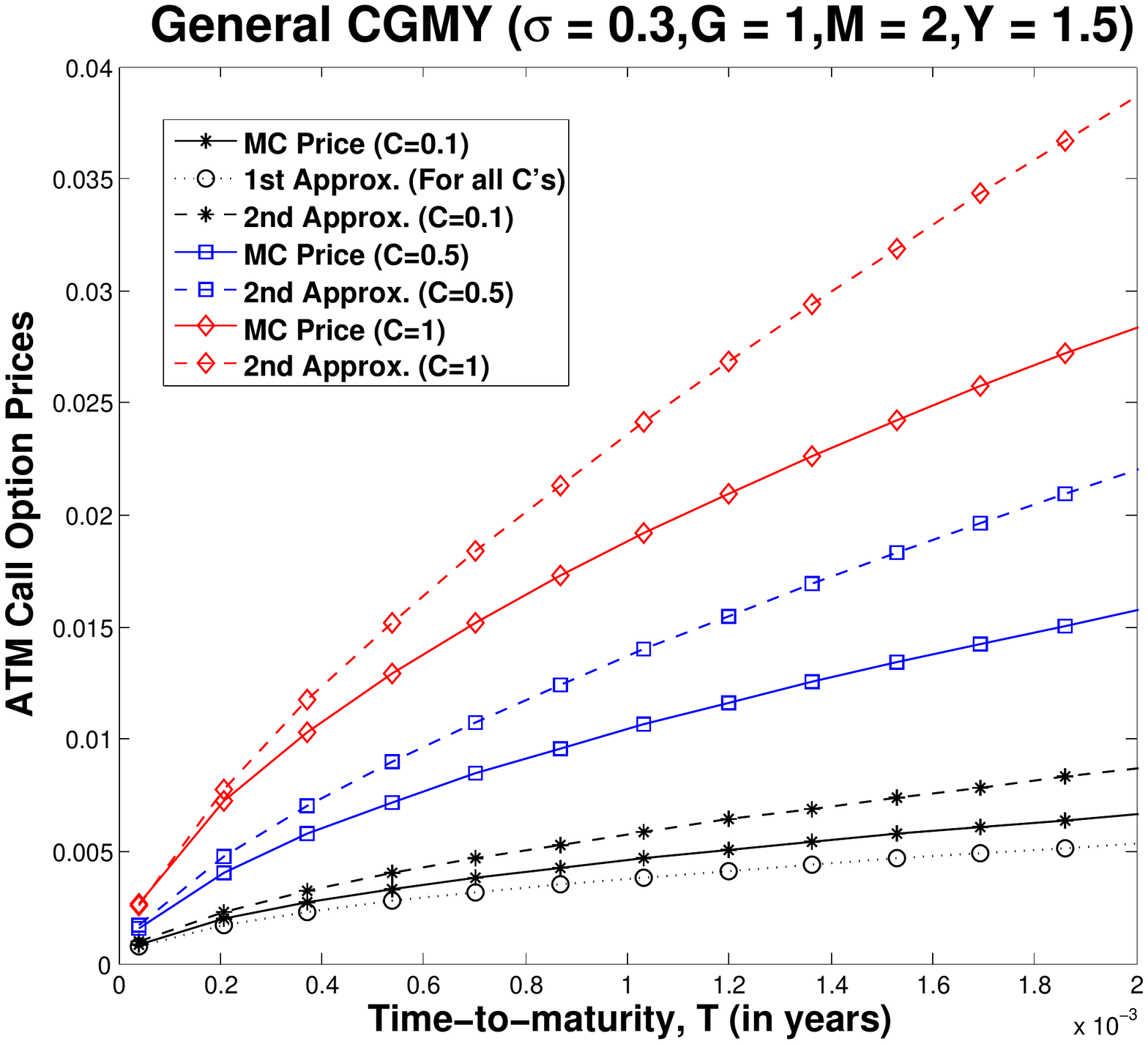}
    \par}\vspace{-1.2 cm}
    \caption{
    Comparisons of ATM call option prices with the short-time approximations for different values of the jump intensity parameter $C$.}\label{Figure2}
\end{figure}

\begin{figure}[htp]
    {\par \centering
    \includegraphics[width=8.0cm,height=9.0cm]{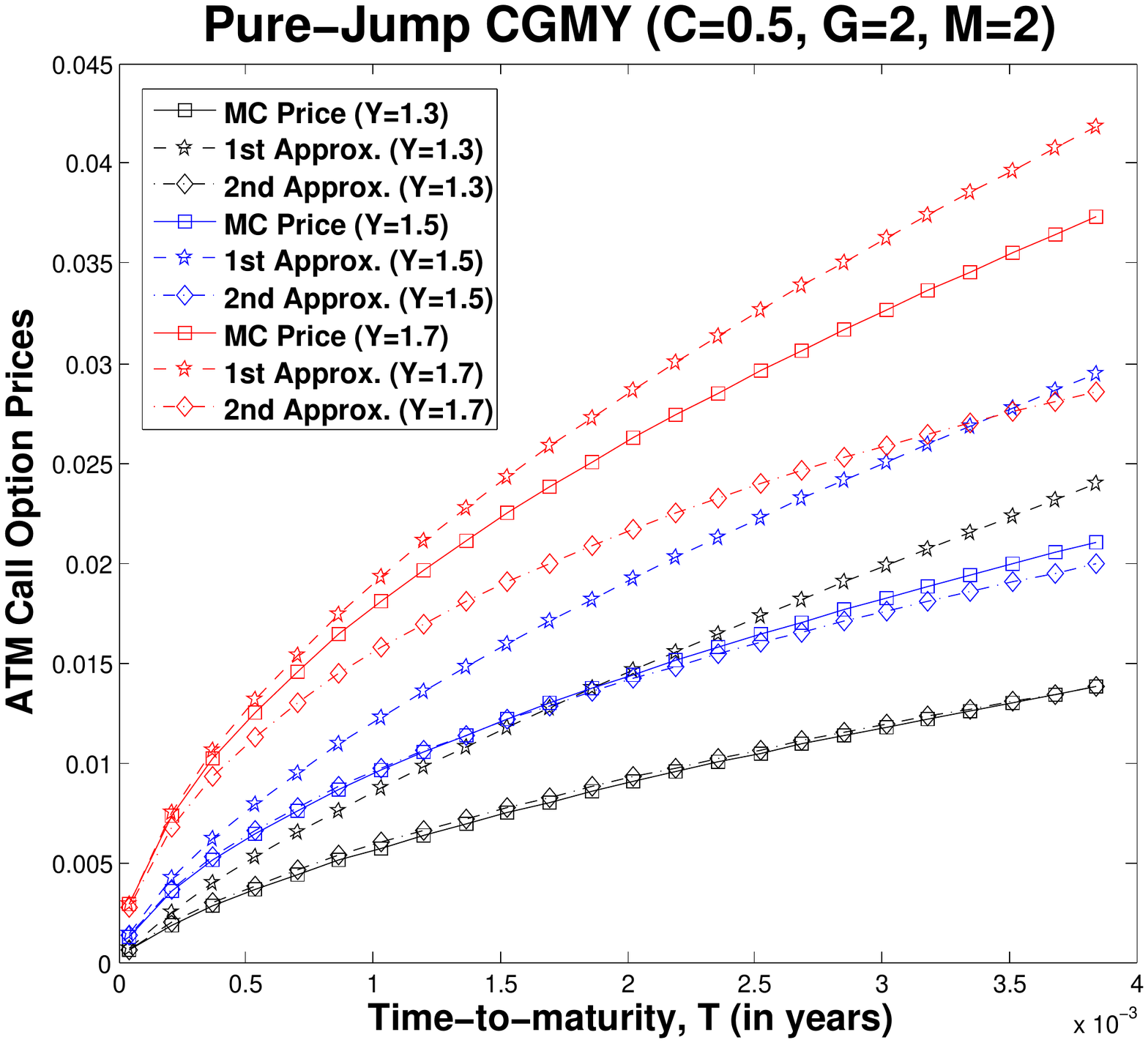}
    \includegraphics[width=8.0cm,height=9.0cm]{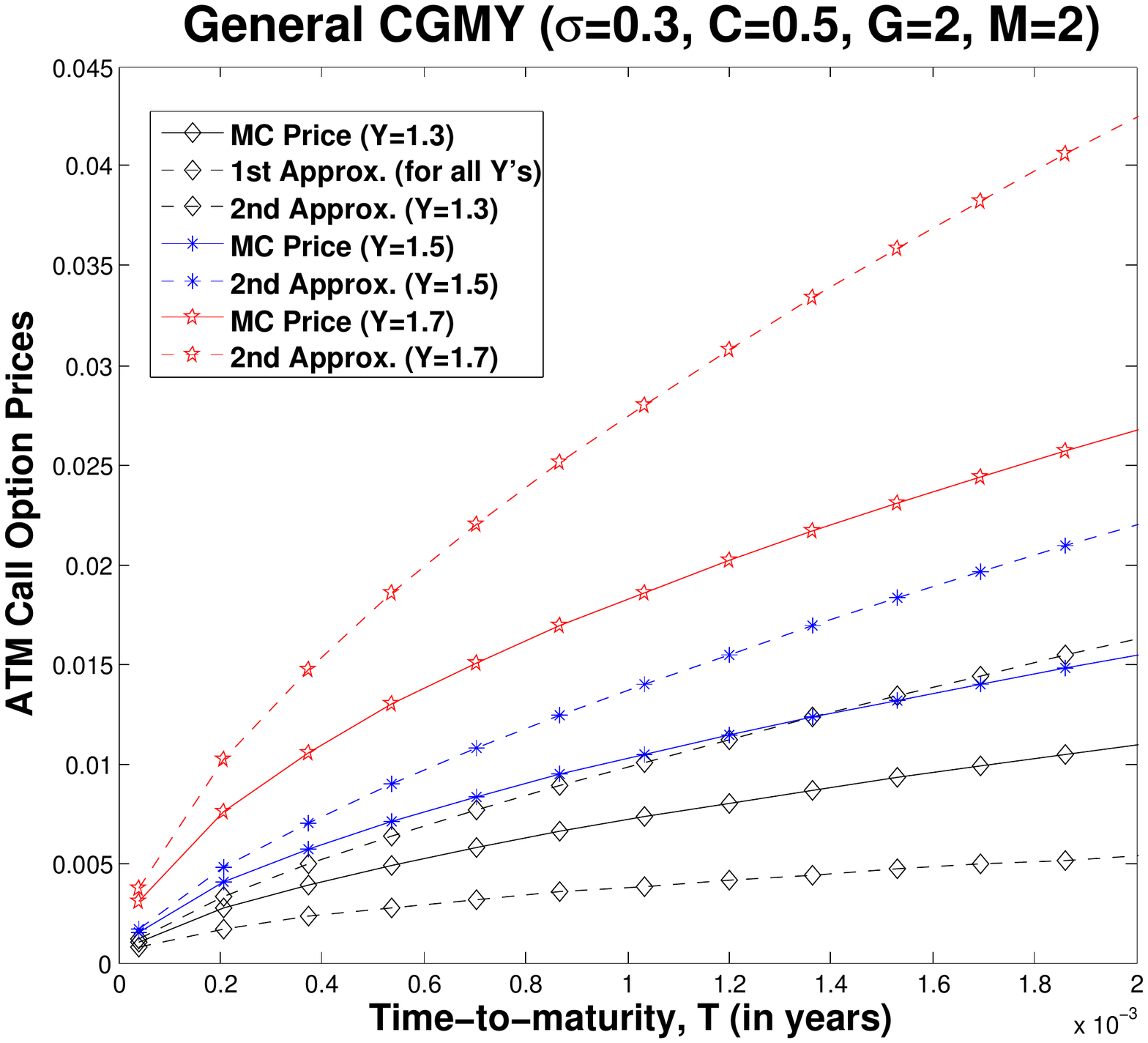}
    \par}\vspace{-1.2 cm}
    \caption{
    Comparisons of ATM call option prices with the short-time approximations for different values of the tail-heaviness parameter $Y$.}\label{Figure3}
\end{figure}

\begin{figure}[htp]
    {\par \centering
    \includegraphics[width=8.0cm,height=9.0cm]{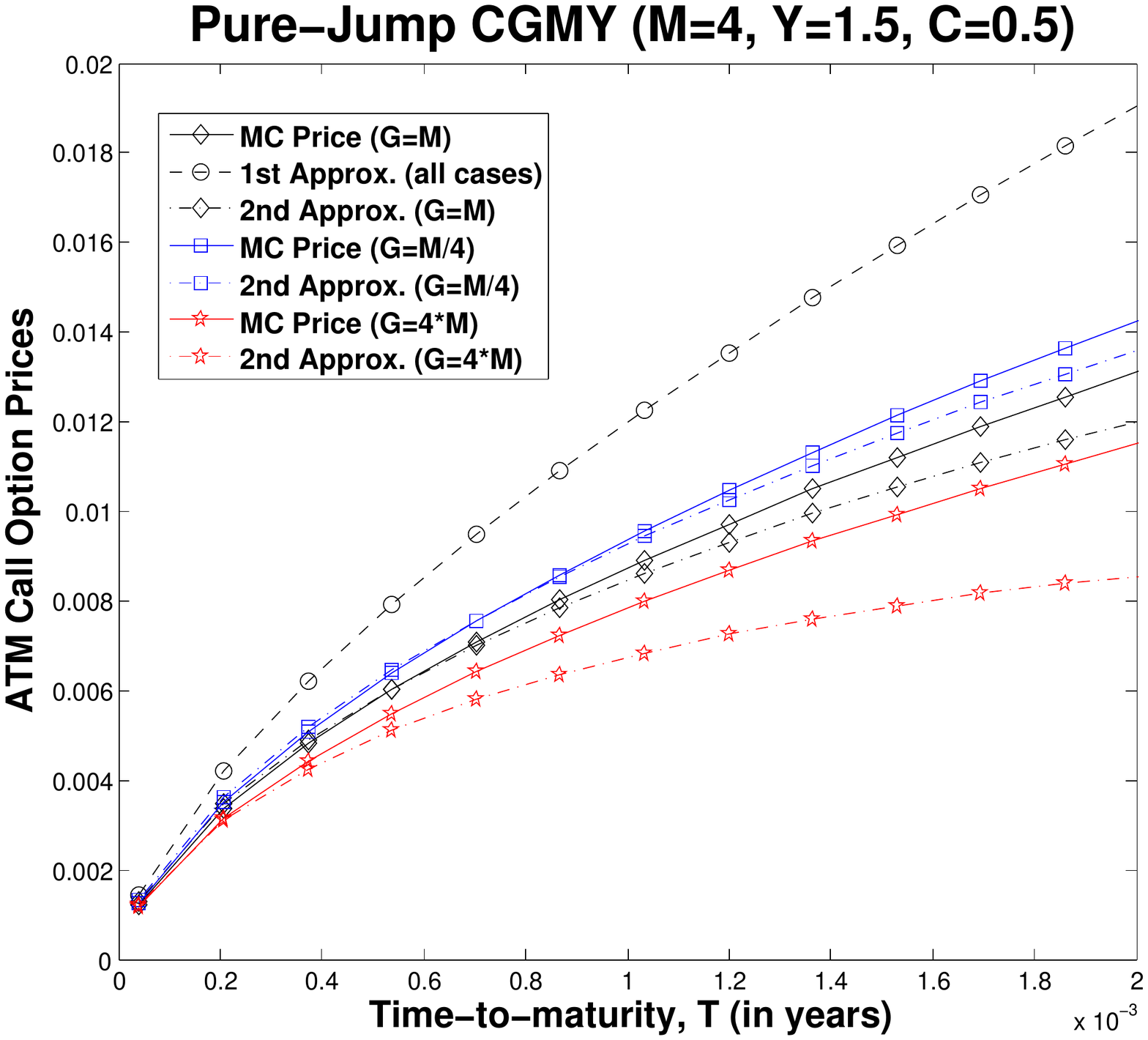}
    \includegraphics[width=8.0cm,height=9.0cm]{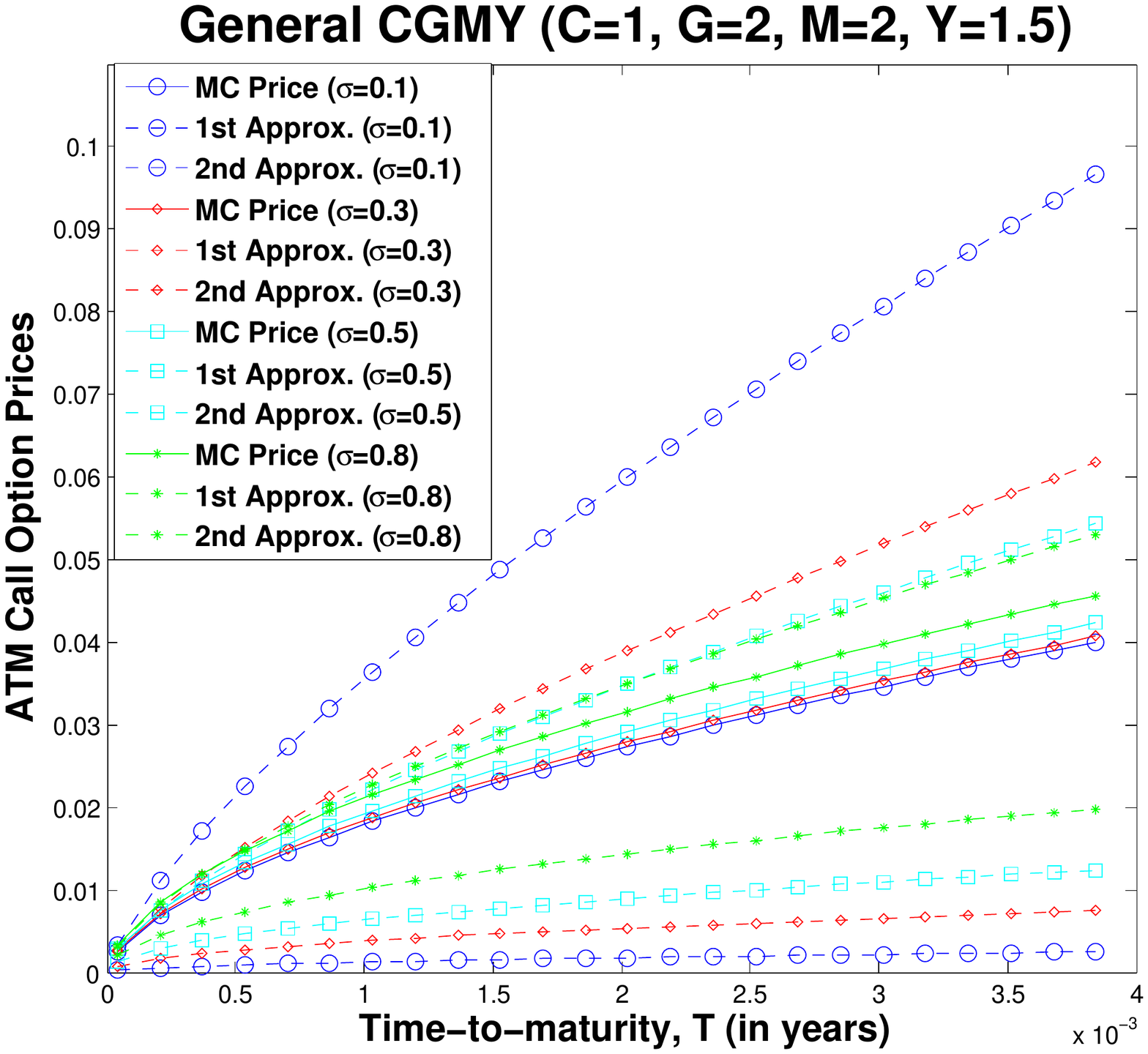}
    \par}\vspace{-1.2 cm}
    \caption{
    Comparisons of ATM call option prices with the short-time approximations for different values of {$G$ or $M$} and different values of the volatility parameter $\sigma$.}\label{Figure4}
\end{figure}

\newpage
\appendix
\section{Proofs of Section \ref{Sec:PureJump}: Pure-jump CGMY model}\label{proofA}

{For simplicity, throughout this section, we fix $S_{0}=1$.}

%\medskip
\noindent
\textbf{Proof of Lemma \ref{Lm:NRATMPJ}.}

\noindent
From (\ref{CMR}), we have
\begin{align}\nn	 t^{-1/Y}\mathbb{E}(S_t-S_{0})^{+}&=t^{-1/Y}\mathbb{P}^{*}(X_{t}\geq{}E)=t^{-1/Y}\int_{0}^{\infty} e^{-x} \mathbb{P}^{*}(X_{t}\geq{}x)dx\\
	&=\int_{0}^{\infty} e^{-t^{1/Y}u} \mathbb{P}^{*}(t^{-1/Y}X_{t}\geq{}u)du.\label{LOP1}
\end{align}	
Next, using {the change of probability measure} (\ref{EMMN}),
\begin{align*}
	\mathbb{P}^{*}(t^{-1/Y}X_{t}\geq{}u)&=\bbe^{*}\left({\bf 1}_{\{t^{-1/Y}X_{t}\geq{}u\}}\right)=\widetilde{\bbe}\left(e^{-U_{t}}{\bf 1}_{\{t^{-1/Y}X_{t}\geq{}u\}}\right),
\end{align*}
and, moreover, since $\sigma=0$, (\ref{SSSP}), and (\ref{DcmLL}),
\begin{align*}
	\mathbb{P}^{*}(t^{-1/Y}X_{t}\geq{}u)&=e^{-\eta t}\widetilde{\bbe}\left(e^{-\widetilde{U}_{t}}{\bf 1}_{\{t^{-1/Y}Z_{t}\geq{}u-\tilde{\gamma} t^{1-1/Y}\}}\right),
\end{align*}
with $\widetilde{U}_{t}:=M^{*}\bar{U}_{t}^{+}-G^{*}\bar{U}_{t}^{-}$.  By the self-similarity property (\ref{SSCN}),
\begin{equation}\label{KR1}
	\mathbb{P}^{*}(t^{-1/Y}X_{t}\geq{}u)=e^{-\eta t}\widetilde{\bbe}\left(e^{-t^{1/Y}\widetilde{U}_{1}}{\bf 1}_{\{Z_{1}\geq{}u-\tilde{\gamma} t^{1-1/Y}\}}	\right).
\end{equation}
{Thus,} plugging (\ref{KR1}) into (\ref{LOP1}),
\begin{align}
	t^{-1/Y}\mathbb{E}(S_t-S_{0})^{+}&=\int_{0}^{\infty} e^{-t^{1/Y}u-\eta t} \,
	\widetilde{\bbe}\left(e^{-t^{1/Y}\widetilde{U}_{1}}{\bf 1}_{\{Z_{1}\geq{}u-\tilde{\gamma} t^{1-1/Y}\}}\right)du,
\end{align}
and, changing variables $\left(v=u-\tilde{\gamma} t^{1-1/Y}\right)$, the result follows. \qed

\subsubsection*{Proof of Theorem \ref{2ndASY}.}
%\noindent\textbf{Proof of Theorem \ref{2ndASY}:}
To begin with, we assume that $M-1=G$, so that $\tilde{\gamma}=0$ (see (\ref{Cent})) and, in light of Lemma \ref{Lm:NRATMPJ},
\begin{align}\label{SmplOP}
	t^{-1/Y}\mathbb{E}(S_t-S_{0})^{+}=e^{-\eta t}\int_{0}^{\infty} e^{-t^{1/Y}v}
	\widetilde{\bbe}\left(e^{-t^{1/Y}\widetilde{U}_{1}}{\bf 1}_{\{Z_{1}\geq{}v\}}\right)dv. %\label{LOP3}
\end{align}
The general case is resolved in Lemma \ref{GenCseGam} below.
Let
	\[
		 D(t):=t^{-1/Y}\mathbb{E}(S_t-S_{0})^{+}-\widetilde{\bbe}(Z_{1}^{+}),
	\]
	which can be written as
	\begin{align}
		D(t)&:=\int_{0}^{\infty} e^{-t^{1/Y}v}
	\widetilde{\bbe}\left(e^{-t^{1/Y}\widetilde{U}_{1}}{\bf 1}_{\{Z_{1}\geq{}v\}}\right)dv-\widetilde{\bbe}(Z_{1}^{+})\nonumber\\
	&\quad\, +(e^{-\eta t}-1) {\widetilde{\bbe}}(Z_{1}^{+})\nonumber\\
	\label{DecomD}&\quad\, +(e^{-\eta t}-1)\left(\int_{0}^{\infty} e^{-t^{1/Y}v}
	\widetilde{\bbe}\left(e^{-t^{1/Y}\widetilde{U}_{1}}{\bf 1}_{\{Z_{1}\geq{}v\}}\right)dv-{\widetilde{\bbe}}(Z_{1}^{+})\right)\\
&=:D_{1}(t)+D_{2}(t)+D_{3}(t).\nonumber
	\end{align}
	We will show that
	\begin{equation}\label{KNLFMP}
	 	t^{1/Y-1}D_{1}(t)\longrightarrow{} \vartheta+ \eta,\qquad {\text{as}\quad t\to{}0},
	\end{equation}
	while it is clear that $D_{3}(t)=o(D_{1}(t))$ and $t^{1/Y-1}D_{2}(t)=o(1)$, as $t\to{}0$.
	First, {note that in light of (\ref{expUt})},
\begin{align}
D_{1}(t)&=\widetilde{\bbe}\left(e^{-t^{1/Y}\widetilde{U}_{1}}\int_{0}^{Z_{1}^{+}}e^{-t^{1/Y}v}dv\right)-\widetilde{\bbe}\big(Z_{1}^{+}\big)\nonumber\\
&{\,=t^{-1/Y}\left(\widetilde{\bbe}\Big(e^{-t^{1/Y}\widetilde{U}_{1}}\Big)-\widetilde{\bbe}\Big(e^{-t^{1/Y}\big(\widetilde{U}_{1}+Z_{1}^{+}\big)}\Big)\right)-\widetilde{\bbe}(Z_{1}^{+})}\nonumber \\
\label{D1}&=t^{-1/Y}\left(e^{\eta t}-\widetilde{\bbe}\Big(e^{-t^{1/Y}\big(\widetilde{U}_{1}+Z_{1}^{+}\big)}\Big)\right)-\widetilde{\bbe}(Z_{1}^{+}).
\end{align}
Thus,
\begin{align}
t^{1/Y-1}D_{1}(t)
%&=t^{1/Y-1}\left[\frac{e^{\eta t}-1}{t^{1/Y}}+\frac{1-\widetilde{\bbe}(e^{-t^{1/Y}(Z_{1}^{+}+\widetilde{U}_{1})})}{t^{1/Y}}-\bbe(Z_{1}^{+})\right]\nonumber\\
&=t^{1/Y-1}\left(\frac{e^{\eta t}-1}{t^{1/Y}}+\frac{1-\widetilde{\bbe}(e^{-t^{1/Y}(Z_{1}^{+}+\widetilde{U}_{1})})}{t^{1/Y}}-{\widetilde{\bbe}}(Z_{1}^{+}+\widetilde{U}_{1})\right)\nonumber\\
&=\frac{e^{\eta t}-1}{t}+\widetilde{\bbe}\int_{0}^{Z_{1}^{+}+\widetilde{U}_{1}}\frac{e^{-t^{1/Y}v}-1}{t^{1-1/Y}}dv{\bf 1}_{\{Z_{1}^{+}+\widetilde{U}_{1}\geq 0\}}-\widetilde{\bbe}\int_{Z_{1}^{+}+\widetilde{U}_{1}}^{0}\frac{e^{-t^{1/Y}v}-1}{t^{1-1/Y}}dv{\bf 1}_{\{Z_{1}^{+}+\widetilde{U}_{1}\leq 0\}}\nonumber\\
&=\underbrace{\frac{e^{\eta t}-1}{t}}_{D_{11}(t)}+\underbrace{\int_{0}^{\infty}\frac{e^{-t^{1/Y}v}-1}{t^{1-1/Y}}\widetilde{\bbp}(Z_{1}^{+}+\widetilde{U}_{1}\geq v)dv}_{D_{12}(t)}-\underbrace{\int_{0}^{\infty}\frac{e^{t^{1/Y}v}-1}{t^{1-1/Y}}\widetilde{\bbp}(Z_{1}^{+}+\widetilde{U}_{1}\leq -v)dv}_{D_{13}(t)}.
\label{DecomD1}
%\\
%&:=D_{11}(t)+D_{12}(t)+D_{13}(t).\nonumber
\end{align}
%We denote $D_{11}(t)$, $D_{12}(t)$, and $D_{13}(t)$ each of the three term on right-hand side of the last equality.
Clearly,
\begin{align}
\label{limD11} D_{11}(t)\rightarrow\eta,\quad \text{ as }\quad t\to 0.
\end{align}
Next, for $D_{13}(t)$, note that
\begin{align}\label{negtailZU}
\widetilde{\bbp}(Z_{1}^{+}+\widetilde{U}_{1}\leq -v)\leq\widetilde{\bbp}(\widetilde{U}_{1}\leq -v)\leq\frac{\widetilde{\bbe}\big(e^{-\widetilde{U}_{1}}\big)}{e^{v}}=e^{\eta-v},
\end{align}
and, {since $0<e^{y}-1\leq ye^{y}$ for $y>0$,}
\begin{align}
\frac{|e^{t^{1/Y}v}-1|}{t^{1-1/Y}}\,\widetilde{\bbp}(Z_{1}^{+}+\widetilde{U}_{1}\leq -v)\leq t^{2/Y-1}e^{\eta}ve^{(t^{1/Y}-1)v}\leq e^{\eta}ve^{-v/2},\nonumber
\end{align}
{for $t>0$ small enough.} The dominated convergence theorem then implies that
\begin{align}
\label{limD13} D_{13}(t)\rightarrow 0,\quad \text{ as } \quad t\to 0.
\end{align}
Finally, let us analyze the term $D_{12}(t)$.
Changing variables $\left(u=t^{1/Y}v\right)$,
\begin{align}
D_{12}(t)=t^{-1}\int_{0}^{\infty}(e^{-u}-1)\widetilde{\bbp}\Big(Z_{1}^{+}+\widetilde{U}_{1}\geq t^{-1/Y}u\Big)du,\nonumber
\end{align}
and, from (\ref{UIn1}) and (\ref{KIn}), there exists, as shown in Appendix \ref{AddtnProofs}, {a} constant $\tilde{\kappa}<\infty$  such that
\begin{equation} \frac{1}{t}\widetilde{\bbp}\Big(Z_{1}^{+}\!+\!\widetilde{U}_{1}\!\geq\!t^{-1/Y}u\Big)\leq\widetilde{\kappa}u^{-Y},\label{TIITI}
\end{equation}
for any $0<t\leq{}1$ and  $u>0$.
Hence, by the dominated convergence theorem,
\begin{align}\label{D12Domi}
\lim_{t\to 0}D_{12}(t)=\int_{0}^{\infty}(e^{-u}-1)\lim_{t\to 0}\left(t^{-1}\widetilde{\bbp}\Big(Z_{1}^{+}+\widetilde{U}_{1}\geq t^{-1/Y}u\Big)\right)du.
\end{align}
To find $\lim_{t\to 0}t^{-1}\widetilde{\bbp}(Z_{1}^{+}+\widetilde{U}_{1}\geq t^{-1/Y}u)$, note that
\begin{align}
\widetilde{\bbp}\left(Z_{1}^{+}+\widetilde{U}_{1}\geq\frac{u}{t^{1/Y}}\right)&=\underbrace{\widetilde{\bbp}\Big(Z_{1}+\widetilde{U}_{1}\geq\frac{u}{t^{1/Y}},Z_{1}\geq 0\Big)}_{I_{1}}+\underbrace{\widetilde{\bbp}\Big(\widetilde{U}_{1}\geq\frac{u}{t^{1/Y}},Z_{1}<0\Big)}_{I_{2}}.\nonumber
\end{align}
Then,
\begin{align}
I_{1}&=\widetilde{\bbp}\Big((M^{*}+1)\bar{U}_{1}^{+}-(G^{*}-1)\bar{U}_{1}^{-}\geq t^{-1/Y}u,\bar{U}_{1}^{+}+\bar{U}_{1}^{-}\geq 0\Big)\nonumber\\
%&=\widetilde{\bbp}\Big(\bar{U}_{1}^{+}\geq\frac{t^{-1/Y}u+(G^{*}-1)\bar{U}_{1}^{-}}{M^{*}+1},\,\bar{U}_{1}^{+}\geq -\bar{U}_{1}^{-}\Big)\nonumber\\
&=\widetilde{\bbp}\Big(\bar{U}_{1}^{+}\geq\frac{t^{-1/Y}u+(G^{*}-1)\bar{U}_{1}^{-}}{M^{*}+1}\geq -\bar{U}_{1}^{-}\Big)+\widetilde{\bbp}\Big(\bar{U}_{1}^{+}\geq -\bar{U}_{1}^{-}\geq\frac{t^{-1/Y}u+(G^{*}-1)\bar{U}_{1}^{-}}{M^{*}+1}\Big)\nonumber\\
&=\widetilde{\bbp}\left(\bar{U}_{1}^{+}\geq\frac{t^{-1/Y}u+(G^{*}-1)\bar{U}_{1}^{-}}{M^{*}+1},\,-\bar{U}_{1}^{-}\leq\frac{t^{-1/Y}u}{M^{*}+G^{*}}\right)+\widetilde{\bbp}\left(\bar{U}_{1}^{+}\geq -\bar{U}_{1}^{-}\geq\frac{t^{-1/Y}u}{M^{*}+G^{*}}\right) \nonumber\\
&:=I_{1,1}(t)+I_{1,2}(t).\nonumber
\end{align}
By the independence of $\bar{U}_{1}^{+}$ and $\bar{U}_{1}^{-}$ and the estimate (\ref{Utail}),
\begin{align}
 I_{1,2}(t)\leq\widetilde{\bbp}\left(\bar{U}_{1}^{+}\geq\frac{t^{-1/Y}u}{M^{*}+G^{*}}\right)\widetilde{\bbp}\left(-\bar{U}_{1}^{-}\geq\frac{t^{-1/Y}u}{M^{*}+G^{*}}\right)=O(t^{2}),
\label{limI12}
\end{align}
as $t\rightarrow 0$. Note also that
\begin{align*}
&\frac{1}{t}\widetilde{\bbp}\left(\bar{U}_{1}^{+}\!\geq\!\frac{t^{-\frac{1}{Y}}u\!-\!(G^{*}\!-\!1)y}{M^{*}+1}\right){\bf 1}_{\{y\leq\frac{t^{-1/Y}u}{M^{*}+G^{*}}\}}\leq{\frac{1}{t}\widetilde{\bbp}\left(\bar{U}_{1}^{+}\!\geq\!\frac{t^{-\frac{1}{Y}}u\!-\!(G^{*}\!-\!1)\frac{t^{-1/Y}u}{M^*+G^*}}{M^{*}+1}\right){\bf 1}_{\{y\leq\frac{t^{-1/Y}u}{M^{*}+G^{*}}\}}\leq\frac{1}{t}\widetilde{\bbp}\left(\bar{U}_{1}^{+}\!\geq\!\frac{t^{-\frac{1}{Y}}u}{M^*\!+\!G^*}\right),}
\end{align*}
and so, recalling that $p(1,y)$ denotes the density of $-\bar{U}^{-}_{1}$, then using (\ref{Utail}), {Lemma \ref{Bnd1TailSt}}, and the dominated convergence theorem,
\begin{align}
\lim_{t\to 0}\frac{1}{t}{I_{1,1}(t)}
%\widetilde{\bbp}\left(\bar{U}_{1}^{+}\geq\frac{t^{-1/Y}u+(G^{*}-1)\bar{U}_{1}^{-}}{M^{*}+1},\,-\bar{U}_{1}^{-}\leq\frac{t^{-1/Y}u}{M^{*}+G^{*}}\right)\nonumber\\
&=\lim_{t\to 0}\frac{1}{t}\int_{\bbr}p(1,y)\widetilde{\bbp}\left(\bar{U}_{1}^{+}\geq\frac{t^{-1/Y}u-(G^{*}-1)y}{M^{*}+1}\right){\bf 1}_{\{y\leq\frac{t^{-1/Y}u}{M^{*}+G^{*}}\}}dy\nonumber\\
&=\int_{\bbr}p(1,y)\lim_{t\to 0}\frac{1}{t}\widetilde{\bbp}\left(\bar{U}_{1}^{+}\geq\frac{t^{-1/Y}u-(G^{*}-1)y}{M^{*}+1}\right){\bf 1}_{\{y\leq\frac{t^{-1/Y}u}{M^{*}+G^{*}}\}}dy\nonumber\\
&=\int_{\bbr}p(1,y)\lim_{t\to 0}\frac{1}{t}\,\frac{t C(M^{*}+1)^{Y}}{Y(u-t^{1/Y}(G^{*}-1)y)^{Y}}{\bf 1}_{\{y\leq\frac{t^{-1/Y}u}{M^{*}+G^{*}}\}}dy\nonumber\\
\label{limI11}&=\frac{C(M^{*}+1)^{Y}}{Y u^{Y}}.
\end{align}
Similarly,
\begin{align}
I_{2}&=\widetilde{\bbp}\Big(M^{*}\bar{U}_{1}^{+}-G^{*}\bar{U}_{1}^{-}\geq t^{-1/Y}u,\bar{U}_{1}^{+}+\bar{U}_{1}^{-}< 0\Big)\nonumber\\
%&=\widetilde{\bbp}\left(-\bar{U}_{1}^{-}\geq\frac{t^{-1/Y}u-M^{*}\bar{U}_{1}^{+}}{G^{*}},-\bar{U}_{1}^{-}>\bar{U}_{1}^{+}\right)\nonumber\\
&=\widetilde{\bbp}\left(-\bar{U}_{1}^{-}\geq\frac{t^{-1/Y}u-M^{*}U_{1}^{+}}{G^{*}}>\bar{U}_{1}^{+}\right)+\widetilde{\bbp}\left(-\bar{U}_{1}^{-}>\bar{U}_{1}^{+}\geq\frac{t^{-1/Y}u-M^{*}\bar{U}_{1}^{+}}{G^{*}}\right)\nonumber\\
&=\widetilde{\bbp}\left(-\bar{U}_{1}^{-}\geq\frac{t^{-1/Y}u-M^{*}\bar{U}_{1}^{+}}{G^{*}},\bar{U}_{1}^{+}<\frac{t^{-1/Y}u}{M^{*}+G^{*}}\right)+\widetilde{\bbp}\left(-\bar{U}_{1}^{-}>\bar{U}_{1}^{+}\geq\frac{t^{-1/Y}u}{M^{*}+G^{*}}\right) \nonumber\\
&:=I_{2,1}(t)+I_{2,2}(t).\nonumber
\end{align}
Again, as in {(\ref{limI12})},
\begin{align}
\widetilde{\bbp}\left(-\bar{U}_{1}^{-}>\bar{U}_{1}^{+}\geq\frac{t^{-1/Y}u}{M^{*}+G^{*}}\right)=O(t^{2}),\qquad (t\to{}0),
\end{align}
{and since
\begin{align*}
\frac{1}{t}\widetilde{\bbp}\bigg(-\bar{U}_{1}^{-}\!\geq\!\frac{t^{-\frac{1}{Y}}u\!-\!M^{*}y}{G^{*}}\bigg){\bf 1}_{\{y\leq\frac{t^{-1/Y}u}{M^{*}+G^{*}}\}}\leq\frac{1}{t}\widetilde{\bbp}\bigg(-\bar{U}_{1}^{-}\!\geq\!\frac{t^{-\frac{1}{Y}}u\!-\!M^{*}\frac{t^{-1/Y}u}{M^{*}+G^{*}}}{G^{*}}\bigg){\bf 1}_{\{y\leq\frac{t^{-1/Y}u}{M^{*}+G^{*}}\}}\leq\frac{1}{t}\widetilde{\bbp}\bigg(-\bar{U}_{1}^{-}\!\geq\!\frac{t^{-\frac{1}{Y}}u}{M^{*}\!+\!G^{*}}\bigg),
\end{align*}
by (\ref{Utail}), Lemma \ref{Bnd1TailSt}, and the dominated convergence theorem,}
\begin{align}
\lim_{t\to 0}\frac{1}{t}{I_{2,1}(t)}
%\widetilde{\bbp}\left(-\bar{U}_{1}^{-}\geq\frac{t^{-1/Y}u-M^{*}\bar{U}_{1}^{+}}{G^{*}},\,\bar{U}_{1}^{+}\leq\frac{t^{-1/Y}u}{M^{*}+G^{*}}\right)\nonumber\\
&=\lim_{t\to 0}t^{-1}\int_{\bbr}p(1,y)\widetilde{\bbp}\left(-\bar{U}_{1}^{-}\geq\frac{t^{-1/Y}u-M^{*}y}{G^{*}}\right){\bf 1}_{\{y\leq\frac{t^{-1/Y}u}{M^{*}+G^{*}}\}}dy\nonumber\\
&= \int_{\bbr}p(1,y)\lim_{t\to 0}t^{-1}\widetilde{\bbp}\left(-\bar{U}_{1}^{-}\geq\frac{t^{-1/Y}u-M^{*}y}{G^{*}}\right){\bf 1}_{\{y\leq\frac{t^{-1/Y}u}{M^{*}+G^{*}}\}}dy\nonumber\\
&=\int_{\bbr}p(1,y)\lim_{t\to 0}\frac{1}{t}\,{\frac{t C(G^{*})^{Y}}{Y(u-t^{1/Y}M^{*}y)^{Y}}}{\bf 1}_{\{y\leq\frac{t^{-1/Y}u}{M^{*}+G^{*}}\}}dy\nonumber\\
\label{limI21}&=\frac{C(G^{*})^{Y}}{Y u^{Y}}.
\end{align}
Combining (\ref{DecomD1}), (\ref{limD11}), (\ref{limD13}), and (\ref{limI12})-(\ref{limI21}) implies that
\[
	\lim_{t\to{}0}t^{1/Y-1}D_{1}(t)=-\frac{C}{Y}\Big((M^{*}+1)^{Y}+(G^{*})^{Y}\Big)\int_{0}^{\infty} \left(1-e^{-v}\right) v^{-Y}dv+\eta.
\]
Finally, we use the following identity (see p. 84 in \cite{Sato:1999}): %$Y=1+\alpha$
\[
	\int_{0}^{\infty}(e^{- v}-1) v^{-Y} dv=\Gamma(1-Y)=-Y\Gamma(-Y).
\]
This concludes the proof.
\qed

\begin{lem}\label{GenCseGam}
	If $\tilde{\gamma} \neq 0$ in (\ref{LOP}), then
\begin{align}\label{2ndASP0}
\lim_{t\to 0}t^{\frac{1}{Y}-1}\left(t^{-\frac{1}{Y}}{\frac{1}{S_{0}}}\bbe(S_t-S_{0})^{+}-  {\widetilde{\bbe}(Z_{1}^{+})}\right)={\vartheta+\eta+ \frac{{\tilde{\gamma}}}{2}}.
\end{align}
\end{lem}
\noindent\textbf{Proof:}
Without loss of generality, fix $S_{0}=1$ and also assume that $\tilde{\gamma}>0$, the case $\tilde\gamma<0$ being similar. Then, using (\ref{LOP}),
\begin{align}
&\lim_{t\to 0}t^{1/Y-1}\left(t^{-1/Y}\bbe(S_t-S_{0})^{+}-{\widetilde{\bbe}(Z_{1}^{+})}\right)\nonumber\\
    & =\lim_{t\to 0}t^{1/Y-1}\left(e^{-({\tilde{\gamma}}+\eta)t}\int_{0}^{\infty} e^{-t^{1/Y}v}
	\widetilde{\bbe}\left(e^{-t^{1/Y}\widetilde{U}_{1}}{\bf 1}_{\{Z_{1}\geq{}v\}}\right)dv-{\widetilde{\bbe}(Z_{1}^{+})}\right)\nonumber\\
    &\quad+\lim_{t\to 0}t^{1/Y-1}e^{-({\tilde{\gamma}}+\eta)t}\int_{-{\tilde{\gamma}} t^{1-1/Y}}^{0} e^{-t^{1/Y}v}
	\widetilde{\bbe}\left(e^{-t^{1/Y}\widetilde{U}_{1}}{\bf 1}_{\{Z_{1}\geq{}v\}}\right)dv\nonumber\\
    &:=\widetilde{D}_{11}(t)+\widetilde{D}_{12}(t).\nonumber
\end{align}
As in the proof of (\ref{KNLFMP}), it can be shown that
\begin{align}\label{limtdD11}
\lim_{t\to 0}\widetilde{D}_{11}(t)={\vartheta+\eta}.
\end{align}
For $\widetilde{D}_{12}(t)$, changing variables $\left(u=t^{1/Y-1}v\right)$, we have
\begin{align}
\widetilde{D}_{12}(t)&\,{=}\,e^{-({\tilde{\gamma}}+\eta)t}\int_{-{\tilde{\gamma}}}^{0}e^{-tu}\widetilde{\bbe}\Big(e^{-t^{1/Y}\widetilde{U}_{1}}{\bf 1}_{\{Z_{1}\geq t^{1-1/Y}u\}}\Big)du=\,e^{-(\tilde{\gamma}+\eta)t}\int_{-\tilde{\gamma}}^{0}g_{t}^{(1)}(u)
%e^{-tu}\widetilde{\bbe}\Big(e^{-t^{1/Y}\widetilde{U}_{1}}{\bf 1}_{\{Z_{1}\geq t^{1-1/Y}u\}}{\bf 1}_{\{\widetilde{U}_{1}\geq 0\}}\Big)
du+e^{-(\tilde{\gamma}+\eta)t}\int_{-\tilde{\gamma}}^{0}
g_{t}^{(2)}(u)
%e^{-tu}\widetilde{\bbe}\Big(e^{-t^{1/Y}\widetilde{U}_{1}}{\bf 1}_{\{Z_{1}\geq t^{1-1/Y}u\}}{\bf 1}_{\{\widetilde{U}_{1}< 0\}}\Big)
du,\nonumber
\end{align}
where
\begin{align}
g_{t}^{(1)}(u):=e^{-tu}\widetilde{\bbe}\Big(e^{-t^{1/Y}\widetilde{U}_{1}}{\bf 1}_{\{Z_{1}\geq t^{1-1/Y}u\}}{\bf 1}_{\{\widetilde{U}_{1}\geq 0\}}\Big),\quad
g_{t}^{(2)}(u):=e^{-tu}\widetilde{\bbe}\Big(e^{-t^{1/Y}\widetilde{U}_{1}}{\bf 1}_{\{Z_{1}\geq t^{1-1/Y}u\}}{\bf 1}_{\{\widetilde{U}_{1}< 0\}}\Big).\nonumber
\end{align}
Since $Y\in(1,2)$, it is easy to see that, for $-{\tilde{\gamma}}\leq u\leq 0$ and $0\leq t\leq 1$,
\begin{align}
\big|\,g_{t}^{(1)}(u)\big|\leq\widetilde{\bbp}\big(Z_{1}\geq{-\tilde{\gamma}},\,\widetilde{U}_{1}\geq 0\big),\qquad
\big|\,g_{t}^{(2)}(u)\big|\leq {e^{\tilde{\gamma}}}\widetilde{\bbe}\Big(e^{-\widetilde{U}_{1}}{\bf 1}_{\{Z_{1}\geq{-\tilde{\gamma}}\}}\Big),\nonumber
\end{align}
and
\begin{align}
e^{-t^{1/Y}\widetilde{U}_{1}}{\bf 1}_{\{Z_{1}\geq t^{1-1/Y}u\}}{\bf 1}_{\{\widetilde{U}_{1}\geq 0\}}\leq {\bf 1}_{\{Z_{1}\geq u\}},\quad
e^{-t^{1/Y}\widetilde{U}_{1}}{\bf 1}_{\{Z_{1}\geq t^{1-1/Y}u\}}{\bf 1}_{\{\widetilde{U}_{1}<0\}}\leq e^{-\widetilde{U}_{1}}{\bf 1}_{\{Z_{1}\geq u\}}.\nonumber
\end{align}
By the dominated convergence theorem, it follows that
\begin{align}\label{NDLGNZ}
\lim_{t\to 0}\widetilde{D}_{12}(t)&=\int_{-\tilde{\gamma}}^{0}\left(\widetilde{\bbp}\big(Z_{1}\geq 0,\,\widetilde{U}_{1}\geq 0\big)+\widetilde{\bbp}\big(Z_{1}\geq 0,\,\widetilde{U}_{1}< 0\big)\right)du=\frac{{\tilde{\gamma}}}{2}. \end{align}
Combining the previous limit with (\ref{limtdD11}) leads to (\ref{2ndASP0}).
%\begin{align}\label{2ndASP}
%\lim_{t\to 0}t^{1/Y-1}\left(t^{-1/Y}\bbe(S_t-S_{0})^{+}-S_{0}\bbe(Z_{1}^{+})\right)={\vartheta+\gamma /2},
%\end{align}
\qed

\noindent
\textbf{Proof of {Proposition} \ref{AsyIVPCGMY}.}

\noindent
The small-time asymptotic behavior of the ATM call-option price $C_{BS}(t,\sigma)$ at maturity $t$ under the Black-Scholes model with volatility $\sigma$ and zero interest rates is known {to be such that (fixing for simplicity $S_{0}=1$)}
\begin{align}\label{AsyIVATMBS}
C_{BS}(t,\sigma)=\frac{\sigma}{\sqrt{2\pi}}t^{1/2}-\frac{\sigma^{3}}{24\sqrt{2\pi}}t^{3/2}+O(t^{5/2}),\qquad t\to 0;
\end{align}
{see, e.g., \cite[Corollary 3.4]{FordeJacLee:2010}).}
To derive the small-time asymptotics for the implied volatility, we need an analogous result to (\ref{AsyIVATMBS}) when $\sigma$ is replaced by $\hat{\sigma}(t)$. {To show such a formula, the following representation due \cite[Lemma 3.1]{RoperRut:2007} will be useful
\begin{align*}
C_{BS}(t,\sigma)=F(\sigma\sqrt{t})\quad\text{with}\quad F(\theta):=\int_{0}^{\theta}\Phi'\left(\frac{v}{2}\right)dv=\frac{1}{\sqrt{2\pi}}\int_{0}^{\theta}\exp\left(-v^{2}/8\right)dv,
\end{align*}
together with the following Taylor expansion for $F$ at $\theta=0$ (see~\cite[Lemma 5.1]{RoperRut:2007})
\begin{align*}
F(\theta)=\frac{1}{\sqrt{2\pi}}\theta-\frac{1}{24\sqrt{2\pi}}\theta^{3}+O(\theta^{5}),\quad\theta\rightarrow 0.
\end{align*}
Then, since $\hat\sigma(t)\to{}0$ as $t\to{}0$ (see, e.g., \cite[Proposition 5]{Tankov}), we conclude that 
\begin{align}\label{eq:BSOptIV}
C_{BS}(t,\hat{\sigma}(t))=
\frac{\hat\sigma(t)}{\sqrt{2\pi}}t^{1/2}-\frac{\hat\sigma(t)^{3}}{24\sqrt{2\pi}}t^{3/2}+O\left(\big(\hat\sigma(t)t^{1/2}\big)^{5}\right),\quad \text{as}\quad t\rightarrow 0.
\end{align}}
%This is the purpose of the next lemma. Note also that, by~\cite[Proposition 5]{Tankov}, $\hat{\sigma}(t)\rightarrow 0$ as $t\rightarrow 0$, {which in turn implies that one can apply (\ref{eq:BSOptIV}) below with $\hat\sigma_{t}$ replaced by $\hat\sigma(t)$.}
%\begin{lem}\label{lem:BSOptIV}
%Let $\hat{\sigma}_{t}$ {be an arbitrary time-dependent positive function such that $\hat{\sigma}_{t}t^{1/2}\to{}0$.} Then, $C_{BS}(t,\hat{\sigma}_{t})$ is such that 
%\begin{align}\label{eq:BSOptIV}
%C_{BS}(t,\hat{\sigma}_{t})=
%\frac{\hat\sigma_{t}}{\sqrt{2\pi}}t^{1/2}-\frac{\hat\sigma_{t}^{3}}{24\sqrt{2\pi}}t^{3/2}+O\left((\hat\sigma_{t}t^{1/2})^{5}\right),\quad \text{as}\quad t\rightarrow 0.
%\end{align}
%\end{lem}
%\noindent
%\begin{proof}[\textbf{\emph{Proof of Lemma:}}] The proof is similar to \cite{RoperRut:2007}. From \cite[Lemma 3.1]{RoperRut:2007},
%\begin{align*}
%C_{BS}(t,\sigma)=F(\theta):=\int_{0}^{\theta}\Phi'\left(\frac{v}{2}\right)dv=\frac{1}{\sqrt{2\pi}}\int_{0}^{\theta}\exp\left(-v^{2}/8\right)dv,
%\end{align*}
%where $\theta=\sigma\sqrt{t}$ and $\Phi$ is the standard normal cumulative distribution function. The Taylor expansion of $F$ at $\theta=0$ yields (see also~\cite[Lemma 5.1]{RoperRut:2007})
%\begin{align*}
%F(\theta)=\frac{1}{\sqrt{2\pi}}\theta-\frac{1}{24\sqrt{2\pi}}\theta^{3}+O(\theta^{5}),\quad\theta\rightarrow 0,
%\end{align*}
%and (\ref{eq:BSOptIV}) follows immediately since {by assumption} $\Theta_{t}\rightarrow 0$ as $t\rightarrow 0$.
%\end{proof}
\noindent
Returning to the proof of {Proposition} \ref{AsyIVPCGMY}, by equating {(\ref{ExpAsymBehCGMY})} and (\ref{eq:BSOptIV}) and comparing the first order terms,
\begin{align*}
\widetilde{\bbe}(Z_{1}^{+})t^{1/Y}\sim\frac{\hat{\sigma}(t)}{\sqrt{2\pi}}t^{1/2},\qquad t\to 0,
\end{align*}
and, therefore, 
\begin{align}\label{1stAsyIVPureCGMY}
\hat{\sigma}(t)\sim\sqrt{2\pi}\widetilde{\bbe}(Z_{1}^{+})t^{\frac{1}{Y}-\frac{1}{2}}:=\sigma_{1}t^{\frac{1}{Y}-\frac{1}{2}},\qquad t\to 0.
\end{align}
Next, set $\tilde{\sigma}(t)=\hat{\sigma}(t)-\sigma_{1}t^{\frac{1}{Y}-\frac{1}{2}}$. {By} comparing the first and second order terms in {(\ref{ExpAsymBehCGMY})} with the first term in {(\ref{eq:BSOptIV})} ({noting} that the second order term in {(\ref{eq:BSOptIV})} is {$o(t)$}),
\begin{align*}
\frac{C\Gamma(-Y)}{2}\left((M-1)^{Y}-M^{Y}-(G+1)^{Y}+G^{Y}\right)t\sim\frac{\tilde{\sigma}(t)}{\sqrt{2\pi}}t^{1/2},\qquad t\to{}0.
\end{align*}
Hence, $\tilde{\sigma}(t)\rightarrow 0$ as $t\rightarrow 0$, and moreover
\begin{align}\label{2ndAsyIVPureCGMY}
\tilde{\sigma}(t)\sim\sqrt{\frac{\pi}{2}}C\Gamma(-Y)\left((M-1)^{Y}-M^{Y}-(G+1)^{Y}+G^{Y}\right)t^{1/2},\qquad t\to{}0.
\end{align}
Combining (\ref{1stAsyIVPureCGMY}) and (\ref{2ndAsyIVPureCGMY}) finishes the proof.
\qed

\section{Proofs of Section \ref{Sect:NonZeroBrwn}: CGMY model with Brownian component}\label{ProofsSectGenCse}

\noindent
\textbf{Proof of Proposition \ref{1stOAsyCGMYB}.}

\noindent
From (\ref{CMR}), note that
\begin{align*}
t^{-1/2}\mathbb{E}(S_t-S_{0})^{+}&=t^{-1/2}S_{0}\bbp^{*}(X_{t}\geq E)=t^{-1/2}S_{0}\int_{0}^{\infty} e^{-x} \bbp^{*}(X_{t}\geq x)dx= S_{0}\int_{0}^{\infty} e^{-\sqrt{t}u} \bbp^{*}(t^{-1/2}X_{t}\geq u)du.
\end{align*}
Now for any $u\geq 0$ and $0<t\leq 1$,
\begin{align}
e^{-\sqrt{t}u}\bbp^{*}(t^{-1/2}X_{t}\geq{}u)&\leq\bbp^{*}(t^{-1/2}\sigma{}{W_{t}^{*}}\geq{u/2})+\bbp^{*}(t^{-1/2}{L_{t}^{*}}\geq{u/2})\nonumber\\
\label{DecomtailL}&=\bbp^{*}(\sigma {W_{1}^{*}}\geq {u/2})+\bbp^{*}(t^{-1/2}{L_{t}^{*}}\geq{u/2}).
\end{align}
Clearly the first term in (\ref{DecomtailL}) is integrable on $[0,\infty)$. To estimate the second term, applying the change of probability measure (\ref{EMMN}) and {using} the self-similarity property (\ref{SSCN}) of $Z$, we obtain
\begin{align*}
\bbp^{*}(t^{-1/2}{L_{t}^{*}}\geq{u/2})=\widetilde{\bbe}\left(e^{-U_{t}}{\bf 1}_{\{t^{-1/2}Z_{t}\geq{u/2}-\tilde{\gamma}\sqrt{t}\}}\right)=\widetilde{\bbe}\left(e^{-U_{t}}{\bf 1}_{\{Z_{1}\geq{t^{\frac{1}{2}-\frac{1}{Y}}u/2}-\tilde{\gamma}t^{1-\frac{1}{Y}}\}}\right).
\end{align*}
Pick $1<q<Y$ and $q'>1$ such that $q^{-1}+{q'}^{-1}=1$, then by H\"{o}lder's inequality {and (\ref{expUt})},
\begin{align*}
\widetilde{\bbe}\left(e^{-U_{t}}{\bf 1}_{\{Z_{1}\geq{t^{\frac{1}{2}-\frac{1}{Y}}u/2}-\tilde{\gamma}t^{1-\frac{1}{Y}}\}}\right)&\leq\left(\widetilde{\bbe}\Big(e^{-q'U_{t}}\Big)\right)^{1/q'} \left(\widetilde{\bbp}\Big(Z_{1}\geq\frac{t^{\frac{1}{2}-\frac{1}{Y}}u}{2}-\tilde{\gamma}t^{1-\frac{1}{Y}}\Big)\right)^{1/q}\nonumber\\
&=e^{-\eta t}\left(\widetilde{\bbe}\Big(e^{-q'\tilde{U}_{t}}\Big)\right)^{1/q'}\left(\widetilde{\bbp}\Big(Z_{1}\geq\frac{t^{\frac{1}{2}-\frac{1}{Y}}u}{2}-\tilde{\gamma}t^{1-\frac{1}{Y}}\Big)\right)^{1/q}\nonumber\\
&{=e^{-\eta t}e^{\eta {q'}^{Y-1}t}}\left(\widetilde{\bbp}\Big(Z_{1}\geq\frac{t^{\frac{1}{2}-\frac{1}{Y}}u}{2}-\tilde{\gamma}t^{1-\frac{1}{Y}}\Big)\right)^{1/q}\nonumber\\
&\leq{\exp{\left\{\eta\Big({q'}^{Y-1}-1\Big)\right\}}}\left(\widetilde{\bbp}\Big(Z_{1}\geq\frac{t^{\frac{1}{2}-\frac{1}{Y}}u}{2}-\tilde{\gamma}t^{1-\frac{1}{Y}}\Big)\right)^{1/q},
\end{align*}
{since as given in (\ref{eta})}, $\eta>0$. If $\tilde{\gamma}\leq 0$, then by (\ref{KIn}), for any $u>0$ and $0<t\leq 1$,
\begin{align*}
\left(\widetilde{\bbp}\Big(Z_{1}\geq\frac{t^{\frac{1}{2}-\frac{1}{Y}}u}{2}-\tilde{\gamma}t^{1-\frac{1}{Y}}\Big)\right)^{1/q} \leq\left(\frac{8\kappa}{\big(\frac{t^{\frac{1}{2}-\frac{1}{Y}}u}{2}-\tilde{\gamma}t^{1-\frac{1}{Y}}\big)^{Y}}\right)^{1/q} \leq\frac{(16\kappa)^{1/q}}{u^{Y/q}},
\end{align*}
and {thus},
\begin{align*}
\left(\widetilde{\bbp}\Big(Z_{1}\geq\frac{t^{\frac{1}{2}-\frac{1}{Y}}u}{2}-\tilde{\gamma}t^{1-\frac{1}{Y}}\Big)\right)^{1/q}\leq\min\left(1,\frac{(16\kappa)^{1/q}}{u^{Y/q}}\right),
\end{align*}
which is integrable on $[0,+\infty)$. On the other hand, if $\tilde{\gamma}>0$,
\begin{align*}
\widetilde{\bbp}\Big(Z_{1}\geq\frac{t^{\frac{1}{2}-\frac{1}{Y}}u}{2}-\tilde{\gamma}t^{1-\frac{1}{Y}}\Big) &=\widetilde{\bbp}\Big(Z_{1}\geq\frac{t^{\frac{1}{2}-\frac{1}{Y}}u}{2}-\tilde{\gamma}t^{1-\frac{1}{Y}}
\Big){{\bf 1}_{\{u>2\tilde{\gamma}\sqrt{t}\}}}+\widetilde{\bbp}\Big(Z_{1}\geq\frac{t^{\frac{1}{2}-\frac{1}{Y}}u}{2}-\tilde{\gamma}t^{1-\frac{1}{Y}}
\Big){{\bf 1}_{\{u\leq 2\tilde{\gamma}\sqrt{t}\}}}\\
&\leq \frac{16\kappa}{u^{Y}}+\frac{2^{Y}\tilde{\gamma}^{Y}}{u^{Y}},
\end{align*}
and thus
\begin{align*}
\left(\widetilde{\bbp}\Big(Z_{1}\geq\frac{t^{\frac{1}{2}-\frac{1}{Y}}u}{2}-\tilde{\gamma}t^{1-\frac{1}{Y}}\Big)\right)^{1/q}\leq\min\left(1,\frac{(16\kappa+2^{Y}\tilde{\gamma}^{Y})^{1/q}}{u^{Y/q}}\right),
\end{align*}
which is also integrable on $[0,+\infty)$. Therefore, by the dominated convergence theorem,
\begin{align*}
\lim_{t\rightarrow 0}t^{-1/2}\bbe(S_{t}-S_{0})_{+}=S_{0}\int_{0}^{\infty}\bbp^{*}({\sigma W_{1}^{*}}\geq u)du=S_{0}\sigma\bbe^{*}{(W_{1}^{*})_{+}}.
\end{align*}
The proof is now complete. \qed

\medskip
\noindent
\textbf{Proof of Theorem \ref{2ndOAsyCGMYB}.}

\noindent For simplicity, we fix $S_{0}=1$.
Recalling that {$X_{t}=L_{t}^{*}+\sigma W_{t}^{*}$ under $\bbp^{*}$} and using (\ref{CMR}), the self-similarity of {$W^{*}$}, and the change of variable $u=t^{-1/2}x$, it follows that
\begin{align*}
	B_{t}&:=t^{-1/2}\bbe(S_{t}-S_{0})_{+}-\sigma\bbe^{*}{(W_{1}^{*})_{+}}=\int_{0}^{\infty}\!\!e^{-\sqrt{t}u}\bbp^{*}\left(\sigma {W_{1}^{*}}\geq{}u-t^{-\frac{1}{2}}L_{t}\right)du-\int_{0}^{\infty}\bbp^{*}(\sigma {W_{1}^{*}}\geq u)du.
\end{align*}
{Next, by changing the probability measure to $\widetilde{\bbp}$ and using} that $ {L_{t}^{*}}=Z_{t}+\tilde{\gamma} t$, $U_{t}=\widetilde{U}_{t}+\eta t$, and the change of variable $y=u-t^{1/2}\tilde{\gamma}$ in the first integral above, {we get}
\begin{align}
	B_{t}&=\int_{-t^{1/2}\tilde\gamma}^{\infty}e^{-\sqrt{t}y-\tilde\gamma t}\,\widetilde{\bbe}\,e^{-\widetilde{U}_{t}-\eta t}{\bf 1}_{\{{\sigma W_{1}^{*}}\geq{}y-t^{-\frac{1}{2}}Z_{t}\}} dy
	- \int_{0}^{\infty} \widetilde\bbe e^{-\widetilde{U}_{t}-\eta t}{\bf 1}_{\{{\sigma W_{1}^{*}}\geq u\}}du \nonumber \\
	&=e^{-(\eta+\tilde\gamma) t}\int_{0}^{\infty}e^{-\sqrt{t}y}\left(\widetilde{\bbe}\,e^{-\widetilde{U}_{t}}{\bf 1}_{\{{\sigma W_{1}^{*}}\geq{}y-t^{-\frac{1}{2}}Z_{t}\}} - \widetilde\bbe e^{-\widetilde{U}_{t}}{\bf 1}_{\{{\sigma W_{1}^{*}}\geq y\}}\right)dy \label{2ndOADecCGMYB}\\
	&\quad +e^{-(\eta+\tilde\gamma) t}\int_{-t^{1/2}\tilde\gamma}^{0}e^{-\sqrt{t}y}\widetilde{\bbe}\,e^{-\widetilde{U}_{t}}{\bf 1}_{\{{\sigma W_{1}^{*}}\geq{}y-t^{-\frac{1}{2}}Z_{t}\}} dy \nonumber
	\\
	&\quad + \int_{0}^{\infty} \left(e^{-\tilde\gamma t-\sqrt{t} y}-1\right)\bbp^{*}(\sigma {W_{1}^{*}}\geq y)
%	\widetilde\bbe e^{-\widetilde{U}_{t}-\eta t}{\bf 1}_{\{\sigma W_{1}\geq y\}}
	dy. \nonumber
\end{align}
The last term above is clearly $O(\sqrt{t})$ as $t\to{}0$, {while the second term above can be shown {to be} asymptotically equivalent to $(\tilde\gamma/2)\sqrt{t}$ {by arguments analogous to those of (\ref{NDLGNZ}).}}
Thus, we only need to analyze the term in (\ref{2ndOADecCGMYB}) that we denote $A_{t}$ and that can be written as follows in light of the self-similarity property of $Z$ and $\widetilde{U}$:
	\begin{align}
		A_{t}
&=e^{-\tilde\eta t}\widetilde{\bbe}\left(e^{-t^{\frac{1}{Y}}\widetilde{U}_{1}}\int_{0}^{\infty}e^{-\sqrt{t}y}\Big({\bf 1}_{\{\sigma {W_{1}^{*}}\geq{}y-t^{\frac{1}{Y}-\frac{1}{2}}Z_{1}\}}-{\bf 1}_{\{\sigma {W_{1}^{*}}\geq y\}}\Big)dy\right),\label{DecomAt}
	\end{align}
where we had denoted $\tilde{\eta}:=\eta+\tilde\gamma$. 	
To analyze the asymptotic behavior of $A_{t}$, we decompose it into the following three {terms:}
\begin{align}
A_{t}&=e^{-\tilde\eta t}\widetilde{\bbe}\left(e^{-t^{\frac{1}{Y}}\widetilde{U}_{1}}{\bf 1}_{\{{W_{1}^{*}\geq 0,\sigma W_{1}^{*}+t^{\frac{1}{Y}-\frac{1}{2}}Z_{1}\geq 0}\}}\int_{{\sigma W_{1}^{*}}}^{{\sigma W_{1}^{*}}+t^{\frac{1}{Y}-\frac{1}{2}}Z_{1}}e^{-\sqrt{t}y}dy\right)\nonumber\\
&\quad -e^{-\tilde\eta t}\widetilde{\bbe}\left(e^{-t^{\frac{1}{Y}}\widetilde{U}_{1}}{\bf 1}_{\{0\leq{\sigma W_{1}^{*}}\leq -t^{\frac{1}{Y}-\frac{1}{2}}Z_{1}\}}\int_{0}^{{\sigma W_{1}^{*}}}e^{-\sqrt{t}y}dy\right)\nonumber\\
&\quad +e^{-\tilde\eta t}\widetilde{\bbe}\left(e^{-t^{\frac{1}{Y}}\widetilde{U}_{1}}{\bf 1}_{\{0\leq {-\sigma W_{1}^{*}}\leq t^{\frac{1}{Y}-\frac{1}{2}}Z_{1}\}}\int_{0}^{{\sigma W_{1}^{*}}+t^{\frac{1}{Y}-\frac{1}{2}}Z_{1}}e^{-\sqrt{t}y}dy\right)\nonumber\\
\label{DecomAt}&:= I_{1}(t)-I_{2}(t)+I_{3}(t).
\end{align}
We analyze each of three terms above in the following three steps:

\smallskip
\noindent
\textbf{Step 1.}
We first analyze the behavior of $I_{1}(t)$. Since $(Z_{t})_{t\geq 0}$ and {$(W_{t}^{*})_{t\geq 0}$} are independent,
\begin{align}
I_{1}(t)&=e^{-\tilde\eta t}\widetilde{\bbe}\left({\bf 1}_{\{{W_{1}^{*}\geq 0,\sigma W_{1}^{*}+t^{\frac{1}{Y}-\frac{1}{2}}Z_{1}\geq{}0}\}}\frac{e^{-t^{\frac{1}{Y}}\widetilde{U}_{1}}-e^{-t^{\frac{1}{Y}}(\widetilde{U}_{1}+Z_{1})}}{\sqrt{t}}{e^{-\sqrt{t}\sigma W_{1}^{*}}}\right)\nonumber\\
&=e^{-\tilde\eta t}\!\!\int_{0}^{\infty}\widetilde{\bbe}\Big({\bf 1}_{\{Z_{1}\geq -t^{\frac{1}{2}-\frac{1}{Y}}y\}}\frac{e^{-t^{\frac{1}{Y}}\widetilde{U}_{1}}(1-e^{-t^{\frac{1}{Y}}Z_{1}})}{\sqrt{t}}\Big)e^{-\sqrt{t}y}\frac{e^{-\frac{y^{2}}{2\sigma^{2}}}}{\sqrt{2\pi\sigma^{2}}}dy\nonumber\\
\label{IntJ1}&:=e^{-\tilde\eta t}\int_{0}^{\infty}J_{1}(t,y)e^{-\sqrt{t}y}\frac{e^{-\frac{y^{2}}{2\sigma^{2}}}}{\sqrt{2\pi\sigma^{2}}}dy.
\end{align}
Using that {the distribution of $Z_{1}$ is symmetric under $\widetilde{\bbp}$ (hence, $\widetilde{\bbe} Z_{1}=0$)}, $J_{1}(t,y)$ is then decomposed as:
\begin{align}
J_{1}(t,y)&=\widetilde{\bbe}\left({\bf 1}_{\{Z_{1}\geq -t^{\frac{1}{2}-\frac{1}{Y}}y\}}\bigg(\frac{e^{-t^{\frac{1}{Y}}\widetilde{U}_{1}}-e^{-t^{\frac{1}{Y}}(\widetilde{U}_{1}+Z_{1})}}{\sqrt{t}}-t^{\frac{1}{Y}-\frac{1}{2}}Z_{1}\bigg)\right)+t^{\frac{1}{Y}-\frac{1}{2}}\widetilde{\bbe}\Big(Z_{1}{\bf 1}_{\{Z_{1}\geq t^{\frac{1}{2}-\frac{1}{Y}}y\}}\Big)\nonumber\\
\label{DecomJ1}&:=J_{11}(t,y)+J_{12}(t,y).
\end{align}
Let us first consider $J_{12}(t,y)$. By (\ref{AsytailZ100}), it follows that, for any $0<t\leq{}1$ and $y>0$,
\begin{equation}\label{Eq:NdBnd1}
	t^{\frac{Y}{2}-1}J_{12}(t,y)\leq{} {\lambda y^{1-Y}},
\end{equation}
for some {$\lambda<\infty$} (see Appendix \ref{AddtnProofs} for {the verification of this claim}).
Moreover, for any fixed $y>0$,
\begin{align*}
t^{\frac{Y}{2}-1}J_{12}(t,y)=t^{\frac{Y}{2}+\frac{1}{Y}-\frac{3}{2}}\int_{t^{\frac{1}{2}-\frac{1}{Y}}y}^{\infty}u p_{Z}(u)du
%&=t^{\frac{Y}{2}+\frac{1}{Y}-\frac{3}{2}}\int_{y}^{\infty}t^{\frac{1}{2}-\frac{1}{Y}}w p_{Z}(t^{\frac{1}{2}-\frac{1}{Y}}w)t^{\frac{1}{2}-\frac{1}{Y}}dw\nonumber\\
=t^{\frac{Y}{2}-\frac{1}{Y}-\frac{1}{2}}\int_{y}^{\infty}wp_{Z}(t^{\frac{1}{2}-\frac{1}{Y}}w)dw.
\end{align*}
{Using (\ref{Asydenpz00})},
\begin{align*}
t^{\frac{Y}{2}-\frac{1}{Y}-\frac{1}{2}}wp_{Z}(t^{\frac{1}{2}-\frac{1}{Y}}w)\leq {(C+1)w^{-Y}},
\end{align*}
for $t$ small enough and all $w\geq y$. Therefore, by the dominated convergence theorem and, in light of (\ref{Asydenpz00}), we get:}
\begin{align}
&\lim_{t\rightarrow 0}t^{\frac{Y}{2}-1}e^{-\tilde\eta t}\int_{0}^{\infty}J_{12}(t,y)e^{-\sqrt{t}y}\frac{e^{-\frac{y^{2}}{2\sigma^{2}}}}{\sqrt{2\pi\sigma^{2}}}dy\nonumber\\
&\quad=\int_{0}^{\infty}\left(\lim_{t\rightarrow 0}t^{\frac{Y}{2}-1}J_{12}(t,y)\right)\frac{e^{-\frac{y^{2}}{2\sigma^{2}}}}{\sqrt{2\pi\sigma^{2}}}dy\nonumber\\
&\quad=\int_{0}^{\infty}\left(\int_{y}^{\infty}w\left(\lim_{t\rightarrow 0}t^{\frac{Y}{2}-\frac{1}{Y}-\frac{1}{2}}p_{Z}(t^{\frac{1}{2}-\frac{1}{Y}}w)\right)dw\right)\frac{e^{-\frac{y^{2}}{2\sigma^{2}}}}{\sqrt{2\pi\sigma^{2}}}dy\nonumber\\
&\quad =\int_{0}^{\infty}\left(\int_{y}^{\infty}{C w^{-Y}}dw\right)\frac{e^{-\frac{y^{2}}{2\sigma^{2}}}}{\sqrt{2\pi\sigma^{2}}}dy={\frac{C}{Y-1}}\int_{0}^{\infty}y^{1-Y}\frac{e^{-\frac{y^{2}}{2\sigma^{2}}}}{\sqrt{2\pi\sigma^{2}}}dy. \label{AsyInJ12}
\end{align}
For $J_{11}(t,y)$, note that
\begin{align}
J_{11}(t,y)&=t^{\frac{1}{Y}-\frac{1}{2}}\widetilde{\bbe}\left({\bf 1}_{\{Z_{1}\geq 0\}}\int_{\widetilde{U}_{1}}^{\widetilde{U}_{1}+Z_{1}}\Big(e^{-t^{\frac{1}{Y}}u}-1\Big)du\right)-t^{\frac{1}{Y}-\frac{1}{2}}\widetilde{\bbe}\left({\bf 1}_{\{-t^{\frac{1}{2}-\frac{1}{Y}}y\leq Z_{1}\leq 0\}}\int_{\widetilde{U}_{1}+Z_{1}}^{\widetilde{U}_{1}}\Big(e^{-t^{\frac{1}{Y}}u}-1\Big)du\right)\nonumber\\
&=t^{\frac{1}{Y}-\frac{1}{2}}\int_{\bbr}\Big(e^{-t^{\frac{1}{Y}}u}-1\Big)\widetilde{\bbp}\big(Z_{1}\geq 0,\widetilde{U}_{1}\leq u\leq\widetilde{U}_{1}+Z_{1}\big)du\nonumber\\
&\quad -t^{\frac{1}{Y}-\frac{1}{2}}\int_{\bbr}\Big(e^{-t^{\frac{1}{Y}}u}-1\Big)\widetilde{\bbp}\big(-t^{\frac{1}{2}-\frac{1}{Y}}y\leq Z_{1}\leq 0,\widetilde{U}_{1}+Z_{1}\leq u\leq\widetilde{U}_{1}\big)du\nonumber\\
&=t^{\frac{1}{Y}-\frac{1}{2}}\int_{0}^{\infty}\Big(e^{-t^{\frac{1}{Y}}u}-1\Big)\Big(\widetilde{\bbp}\big(Z_{1}\geq 0,\widetilde{U}_{1}\leq u\leq\widetilde{U}_{1}+Z_{1}\big)-\widetilde{\bbp}\big(-t^{\frac{1}{2}-\frac{1}{Y}}y\leq Z_{1}\leq 0,\widetilde{U}_{1}+Z_{1}\leq u\leq\widetilde{U}_{1}\big)\Big)du\nonumber\\
&\quad +t^{\frac{1}{Y}-\frac{1}{2}}\int_{-\infty}^{0}\Big(e^{-t^{\frac{1}{Y}}u}-1\Big)\Big(\widetilde{\bbp}\big(Z_{1}\geq 0,\widetilde{U}_{1}\leq u\leq\widetilde{U}_{1}+Z_{1}\big) -\widetilde{\bbp}\big(-t^{\frac{1}{2}-\frac{1}{Y}}y\leq Z_{1}\leq 0,\widetilde{U}_{1}+Z_{1}\leq u\leq\widetilde{U}_{1}\big)\Big)dx.\nonumber
\end{align}
Next, change variable back to $x=t^{1/Y}u$:
\begin{align}
J_{11}(t,y)&=\frac{1}{\sqrt{t}}\int_{0}^{\infty}\big(e^{-x}-1\big)\Bigg(\widetilde{\bbp}\big(Z_{1}\geq 0,\widetilde{U}_{1}\leq t^{-\frac{1}{Y}}x\leq\widetilde{U}_{1}+Z_{1}\big)\nonumber\\
&\quad -\widetilde{\bbp}\big(-t^{\frac{1}{2}-\frac{1}{Y}}y\leq Z_{1}\leq 0,\widetilde{U}_{1}+Z_{1}\leq t^{-\frac{1}{Y}}x\leq\widetilde{U}_{1}\big)\Bigg)du\nonumber\\
&\quad +\frac{1}{\sqrt{t}}\int_{-\infty}^{0}\big(e^{-x}-1\big)\Bigg(\widetilde{\bbp}\big(Z_{1}\geq 0,\widetilde{U}_{1}\leq t^{-\frac{1}{Y}}x\leq\widetilde{U}_{1}+Z_{1}\big)\nonumber\\
\label{DecomJ11}&\quad -\widetilde{\bbp}\big(-t^{\frac{1}{2}-\frac{1}{Y}}y\leq Z_{1}\leq 0,\widetilde{U}_{1}+Z_{1}\leq t^{-\frac{1}{Y}}x\leq\widetilde{U}_{1}\big)\Bigg)dx.
\end{align}
For $t>0$ and $y>0$, set
\begin{align*}
T_{1}(t,x,y):=\widetilde{\bbp}\big(Z_{1}\geq 0,\widetilde{U}_{1}\leq t^{-\frac{1}{Y}}x\leq\widetilde{U}_{1}+Z_{1}\big),\quad T_{2}(t,x,y):=\widetilde{\bbp}\big(-t^{\frac{1}{2}-\frac{1}{Y}}y\leq Z_{1}\leq 0,\widetilde{U}_{1}+Z_{1}\leq t^{-\frac{1}{Y}}x\leq\widetilde{U}_{1}\big).
\end{align*}
By (\ref{TIITI}), there exists {$\tilde{\kappa}>0$} such that for any $x>0$ and $0<t\leq{}1$,
\begin{align}\label{NEFS0}
T_{1}(t,x,y)\leq\widetilde{\bbp}\big( t^{-\frac{1}{Y}}x\leq\widetilde{U}_{1}+Z_{1}\big)\leq {\tilde{\kappa} tx^{-Y}}.
\end{align}
Hence,
\begin{align}
0&\leq e^{-\eta t}t^{\frac{Y}{2}-1}\int_{0}^{\infty}\int_{0}^{\infty}\frac{(1- e^{-x})}{\sqrt{t}}T_{1}(t,x,y)dx\frac{e^{-\sqrt{t}y}e^{-\frac{y^{2}}{2\sigma^{2}}}}{\sqrt{2\pi\sigma^{2}}}dy\nonumber\\
&\leq{}t^{\frac{Y-3}{2}}\int_{0}^{\infty}\int_{0}^{\infty}(1-e^{-x})T_{1}(t,x,y)dx\frac{e^{-\frac{y^{2}}{2\sigma^{2}}}}{\sqrt{2\pi\sigma^{2}}}dy\nonumber\\
\label{AsyDecomJ111}&\leq{}{{}\kappa{}t^{\frac{Y-1}{2}}}\int_{0}^{\infty}\int_{0}^{\infty}(1-e^{-x})x^{-Y}dx\frac{e^{-\frac{y^{2}}{2\sigma^{2}}}}{\sqrt{2\pi\sigma^{2}}}dy\longrightarrow 0,
\end{align}
as $t\rightarrow 0$, since $Y>1$. {Similarly, using $\widetilde{U}_{1}=M^{*} \bar{U}_{1}^{+}- G^{*} \bar{U}_{1}^{-}$ and Lemma \ref{Bnd1TailSt}, {for any $0<t\leq 1$ and $x,y>0$}, we have
\begin{align}
T_{2}(t,x,y)&\leq\widetilde{\bbp}\big(t^{-\frac{1}{Y}}x\leq\widetilde{U}_{1}\big)\nonumber\\
&\leq\widetilde{\bbp}\left(\bar{U}_{1}^{+}\geq \frac{t^{-\frac{1}{Y}}x}{2M^{*}}\right)+\widetilde{\bbp}\left(-\bar{U}_{1}^{-}\geq \frac{t^{-\frac{1}{Y}}x}{2G^{*}}\right)\nonumber\\
&\leq 2\widetilde{\bbp}\left(\bar{U}_{1}^{+}\geq \frac{t^{-\frac{1}{Y}}x}{2(M^{*}+G^{*})}\right)\nonumber\\
&\leq 8\kappa t(M^{*}+G^{*})^{Y}x^{-Y}.
\label{EstT2post}
\end{align}}
Therefore,
\begin{align}\label{AsyDecomJ112}
\lim_{t\rightarrow 0}\frac{e^{-\eta t}}{t^{1-\frac{Y}{2}}}\int_{0}^{\infty}\int_{0}^{\infty}\frac{(1- e^{-x})}{\sqrt{t}}T_{2}(t,x,y)dx\frac{e^{-\sqrt{t}y}e^{-\frac{y^{2}}{2\sigma^{2}}}}{\sqrt{2\pi\sigma^{2}}}dy=0.
\end{align}
For $x<0$, since $\bar{U}_{1}^{+}$ and $-\bar{U}_{1}^{-}$ are identically distributed, we proceed in the proof as follows:
\begin{align}
T_{1}(t,x,y)&\leq \widetilde{\bbp}\big(\widetilde{U}_{1}\leq t^{-\frac{1}{Y}}x\big)=\widetilde{\bbp}\big(M^* \bar{U}_{1}^ {+}-G^{*}\bar{U}^{-}_{1}\leq t^{-\frac{1}{Y}}x \big)\nonumber\\
&\leq \widetilde{\bbp}\big(\bar{U}_{1}^{+}\leq(2M^{*})^{-1}t^{-\frac{1}{Y}}x\big)+ \widetilde{\bbp}\big(-\bar{U}^{-}_{1}\leq (2G^*)^{-1}t^{-\frac{1}{Y}}x \big) \nonumber\\
&\leq 2\widetilde{\bbp}\big(\bar{U}_{1}^ {+}\leq (2M^{*}+2G^{*})^{-1}t^{-\frac{1}{Y}}x\big)\nonumber\\
\label{EstT1neq}&\leq 2\widetilde{\bbe}(e^{-\bar{U}_{1}^{+}})\exp{\left(\frac{t^{-\frac{1}{Y}}x}{2(M^{*}+G^{*})}\right)},
\end{align}
which is again independent of $y$ and integrable on $[0,\infty)$ when multiplied by $e^{-x}-1$. Moreover,
\begin{align*}
0&\leq t^{\frac{Y-3}{2}}\int_{-\infty}^{0}(e^{-x}-1) T_{1}(t,x,y)dx\\
&\leq{}2\widetilde{\bbe}(e^{-\bar{U}_{1}^{+}})t^{\frac{Y-3}{2}}\int_{-\infty}^{0}\left(\exp{\left(\frac{t^{-\frac{1}{Y}}x}{2(M^{*}+ G^{*})}-x\right)}-\exp{\left(\frac{t^{-\frac{1}{Y}}x}{2(M^{*}+G^{*})}\right)}\right)dx\\
&= 2\widetilde{\bbe}(e^{-\bar{U}_{1}^{+}})t^{\frac{Y-3}{2}}\frac{t^{\frac{2}{Y}}(M^{*}+G^{*})^{2}}{1-2 t^{\frac{1}{Y}}(M^{*}+G^{*})}.
\end{align*}
Hence, by the dominated convergence theorem,
\begin{align}
0&\leq e^{-\eta t} t^{\frac{Y}{2}-1}\int_{0}^{\infty}\int_{-\infty}^{0}\frac{e^{-x}-1}{\sqrt{t}}T_{1}(t,x,y) dx\frac{e^{-\sqrt{t}y}e^{-\frac{y^{2}}{2\sigma^{2}}}}{\sqrt{2\pi\sigma^{2}}}dy\nonumber\\
\label{AsyDecomJ113}&\leq{}2\widetilde{\bbe}(e^{-\bar{U}_{1}^{+}})t^{\frac{Y-3}{2}}\frac{t^{\frac{2}{Y}}(M^{*}+G^{*})^{2}}{1-2 t^{\frac{1}{Y}}(M^{*}+G^{*})}\int_{0}^{\infty}\frac{e^{-\frac{y^{2}}{2\sigma^{2}}}}{\sqrt{2\pi\sigma^{2}}}dy\longrightarrow {}0,
\end{align}
as $t\rightarrow 0$ since $2/Y>1>(3-Y)/2$, for $1<Y<2$. Similarly, since $\widetilde{U}_{1}+Z_{1}=M\bar{U}^{+}_{1}-G\bar{U}^{-}_{1}$, it follows {from (\ref{EstT1neq})} that
\begin{align*}
T_{2}(t,x,y)\leq\widetilde{\bbp}\big(M \bar{U}^{+}_{1}-G \bar{U}^{-}_{1}\leq t^{-\frac{1}{Y}}x\big)\leq2\widetilde{\bbe}(e^{-\bar{U}_{1}^{+}})\exp{\left(\frac{t^{-\frac{1}{Y}}x}{2(M+G)}\right)}.
\end{align*}
Therefore,
\begin{align}\label{AsyDecomJ114}
\lim_{t\rightarrow 0}e^{-\eta t}t^{\frac{Y}{2}-1}\int_{0}^{\infty}\int_{-\infty}^{0}\frac{(1- e^{-x})}{\sqrt{t}}T_{2}(t,x,y)dx\frac{e^{-\sqrt{t}y}e^{-\frac{y^{2}}{2\sigma^{2}}}}{\sqrt{2\pi\sigma^{2}}}dy=0.
\end{align}
{Combining (\ref{AsyInJ12}), (\ref{AsyDecomJ111}), (\ref{AsyDecomJ112}), (\ref{AsyDecomJ113}) and (\ref{AsyDecomJ114}),} we obtain
\begin{align}\label{AsyI1}
\lim_{t\rightarrow 0}t^{\frac{Y}{2}-1}I_{1}(t)={\frac{C}{Y-1}}\int_{0}^{\infty}y^{1-Y}\frac{e^{-\frac{y^{2}}{2\sigma^{2}}}}{\sqrt{2\pi\sigma^{2}}}dy.
\end{align}

\smallskip
\noindent
\textbf{Step 2.}
Next, we analyze the asymptotic behavior of $I_{2}(t)$. Using the independence of $(Z_{t})_{t\geq 0}$ and {$(W_{t}^{*})_{t\geq 0}$},
\begin{align}
I_{2}(t)&=e^{-\tilde\eta t}\widetilde{\bbe}\left(e^{-t^{\frac{1}{Y}}\widetilde{U}_{1}}{\bf 1}_{\{0\leq{\sigma W_{1}^{*}}\leq -t^{\frac{1}{Y}-\frac{1}{2}}Z_{1}\}}\frac{1-{e^{-\sqrt{t}\sigma W_{1}^{*}}}}{\sqrt{t}}\right)\nonumber\\
&=e^{-\tilde\eta{}t}\int_{0}^{\infty}\widetilde{\bbe}\Big(e^{-t^{\frac{1}{Y}}\widetilde{U}_{1}}{\bf 1}_{\{Z_{1}\leq-t^{\frac{1}{2}-\frac{1}{Y}}y\}}\Big)\frac{1-e^{-\sqrt{t}y}}{\sqrt{t}}\frac{e^{-\frac{y^{2}}{2\sigma^{2}}}}{\sqrt{2\pi\sigma^{2}}}dy\nonumber\\
&=e^{-\tilde\eta{}t}\int_{0}^{\infty}\widetilde{\bbe}\Big(\big(e^{-t^{\frac{1}{Y}}\widetilde{U}_{1}}-1\big){\bf 1}_{\{Z_{1}\leq-t^{\frac{1}{2}-\frac{1}{Y}}y\}}\Big)\frac{1-e^{-\sqrt{t}y}}{\sqrt{t}}\frac{e^{-\frac{y^{2}}{2\sigma^{2}}}}{\sqrt{2\pi\sigma^{2}}}dy\nonumber\\
\label{DecomI2}&\quad +e^{-\tilde\eta{}t}\int_{0}^{\infty}\widetilde{\bbp}\Big(Z_{1}\leq-t^{\frac{1}{2}-\frac{1}{Y}}y\Big)\frac{1-e^{-\sqrt{t}y}}{\sqrt{t}}\frac{e^{-\frac{y^{2}}{2\sigma^{2}}}}{\sqrt{2\pi\sigma^{2}}}dy.
\end{align}
By (\ref{KIn}) and the {symmetry} of $Z_{1}$,
\begin{align*}
\widetilde{\bbp}\Big(Z_{1}\!\leq\!-t^{\frac{1}{2}\!-\!\frac{1}{Y}}y\Big)\frac{1\!-\!e^{-\sqrt{t}y}}{\sqrt{t}}\!=\!\widetilde{\bbp}\Big(Z_{1}\!\geq\! t^{\frac{1}{2}\!-\!\frac{1}{Y}}y\Big)\frac{1\!-\!e^{-\sqrt{t}y}}{\sqrt{t}}\!\leq\!{8\kappa{}t^{1-\frac{Y}{2}}y^{1-Y}\leq{}8\kappa{}y^{1-Y}},
\end{align*}
which, when multiplied by {$\exp(-y^{2}/(2\sigma^{2}))$}, becomes integrable on $[0,\infty)$. Hence, by (\ref{AsytailZ100}) and the dominated convergence theorem,
\begin{align}
&\lim_{t\rightarrow 0}t^{\frac{Y}{2}-1}e^{-\tilde\eta{}t}\int_{0}^{\infty}\widetilde{\bbp}\Big(Z_{1}\leq-t^{\frac{1}{2}-\frac{1}{Y}}y\Big)\frac{1-e^{-\sqrt{t}y}}{\sqrt{t}}\frac{e^{-\frac{y^{2}}{2\sigma^{2}}}}{\sqrt{2\pi\sigma^{2}}}dy\nonumber\\
&\quad=\int_{0}^{\infty}\lim_{t\rightarrow 0}t^{\frac{Y}{2}-1}\widetilde{\bbp}\Big(Z_{1}\leq-t^{\frac{1}{2}-\frac{1}{Y}}y\Big)y\frac{e^{-\frac{y^{2}}{2\sigma^{2}}}}{\sqrt{2\pi\sigma^{2}}}dy\nonumber\\
&\quad=\int_{0}^{\infty}\lim_{t\rightarrow 0}t^{\frac{Y}{2}-1}\widetilde{\bbp}\Big(Z_{1}\geq{}t^{\frac{1}{2}-\frac{1}{Y}}y\Big)y\frac{e^{-\frac{y^{2}}{2\sigma^{2}}}}{\sqrt{2\pi\sigma^{2}}}dy\nonumber\\
&\quad=\int_{0}^{\infty}\lim_{t\rightarrow 0}t^{\frac{Y}{2}-1}{\frac{C}{Y}}(t^{\frac{1}{2}-\frac{1}{Y}}y)^{-Y}y\frac{e^{-\frac{y^{2}}{2\sigma^{2}}}}{\sqrt{2\pi\sigma^{2}}}dy\nonumber \\
&\quad ={\frac{C}{Y}}\int_{0}^{\infty}y^{1-Y}\frac{e^{-\frac{y^{2}}{2\sigma^{2}}}}{\sqrt{2\pi\sigma^{2}}}dy.
\label{AsyI2second}
\end{align}
To find the asymptotic behavior of the first integral in (\ref{DecomI2}), we decompose it as:
\begin{align}
\widetilde{\bbe}\Big(\big(e^{-t^{\frac{1}{Y}}\widetilde{U}_{1}}-1\big){\bf 1}_{\{Z_{1}\leq-t^{\frac{1}{2}-\frac{1}{Y}}y\}}\Big)&=\widetilde{\bbe}\Big({\bf 1}_{\{Z_{1}\leq -t^{\frac{1}{2}-\frac{1}{Y}}y,\widetilde{U}_{1}<0\}}\int_{t^{\frac{1}{Y}}\widetilde{U}_{1}}^{0}e^{-u}du\Big) \nonumber \\
&\quad -\widetilde{\bbe}\Big({\bf 1}_{\{Z_{1}\leq -t^{\frac{1}{2}-\frac{1}{Y}}y,\widetilde{U}_{1}\geq 0\}}\int_{0}^{t^{\frac{1}{Y}}\widetilde{U}_{1}}e^{-u}du\Big)\nonumber\\
\label{DocomFirIntI2}& := J_{21}(t,y)+J_{22}(t,y).
\end{align}
For $J_{21}(t,y)$, note that for any $0<t\leq 2^{-Y}(M^{*}+G^{*})^{-Y}$ and $y\geq 0$,
\begin{align*}
J_{21}(t,y)&=\int_{-\infty}^{0}e^{-x}\widetilde{\mathbb{P}}\Big(\bar{U}_{1}^{+}+\bar{U}_{1}^{-}\leq-t^{\frac{1}{2}-\frac{1}{Y}}y,M^{*}\bar{U}_{1}^{+}-G^{*}\bar{U}_{1}^{-}\leq t^{-\frac{1}{Y}}x\Big)dx\\
&\leq\int_{-\infty}^{0}e^{-x}\widetilde{\mathbb{P}}\Big(M^{*}\bar{U}_{1}^{+}-G^{*}\bar{U}_{1}^{-}\leq t^{-\frac{1}{Y}}x\Big)dx\\
&\leq 2\int_{-\infty}^{0}e^{-x}\widetilde{\mathbb{P}}\Big(\bar{U}_{1}^{+}\leq\frac{t^{-\frac{1}{Y}}x}{M^{*}+G^{*}}\Big)dx\\
&\leq 2\widetilde{\mathbb{E}}\big(e^{-\bar{U}_{1}^{+}}\big)\int_{-\infty}^{0}e^{-x}\exp{\left(\frac{t^{-\frac{1}{Y}}x}{M^{*}+G^{*}}\right)}dx\\
&=2\widetilde{\mathbb{E}}\big(e^{-\bar{U}_{1}^{+}}\big)\frac{t^{\frac{1}{Y}}(M^{*}+G^{*})}{1-t^{\frac{1}{Y}}(M^{*}+G^{*})},
\end{align*}
which is independent of $y$. Since $1-Y/2<1/2<1/Y$, for $1<Y<2$, by the dominated convergence theorem,
\begin{align}
0&\leq{}t^{\frac{Y}{2}-1}e^{-\eta t}\int_{0}^{\infty}J_{21}(t,y)\frac{1-e^{-\sqrt{t}y}}{\sqrt{t}}\frac{e^{-\frac{y^{2}}{2\sigma^{2}}}}{\sqrt{2\pi\sigma^{2}}}dy\nonumber\\
\label{AsyInJ21}&\leq{}t^{\frac{Y}{2}-1}2\widetilde{\bbe}\big(e^{-\bar{U}_{1}^{+}}\big)\frac{t^{\frac{1}{Y}}(M^{*}+G^{*})}{1-t^{\frac{1}{Y}}(M^{*}+G^{*})}\int_{0}^{\infty}y\frac{e^{-\frac{y^{2}}{2\sigma^{2}}}}{\sqrt{2\pi\sigma^{2}}}dy\longrightarrow{}0.
\end{align}
We further decompose the second term $J_{22}(t,y)$ in (\ref{DocomFirIntI2}) as:
\begin{align}
J_{22}(t,y)&=\widetilde{\bbe}\Big(\big(e^{-t^{\frac{1}{Y}}\widetilde{U}_{1}}-1+t^{\frac{1}{Y}}\widetilde{U}_{1}\big){\bf 1}_{\{Z_{1}\leq-t^{\frac{1}{2}-\frac{1}{Y}}y,\widetilde{U}_{1}\geq 0\}}\Big)\nonumber\\
&\quad-t^{\frac{1}{Y}}\widetilde{\bbe}\Big(\widetilde{U}_{1}{\bf 1}_{\{Z_{1}\leq -t^{\frac{1}{2}-\frac{1}{Y}}y,\widetilde{U}_{1}\geq 0\}}\Big)\nonumber\\
\label{DecomJ22}&:=J_{22}^{(1)}(t,y)-J_{22}^{(2)}(t,y).
\end{align}
Since $1-Y/2<1/2<1/Y$, for $1<Y<2$, it is easy to see that
\begin{align}\label{AsyInJ222}
J_{22}^{(2)}(t,y)\leq t^{\frac{1}{Y}}\widetilde{\bbe}\big|\widetilde{U}_{1}\big|=O(t^{\frac{1}{Y}})=o(t^{1-\frac{Y}{2}}).
\end{align}
Moreover,
\begin{align*}
J_{22}^{(1)}(t,y)&=\widetilde{\bbe}\left(\int_{0}^{t^{\frac{1}{Y}}\widetilde{U}_{1}}(1-e^{-w})dw\cdot{\bf 1}_{\{Z_{1}\leq-t^{\frac{1}{2}-\frac{1}{Y}}y,\widetilde{U}_{1}\geq 0\}}\right)\\
&=\int_{0}^{\infty}(1-e^{-w})\widetilde{\bbp}\Big(\widetilde{U}_{1}\geq t^{-\frac{1}{Y}}w,Z_{1}\leq-t^{\frac{1}{2}-\frac{1}{Y}}y\Big)dw\\
&\leq\int_{0}^{\infty}(1-e^{-w})\widetilde{\bbp}\Big(\widetilde{U}_{1}\geq t^{-\frac{1}{Y}}w\Big)dw.
\end{align*}
%Using {(\ref{EstT2post})} as well as Lemma \ref{Bnd1TailSt},  there exists {$\kappa>0$} such that, for any $w\geq 0$ and $0<t\leq 1$, {$\widetilde{\bbp}\Big(\widetilde{U}_{1}\geq t^{-\frac{1}{Y}}w\Big)\leq 8\kappa t(M^{*}+G^{*})^{Y}w^{-Y}$}, which is independent of $y$ and is integrable on $[0,\infty)$ when multiplied by $1-e^{-w}$.
{Hence by {(\ref{EstT2post})} and the dominated convergence theorem,}
\begin{align}
0&\leq{}t^{\frac{Y}{2}\!-\!1}e^{-\tilde\eta t}\int_{0}^{\infty}J_{22}^{(1)}(t,y)\frac{1-e^{-\sqrt{t}y}}{\sqrt{t}}\frac{e^{-\frac{y^{2}}{2\sigma^{2}}}}{\sqrt{2\pi\sigma^{2}}}dy\leq {{}8\kappa(M^{*}\!+\!G^{*})^{Y}t^{\frac{Y}{2}}}\int_{0}^{\infty}\frac{1-e^{-w}}{w^{-Y}}dw\int_{0}^{\infty}y\frac{e^{-\frac{y^{2}}{2\sigma^{2}}}}{\sqrt{2\pi\sigma^{2}}}dy\rightarrow 0,
\label{AsyInJ221}
\end{align}
as $t\to{}0$. {Combining (\ref{AsyI2second}), (\ref{AsyInJ21}), (\ref{AsyInJ222}) and (\ref{AsyInJ221}),} we obtain
\begin{align}\label{AsyI2}
\lim_{t\rightarrow 0}t^{\frac{Y}{2}-1}I_{2}(t)={\frac{C}{Y}}\int_{0}^{\infty}y^{1-Y}\frac{e^{-\frac{y^{2}}{2\sigma^{2}}}}{\sqrt{2\pi\sigma^{2}}}dy.
\end{align}

\smallskip
\noindent
\textbf{Step 3.}
We finally analyze the behavior of $I_{3}(t)$. Note that
\begin{align}
I_{3}(t)&=e^{-\tilde\eta t}\int_{0}^{\infty}\widetilde{\bbe}\Bigg(e^{-t^{\frac{1}{Y}}\widetilde{U}_{1}}{\bf 1}_{\{Z_{1}\geq t^{\frac{1}{2}-\frac{1}{Y}}y\}}\frac{1-e^{\sqrt{t}y}e^{-t^{\frac{1}{Y}}Z_{1}}}{\sqrt{t}}\Bigg)\frac{e^{-\frac{y^{2}}{2\sigma^{2}}}}{\sqrt{2\pi\sigma^{2}}}dy\nonumber\\
&=e^{-\tilde\eta t}\int_{0}^{\infty}\widetilde{\bbe}\Big(e^{-t^{\frac{1}{Y}}\widetilde{U}_{1}}{\bf 1}_{\{Z_{1}\geq t^{\frac{1}{2}-\frac{1}{Y}}y\}}\Big)\frac{1-e^{\sqrt{t}y}}{\sqrt{t}}\frac{e^{-\frac{y^{2}}{2\sigma^{2}}}}{\sqrt{2\pi\sigma^{2}}}dy\nonumber\\
&\quad+e^{-\tilde\eta t}\int_{0}^{\infty}\widetilde{\bbe}\Bigg(e^{-t^{\frac{1}{Y}}\widetilde{U}_{1}}{\bf 1}_{\{Z_{1}\geq t^{\frac{1}{2}-\frac{1}{Y}}y\}}\frac{1-e^{-t^{\frac{1}{Y}}Z_{1}}}{\sqrt{t}}\Bigg)e^{\sqrt{t}y}\frac{e^{-\frac{y^{2}}{2\sigma^{2}}}}{\sqrt{2\pi\sigma^{2}}}dy\nonumber\\
&:= e^{-\tilde\eta{}t}\!\!\int_{0}^{\infty}\!\!\!J_{31}(t,y)\frac{1\!-\!e^{\sqrt{t}y}}{\sqrt{t}}\frac{e^{-\!\frac{y^{2}}{2\sigma^{2}}}}{\sqrt{2\pi\sigma^{2}}}dy+e^{-\tilde\eta{}t}\int_{0}^{\infty}\!\!\!J_{32}(t,y)\frac{e^{\sqrt{t}y}e^{-\!\frac{y^{2}}{2\sigma^{2}}}}{\sqrt{2\pi\sigma^{2}}}dy. \label{DefnJ31J32}
\end{align}
We first investigate the asymptotic of $J_{31}(t,y)$ by decomposing it as:
\begin{align}
J_{31}(t,y)&=\widetilde{\bbe}\Big(\big(e^{-t^{\frac{1}{Y}}\widetilde{U}_{1}}-1\big){\bf 1}_{\{Z_{1}\geq t^{\frac{1}{2}-\frac{1}{Y}}y\}}\Big)+\widetilde{\bbp}\Big(Z_{1}\geq t^{\frac{1}{2}-\frac{1}{Y}}y\Big)\nonumber\\
\label{DecomJ31}&:=J_{31}^{(1)}(t,y)+J_{31}^{(2)}(t,y).
\end{align}
By (\ref{KIn}), it is easy to see that {$J_{31}^{(2)}(t,y)\leq{}8\kappa t^{1-\frac{Y}{2}}y^{-Y}$}, for any $0<t\leq 1$ and $y\geq 0$.
Hence, by (\ref{AsytailZ100}) and the dominated convergence theorem,
\begin{align}\label{AsyInJ312}
\lim_{t\rightarrow 0}\!\frac{e^{-\tilde\eta{}t}}{t^{1\!-\!\frac{Y}{2}}}\!\!\int_{0}^{\infty}\!\!\!J_{31}^{(2)}(t,y)\frac{1\!-\!e^{\sqrt{t}y}}{\sqrt{t}}\frac{e^{-\frac{y^{2}}{2\sigma^{2}}}}{\sqrt{2\pi\sigma^{2}}}dy={-\frac{C}{Y}}\int_{0}^{\infty}\!\!\!y^{1\!-\!Y}\!\frac{e^{-\frac{y^{2}}{2\sigma^{2}}}}{\sqrt{2\pi\sigma^{2}}}dy.
\end{align}
Also, $J_{31}^{(1)}(t,y)$ can be further decomposed as:
\begin{align*}
{\widetilde{\bbe}\Big(\big(e^{-t^{\frac{1}{Y}}\widetilde{U}_{1}}-1\big){\bf 1}_{\{Z_{1}\geq t^{\frac{1}{2}-\frac{1}{Y}}y,\widetilde{U}_{1}\geq{}0\}}\Big)+
\int_{0}^{\infty}e^{u}\widetilde{\bbp}\Big(Z_{1}\geq t^{\frac{1}{2}-\frac{1}{Y}}y,\widetilde{U}_{1}\leq-t^{-\frac{1}{Y}}u\Big)du.}
%-\int_{0}^{\infty}e^{-u}\widetilde{\bbp}\Big(Z_{1}\geq t^{\frac{1}{2}-\frac{1}{Y}}y,\widetilde{U}_{1}\geq t^{\frac{1}{Y}}u\Big)du.
\end{align*}
For any $u>0$, $y>0$ and $t>0$, let $T_{3}(t,u,y):=\widetilde{\bbp}\Big(Z_{1}\geq{}t^{\frac{1}{2}-\frac{1}{Y}}y,\widetilde{U}_{1}\leq-t^{-\frac{1}{Y}}u\Big)$.
It is easily seen that
\begin{align}
T_{3}(t,u,y)
%&\leq\widetilde{\bbp}\Big(\widetilde{U}_{1}\leq-t^{\frac{1}{Y}}u\Big)\nonumber\\
&\leq{}\widetilde{\bbp}\Big(\bar{U}_{1}^{+}\leq\frac{-t^{-\frac{1}{Y}}u}{2M^{*}}\Big)+\widetilde{\bbp}\Big(-\bar{U}_{1}^{-}\leq\frac{-t^{-\frac{1}{Y}}u}{2G^{*}}\Big)\nonumber\\
&\leq{}2\widetilde{\bbp}\Big({-\bar{U}_{1}^{-}}\leq\frac{-t^{-\frac{1}{Y}}u}{2(M^{*}+G^{*})}\Big)\nonumber \\
&\leq{}2\widetilde{\bbe}e^{\bar{U}_{1}^{-}}\exp{\bigg(\frac{-t^{-\frac{1}{Y}}u}{2(M^{*}+G^{*})}\bigg)}. \label{EstJ31first}
\end{align}
Moreover,
\begin{align}
{0\leq \widetilde{\bbe}\Big(\big(1-e^{-t^{\frac{1}{Y}}\widetilde{U}_{1}}\big){\bf 1}_{\{Z_{1}\geq t^{\frac{1}{2}-\frac{1}{Y}}y,\widetilde{U}_{1}\geq 0\}}\Big)\leq t^{\frac{1}{Y}}\widetilde{\bbe}\big|\widetilde{U}_{1}\big|.} \label{EstJ31sec}
\end{align}
Hence, for any $y\geq 0$ and $0<t\leq 2^{-Y}(M^{*}+G^{*})^{-Y}$, and since $1-Y/2<1/2<1/Y$, for $1<Y<2$,
\begin{align}
t^{\frac{Y}{2}-1}|J_{31}^{(1)}(t,y)|&\leq{}t^{\frac{Y}{2}-1}2\widetilde{\bbe}e^{\bar{U}_{1}^{-}}\int_{0}^{\infty}\exp{\bigg(\frac{-t^{-\frac{1}{Y}}u}{2(M^{*}+G^{*})}+u\bigg)}du+t^{\frac{1}{Y}+\frac{Y}{2}-1}\widetilde{\bbe}\big|\widetilde{U}_{1}\big|\nonumber\\
&=t^{\frac{Y}{2}-1}\frac{4\widetilde{\bbe}\Big(e^{\bar{U}_{1}^{-}}\Big)t^{\frac{1}{Y}}(M^{*}+G^{*})}{1-2t^{\frac{1}{Y}}(M^{*}+G^{*})}+t^{\frac{1}{Y}+\frac{Y}{2}-1}\widetilde{\bbe}\big|\widetilde{U}_{1}\big|\longrightarrow 0,\qquad \text{ as } t\rightarrow 0.
\label{AsyJ311}
\end{align}
Since both control functions in (\ref{EstJ31first}) and (\ref{EstJ31sec}) are independent of $y$, combining (\ref{AsyInJ312}) and (\ref{AsyJ311}), and by the dominated convergence theorem,
\begin{align}\label{AsyInJ31}
\lim_{t\rightarrow 0}\!\frac{e^{-\tilde\eta{}t}}{t^{1\!-\!\frac{Y}{2}}}\!\!\int_{0}^{\infty}\!\!\!J_{31}(t,y)\frac{1\!-\!e^{\sqrt{t}y}}{\sqrt{t}}\frac{e^{-\frac{y^{2}}{2\sigma^{2}}}}{\sqrt{2\pi\sigma^{2}}}dy={-\frac{C}{Y}}\int_{0}^{\infty}\!\!\!y^{1\!-\!Y}\!\frac{e^{-\frac{y^{2}}{2\sigma^{2}}}}{\sqrt{2\pi\sigma^{2}}}dy.
\end{align}
Next, we decompose the quantity $J_{32}(t,y)$ defined in (\ref{DefnJ31J32}) as:
\begin{align}
J_{32}(t,y)&=\widetilde{\bbe}\Bigg({\bf 1}_{\{Z_{1}\geq t^{\frac{1}{2}-\frac{1}{Y}}y\}}\Big(\frac{e^{-t^{\frac{1}{Y}}\widetilde{U}_{1}}-e^{-t^{\frac{1}{Y}}(Z_{1}+\widetilde{U}_{1})}}{\sqrt{t}}-t^{\frac{1}{Y}-\frac{1}{2}}Z_{1}\Big)\Bigg)\nonumber\\
&\quad+\widetilde{\bbe}\Big(t^{\frac{1}{Y}-\frac{1}{2}}Z_{1}{\bf 1}_{\{Z_{1}\geq t^{\frac{1}{2}-\frac{1}{Y}}y\}}\Big)\nonumber\\
\label{DecomJ32}&:= J_{32}^{(1)}(t,y)+J_{32}^{(2)}(t,y).
\end{align}
Note that $J_{32}^{(2)}(t,y)$ is the same as $J_{12}(t,y)$ in (\ref{DecomJ1}), and thus the corresponding integral has an asymptotic behavior similar to (\ref{AsyInJ12}):
\begin{align}
\lim_{t\rightarrow 0}t^{\frac{Y}{2}-1}e^{-\tilde\eta t}\int_{0}^{\infty}\widetilde{\bbe}\Big(t^{\frac{1}{Y}-\frac{1}{2}}Z_{1}{\bf 1}_{\{Z_{1}\geq t^{\frac{1}{2}-\frac{1}{Y}}y\}}\Big)e^{\sqrt{t}y}\frac{e^{-\frac{y^{2}}{2\sigma^{2}}}}{\sqrt{2\pi\sigma^{2}}}dy={\frac{C}{Y-1}}
%\frac{Y\Gamma(Y)}{\pi (Y-1)}\sin{\Big(\frac{\pi Y}{2}\Big)}
\int_{0}^{\infty}y^{1-Y}\frac{e^{-\frac{y^{2}}{2\sigma^{2}}}}{\sqrt{2\pi\sigma^{2}}}dy.
\label{AsyInJ32sec}
\end{align}
Next, we further decompose $J_{32}^{(1)}(t,y)$ as:
\begin{align*}
J_{32}^{(1)}(t,y)&=t^{\frac{1}{Y}-\frac{1}{2}}\widetilde{\bbe}\Bigg({\bf 1}_{\{Z_{1}\geq t^{\frac{1}{2}-\frac{1}{Y}}y\}}\int_{\widetilde{U}_{1}}^{Z_{1}+\widetilde{U}_{1}}\big(e^{-t^{\frac{1}{Y}}u}-1\big)du\Bigg)\\
&=t^{\frac{1}{Y}-\frac{1}{2}}\int_{\bbr}(e^{-t^{\frac{1}{Y}}u}-1)\widetilde{\bbp}\Big(Z_{1}\geq t^{\frac{1}{2}-\frac{1}{Y}}y,\widetilde{U}_{1}\leq u\leq Z_{1}+\widetilde{U}_{1}\Big)du\\
&=\frac{1}{\sqrt{t}}\int_{-\infty}^{0}(e^{-x}-1)\widetilde{\bbp}\Big(Z_{1}\geq t^{\frac{1}{2}-\frac{1}{Y}}y,\widetilde{U}_{1}\leq t^{-\frac{1}{Y}}x\leq Z_{1}+\widetilde{U}_{1}\Big)dx\\
&\quad+\frac{1}{\sqrt{t}}\int_{0}^{\infty}(e^{-x}-1)\widetilde{\bbp}\Big(Z_{1}\geq t^{\frac{1}{2}-\frac{1}{Y}}y,\widetilde{U}_{1}\leq t^{-\frac{1}{Y}}x\leq Z_{1}+\widetilde{U}_{1}\Big)dx.
\end{align*}
Note that for $x>0$,
\begin{align*}
\widetilde{\bbp}\Big(Z_{1}\geq t^{\frac{1}{2}-\frac{1}{Y}}y,\widetilde{U}_{1}\leq t^{-\frac{1}{Y}}x\leq Z_{1}+\widetilde{U}_{1}\Big)\leq\widetilde{\bbp}\Big(t^{-\frac{1}{Y}}x\leq Z_{1}+\widetilde{U}_{1}\Big),
\end{align*}
while for $x<0$,
\begin{align*}
\widetilde{\bbp}\Big(Z_{1}\geq t^{\frac{1}{2}-\frac{1}{Y}}y,\widetilde{U}_{1}\leq t^{-\frac{1}{Y}}x\leq Z_{1}+\widetilde{U}_{1}\Big)\leq\widetilde{\bbp}\Big(\widetilde{U}_{1}\leq{}t^{-\frac{1}{Y}}x\Big).
\end{align*}
Using the estimates (\ref{NEFS0}) and (\ref{EstT1neq}), a proof as in getting (\ref{AsyDecomJ111}) and (\ref{AsyDecomJ113}) gives
\begin{align}\label{AsyInJ321}
\lim_{t\rightarrow{}0}\frac{e^{-\tilde\eta{}t}}{t^{1-\frac{Y}{2}}}\int_{0}^{\infty}J_{32}^{(1)}(t,y)e^{\sqrt{t}y}\frac{e^{-\frac{y^{2}}{2\sigma^{2}}}}{\sqrt{2\pi\sigma^{2}}}dy=0.
\end{align}
Combining (\ref{AsyInJ31}), (\ref{AsyInJ32sec}) and (\ref{AsyInJ321}), we have
\begin{align}\label{AsyI3}
\lim_{t\rightarrow{}0}t^{\frac{Y}{2}-1}I_{3}(t)={\frac{C}{Y(Y-1)}}\int_{0}^{\infty}y^{1-Y}\frac{e^{-\frac{y^{2}}{2\sigma^{2}}}}{\sqrt{2\pi\sigma^{2}}}dy.
\end{align}
Finally, from (\ref{DecomAt}), (\ref{AsyI1}), (\ref{AsyI2}) and (\ref{AsyI3}), and since $1-Y/2<1/2$, for $1<y<2$, we obtain (\ref{2ndAsyCGMYB}), therefore finishing the proof. \qed

\noindent
\textbf{Proof of {Proposition} \ref{AsyIVCGMYB}.}\\
\noindent
When the diffusion component exists, \cite[Proposition 5]{Tankov} implies that $\hat{\sigma}(t)\rightarrow\sigma$ as $t\rightarrow 0$. In particular, $\hat{\sigma}_{t}t^{1/2}\rightarrow 0$ as $t\rightarrow 0$ and, thus, (\ref{eq:BSOptIV}) above still holds. Let $\tilde{\sigma}(t):=\hat{\sigma}_{t}-\sigma$, then $\tilde{\sigma}(t)\rightarrow 0$ as $t\rightarrow 0$, and (\ref{eq:BSOptIV}) can be written as
\begin{align}\label{eq:BSOptIVBC}
C_{BS}(t,\hat{\sigma}(t))=\frac{\sigma}{\sqrt{2\pi}}t^{1/2}+\frac{\tilde{\sigma}(t)}{\sqrt{2\pi}}t^{1/2}-\frac{\hat\sigma(t)^{3}}{24\sqrt{2\pi}}t^{3/2}+O\left(\big(\hat\sigma(t)t^{1/2}\big)^{5}\right)=\frac{\sigma}{\sqrt{2\pi}}t^{1/2}+\frac{\tilde{\sigma}(t)}{\sqrt{2\pi}}t^{1/2}+O(t^{3/2}).
\end{align}
By comparing {(\ref{ExpAsymBehCGMYBM})-(\ref{PrcDefn2ndTrm})} and (\ref{eq:BSOptIVBC}), we have
\begin{align*}
\frac{C2^{1-Y}\sigma^{1-Y}}{Y(Y-1)\sqrt{\pi}}\Gamma\left(1-\frac{Y}{2}\right)t^{\frac{3-Y}{2}}\sim\frac{\tilde{\sigma}(t)}{\sqrt{2\pi}}t^{1/2},\qquad t\to{}0,
\end{align*}
and, therefore, 
\begin{align*}
\tilde{\sigma}(t)\sim\frac{C2^{\frac{3}{2}-Y}\sigma^{1-Y}}{Y(Y-1)}\Gamma\left(1-\frac{Y}{2}\right)t^{1-\frac{Y}{2}},\qquad t\rightarrow 0.
\end{align*}
The proof is now complete. \qed

\section{Additional proofs}\label{AddtnProofs}
\noindent
\textbf{Verification of (\ref{CFCGMYShare}).}

\noindent  By the very definition of $\bbp^*$ and (\ref{CFVCGMY}), we have
	\begin{align*}
		\mathbb{E}^*(e^{iu X_t})&%=\mathbb{E}(e^{iu X_t+ X_{t}})
		=\mathbb{E}(e^{(iu +1)X_t})=\phi_t(u-i)=e^{t \,C \Gamma(-Y)\left((M-i(u-i))^Y+(G+i(u-i))^Y-M^Y-G^Y\right)+i  \c (u-i)t-{\frac{\sigma^{2}}{2}(u-i)^{2}t}}\\
		&=e^{t \,C \Gamma(-Y)\left((M^*-iu))^Y+(G^*+iu)^Y-M^Y-G^Y\right)+i  {(\c+\sigma^{2})} ut+ct-{\frac{\sigma^{2}u^{2}}{2}t+\frac{\sigma^{2}}{2}t}}
	\end{align*}
	with $M^*=M-1$ and $G^*=G+1$.
Next, using (\ref{cVValMartM}), we clearly have
\[
	t C\Gamma(-Y)(-M^Y-G^Y)+ct{+t\sigma^{2}/2=-t}\C\Gamma(-Y)\left((M-1)^Y-(G+1)^Y\right),
\]
and thus,
\[
	\mathbb{E}^*(e^{iu X_t})=e^{t \,C \Gamma(-Y)\left((M^*-iu)^Y+(G^*+iu)^Y-{M^*}^Y-{G^*}^Y\right)+i  \c^* ut-{\frac{\sigma^{2}u^{2}}{2}t}},
\]
with {$c^*=c+\sigma^{2}$}.
\qed

\medskip
\noindent
\textbf{Verification of (\ref{Cent}).}

\noindent Using (\ref{b*TripletCGMY}),
\begin{align*}
\widetilde{\bbe}X_{1}&=\tilde{b}+\int_{\{|x|>1\}}x\tilde{\nu}(dx)=b^{*}+\int_{|x|\leq 1}x(\tilde{\nu}-\nu^{*})(dx)+\int_{|x|>1}x\tilde{\nu}(dx)\\
&={c^{*}}+\int_{\bbr}x(\tilde\nu-{\nu^{*}})(dx)- CY\Gamma(-Y)((M^*)^{Y-1}-(G^*)^{Y-1}).
\end{align*}
On the other hand,
\begin{align*}
&\int_{\bbr}x(\tilde\nu-{\nu^{*}})(dx)=\int_{\bbr}x(e^{{\varphi(x)}}-1)\nu^{*}(dx)\\
&\quad =\int_{0}^{\infty}x(e^{M^{*}x}-1)\frac{Ce^{-M^{*}x}}{x^{1+Y}}dx+\int_{-\infty}^{0}x(e^{-G^{*}x}-1)\frac{Ce^{G^{*}x}}{|x|^{Y+1}}dx\\
&\quad =\sum_{n=1}^{\infty}\frac{C(M^{*})^{n}}{n!}\int_{0}^{\infty}x^{n-Y}e^{-M^{*}x}dx-\sum_{m=1}^{\infty}\frac{C(G^{*})^{m}}{m!}\int_{0}^{\infty}x^{m-Y}e^{-G^{*}x}dx\\
%=&C(M^{*})^{Y-1}\sum_{n=1}^{\infty}\frac{1}{n!}\int_{0}^{\infty}z^{n-Y}e^{-z}dz-C(G^{*})^{Y-1}\sum_{m=1}^{\infty}\frac{1}{m!}\int_{0}^{\infty}z^{m-Y}e^{-z}dz\\
&\quad =C(M^{*})^{Y-1}\sum_{n=1}^{\infty}\frac{\Gamma(n-Y+1)}{n!}-C(G^{*})^{Y-1}\sum_{m=1}^{\infty}\frac{\Gamma(m-Y+1)}{m!}\\
&\quad =-C(M^{*})^{Y-1}\Gamma(1-Y)+C(G^{*})^{Y-1}\Gamma(1-Y).
\end{align*}
Hence, {$\widetilde{\bbe}X_{1}=c^{*}$} and (\ref{Cent}) follows.
\qed

\medskip
\noindent
\textbf{Proof of Lemma \ref{Bnd1TailSt}.}

\noindent
By (\ref{UIn1}), for any $0<t\leq 1$ and $v>0$, we have
\begin{align*}
\frac{1}{t}\widetilde{\bbp}\left(\bar{U}_{1}^{+}\geq t^{-1/Y} v\right)&=\frac{1}{t}\widetilde{\bbp}\left(\bar{U}_{1}^{+}\geq
t^{-1/Y} v \right)\left({\bf 1}_{\{t^{-1/Y} v\geq N\}}+{\bf 1}_{\{t^{-1/Y} v<N\}}\right)\\
&\leq\frac{1}{t}\!\left( {\frac{2C}{Y}} t {v^{-Y}}{\bf 1}_{\{vt^{-1/Y}\geq N\}}+\frac{N^{Y}}{(t^{-1/Y}v)^{Y}}{\bf 1}_{\{t^{-1/Y}v< N\}}\right)\\
&\leq\left({2CY^{-1}} + N^{Y}\right)v^{-Y}.
\end{align*}
{The result follows by setting $\kappa=2CY^{-1}+N^{Y}$.}
\qed

\medskip
\noindent
\textbf{Verification of (\ref{TIITI}).}

\noindent
Note that
\begin{align*}\nonumber
t^{-1}\widetilde{\bbp}\Big(Z_{1}^{+}\!+\!\widetilde{U}_{1}\!\geq\!t^{-1/Y}u\Big)&\leq t^{-1}\widetilde{\bbp}\left(Z_{1}\geq \frac{t^{-1/Y}u}{2}\right)+t^{-1}\widetilde{\bbp}\left(\bar{U}_{1}^{+}\geq\frac{t^{-1/Y}u}{4M^{*}}\right)+t^{-1}\widetilde{\bbp}\left(-\bar{U}_{1}^{-}\geq\frac{t^{-1/Y}u}{4G^{*}}\right).\end{align*}
Using Lemma \ref{Bnd1TailSt}, there exists a constant {$0<\kappa<\infty$} such that
\begin{align*}
t^{-1}\widetilde{\bbp}\left(\bar{U}_{1}^{+}\geq\frac{t^{-1/Y}u}{4M^{*}}\right)+t^{-1}\widetilde{\bbp}\left(-\bar{U}_{1}^{-}\geq\frac{t^{-1/Y}u}{4G^{*}}\right)\leq 2\kappa (4M^{*}+4G^{*})^{Y}u^{-Y},
\end{align*}
for all $u>0$ and $0<t\leq{}1$.
Clearly, (\ref{KIn}) implies that, for any $0<t\leq 1$ and $u>0$,
\begin{equation*}
\frac{1}{t}\widetilde{\bbp}\left(Z_{1}\geq\frac{u}{2t^{\frac{1}{Y}}}\right)\leq 2^{2Y+1}\kappa u^{-Y}\leq 32\kappa u^{-Y},
\end{equation*}
and (\ref{TIITI}) follows.
\qed

\medskip
\noindent
\textbf{Verification of (\ref{Eq:NdBnd1})}

\noindent
In light of (\ref{AsytailZ100}), there exist {$R>0$} and {$H>0$}, such that for any {$u\geq H$},
\begin{align}\label{AsyInqdenpz}
p_{Z}(u)\leq {R u^{-Y-1}}.
\end{align}
Now for $J_{12}(t,y)$, using (\ref{AsyInqdenpz}), for any {$0<t\leq 1$} and $y>0$,
\begin{align*}
t^{\frac{Y}{2}-1}J_{12}(t,y)&=t^{\frac{Y}{2}-1} t^{\frac{1}{Y}-\frac{1}{2}}\int_{t^{\frac{1}{2}-\frac{1}{Y}}y}^{\infty}up_{Z}(u)du\\
&= t^{(\frac{1}{Y}-\frac{1}{2})(1-Y)}{\bf 1}_{\{{t^{\frac{1}{2}-\frac{1}{Y}}y\geq H}\}}\int_{t^{\frac{1}{2}-\frac{1}{Y}}y}^{\infty}up_{Z}(u)du\\
&\quad +t^{(\frac{1}{Y}-\frac{1}{2})(1-Y)}{\bf 1}_{\{{t^{\frac{1}{2}-\frac{1}{Y}}y< H}\}}\left(\int_{{{}H}}^{\infty}up_{Z}(u)du+\int_{t^{\frac{1}{2}-\frac{1}{Y}}y}^{{{}H}}up_{Z}(u)du\right)\\
&\leq t^{(\frac{1}{Y}-\frac{1}{2})(1-Y)}{\bf 1}_{\{{t^{\frac{1}{2}-\frac{1}{Y}}y\geq{}H}\}}\int_{t^{\frac{1}{2}-\frac{1}{Y}}y}^{\infty}{Ru^{-Y}}du\\
&\quad +t^{(\frac{1}{Y}-\frac{1}{2})(1-Y)}{\bf 1}_{\{{t^{\frac{1}{2}-\frac{1}{Y}}y<H}\}}\left(\int_{{{}H}}^{\infty}{Ru^{-Y}}du+{{}H}\widetilde{\bbp}(Z_{1}\geq t^{\frac{1}{2}-\frac{1}{Y}}y)\right)\\
&{\leq t^{(\frac{1}{Y}-\frac{1}{2})(1-Y)}\int_{t^{\frac{1}{2}-\frac{1}{Y}}y}^{\infty}{Ru^{-Y}}du+t^{(\frac{1}{Y}-\frac{1}{2})(1-Y)}{\bf 1}_{\{t^{\frac{1}{2}-\frac{1}{Y}}y< H\}}H\left(\frac{H}{t^{\frac{1}{2}-\frac{1}{Y}}y}\right)^{Y-1}}\\
&{\leq y^{1-Y}\left(\frac{R}{Y-1}+H^{Y}\right)},
\end{align*}
 and (\ref{Eq:NdBnd1}) follows.
\qed

\vspace{.5 cm}
\noindent
{
{\bf Acknowledgments:}
It is a pleasure to thank Peter Tankov and other participants of the 2012 SIAM Conference on Financial Mathematics and Engineering for some useful comments. }

%\newpage
\bibliographystyle{plain}

\begin{thebibliography}{1}

\bibitem{BNS:1998}
O. E. Barndorff-Nielsen,
\newblock {Processes of Normal Inverse Gaussian type},
\newblock {\em {Finance and Stochastics}}, 2, 41-68, {1997}.


\bibitem{BBF:2002}
H. Berestycki, J. Busca, and I. Florent.
\newblock{Asymptotics and calibration of local volatility models},
\newblock{\em {Quantitative Finance}}, 2, 61-69, 2002.

\bibitem{BBF2:2004}
H. Berestyki, J. Busca, and I. Florent.
\newblock{Computing the implied volatility in stochastic volatility models},
\newblock{\em {Communications on Pure and Applied Mathematics}}, Vol LVII, 1352-1373, 2004.

\bibitem{BL02} Boyarchenko, S.I., and S.Z. Levendorksii. \newblock {Non-Gaussian Merton-Black- Scholes theory}, \newblock {\em {Adv. Ser. Stat. Sci. Appl. Probab.}} 9. World ScientiÞc Publishing Co., Inc., River Edge, NJ, 2002.

\bibitem{CGMY:2002}
P. Carr, H. Geman, D. Madan, and M. Yor.
\newblock {The fine structure of asset returns: an empirical investigation},
\newblock {\em Journal of Business}, 75, 305-332, 2002.

\bibitem{CM09}
P. Carr and D. Madan.
\newblock {Saddle point methods for option pricing},
\newblock {\em {The Journal of Computational Finance}}, 13(1), 49-61, 2009.
%
\bibitem{Cont:1997}
R.~Cont, J.~Bouchaud, and M.~Potters.
\newblock {Scaling in financial data: stable laws and beyond},
\newblock {\em {in Scale Invariance and Beyond, Dubrulle, B., Graner, F., and
  Sornette, D., eds.}}, 1997.

\bibitem{CT04}
R.~Cont and P.~Tankov,
\newblock {\em Financial modelling with jump processes},
\newblock Chapman \& Hall, 2004.

\bibitem{CW03}
P. Carr and L. Wu.
\newblock {What type of process underlies options? A simple robust test},
\newblock {\em Journal of Finance}, 58(6), 2581-2610, 2003.

\bibitem{EKP:1998}
E. Eberlein, U. Keller, and K. Prause.
\newblock {New insights into smile, mispricing and value at risk},
\newblock {\em {Journal of Bussiness}}, 71, 371-406, 1998.

\bibitem{FFF:2010}
J. Feng, M. Forde, and J.P. Fouque.
\newblock {Short maturity asymptotics for a fast mean reverting Heston stochastic volatility model},
\newblock {\em SIAM Journal on Financial Mathematics}, 1, 126-141, 2010.

\bibitem{FFK:2012}
J. Feng, J.P. Fouque, and R. Kumar,
\newblock {Small-time Asymptotics for Fast Mean-Reverting Stochastic Volatility Models},
\newblock {Forthcoming in \em {The Annals of Applied Probability}}, 2012.

\bibitem{FigForde:2012}
J.E. Figueroa-L\'opez and M. Forde.
\newblock {The small-maturity smile for exponential L\'evy models}, 
\newblock{\em SIAM Journal on Financial Mathematics} 3(1), 33-65, 2012.

\bibitem{FigHou:2008}
J.E. Figueroa-L\'opez and C.~Houdr\'e.
\newblock {Small-time expansions for the transition distributions of L\'evy
  processes},
\newblock {\em Stochastic Processes and their Applications}, 119, 3862--3889,
  2009.

\bibitem{FigGH:2011}
J.E. Figueroa-L\'opez, R. Gong, and C. Houdr\'e.
\newblock {Small-time expansions of the distributions, densities, and option prices under stochastic volatility models with L\'{e}vy jumps},
\newblock {{\em Stochastic Processes and their Applications}, 122, 1808-1839, 2012.}
%
\bibitem{FordeJac:2009}
M. Forde and A. Jacquier.
\newblock{Small-time asymptotics for implied volatility under {the} Heston model},
\newblock{\em Int. J. Theor. Appl. Finance}, 12(6), 861-876, 2009.

\bibitem{FordeJac:2010}
M. Forde and A. Jacquier.
\newblock{Small time asymptotics for an uncorrelated local-stochastic volatility model},
{\newblock{\em {Applied Mathematical Finance}}, 18(6), 517-535, 2011.}

\bibitem{FordeJacLee:2010}
M. Forde, A. Jacquier, and R. Lee.
\newblock{The small-time smile and term structure for implied volatility under the Heston model},
\newblock{\em {preprint}}, {2011}.


\bibitem{Gatheral:2009}
J. Gatheral, E. Hsu, P. Laurence, C. Ouyang, and T-H. Wang.
\newblock {Asymptotics of implied volatility in local volatility models},
{\newblock {Forthcoming in \em {Mathematical Finance}}, 2012.}


\bibitem{Henry:2009}
P. Henry-Labord\`{e}re.
\newblock{Analysis, geometry, and modeling in finance: advanced methods in option pricing},
\newblock{\em {Chapman \& Hall}}, 2009.

%\bibitem{Hou:2002}
%C. Houdr\'{e}.
%\newblock {Remarks on deviation inequalities for functions of infinitely divisible random vectors},
%\newblock {\em {Annals of Probability}}, 30(3), 1223-1237, 2002.
%
%\bibitem{HouMar:2004}
%C. Houdr\'{e} and P. Marchal.
%\newblock {On the concentration of measure phenomenon for stable and related random vectors},
%\newblock {\em {Annals of Probability}}, 32(2), 1496-1508, 2004.


\bibitem{Koponen:1995}
I. Koponen.
\newblock{Analytic approach to the problem of convergence of truncated L\'{e}vy flights towards the Gaussian stochastic process},
\newblock{\em {Physical Review E}}, 52, 1197-1199, 1995.

\bibitem{Kou:2002}
S. Kou.
\newblock {A jump-diffusion model for option pricing},
\newblock {\em {Management Science}}, 48, 1086-1101, 2002.


\bibitem{MCC:1998}
D. B. Madan, P. Carr, and E. Chang.
\newblock {The variance gamma process and option pricing},
\newblock {\em {European Finance Review}}, 2, 79-105, 1998.

\bibitem{MadMil:1991}
D. B. Madan and F. Milne.
\newblock {Option pricing with VG martingale components},
\newblock {\em {Mathematical Finance}}, 1, 39-56, 1991.

\bibitem{MadSen:1990}
D. B. Madan and E. Seneta.
\newblock {The variance gamma (VG) model for share market {returns}},
\newblock {\em {Journal of Business}}, 63, 511-524, 1990.

\bibitem{Mandelbrot}
B.~Mandelbrot.
\newblock {The variation of certain speculative prices}.
\newblock {\em The Journal of Business}, 36:394--419, 1963.


\bibitem{Matacz}
A.~Matacz.
\newblock {Financial modeling and option theory with the truncated L\'evy
  process},
\newblock {\em Int. J. Theor. Appl. Finance}, 3:143--160, 2000.

\bibitem{Merton:76}
R. Merton.
\newblock{Option pricing when underlying stock returns are discontinuous},
\newblock{\em {Journal of Financial Economics}}, 3, 125-144, 1976.

\bibitem{MuhNut:2009}
J. Muhle-Karbe and M. Nutz.
\newblock {Small-time asymptotics of option prices and first absolute moments},
{\newblock {\em Journal of Applied Probability}, 48(4), 1003-1020, 2011.}

\bibitem{Paulot:2009}
L. Paulot.
\newblock{Asymptotic implied volatility at the second order with application to the SABR model},
\newblock{\em {Preprint}}, 2009.

\bibitem{Press}
S.J.~Press.
\newblock {A compound event model for security prices}.
\newblock {\em The Journal of Business}, 40:317--335, 1967.

\bibitem{Rop10}
M.~Roper.
\newblock {Implied volatility: small time to expiry asymptotics in exponential L\'{e}vy models},
\newblock {\em Thesis, University of New South Wales, 2009}.

\bibitem{RoperRut:2007}
M.~Roper and M.~Rutkowski.
\newblock {A note on the behaviour of the Black-Scholes implied volatility close to expiry},
\newblock {\em preprint}, 2007.

\bibitem{Ruschendorf}
L.~R\"uschendorf and J.~Woerner.
\newblock {Expansion of transition distributions of L\'evy processes in small
  time}.
\newblock {\em Bernoulli}, 8, 81-96, 2002.

\bibitem{Rosinski:2007}
J.~Rosi\'nski.
\newblock {Tempering stable processes}.
\newblock {\em {Stochastic processes and their applications}}, 117:\penalty0 677--707,
  2007.

\bibitem{Rosinski:2010}
J. Rosi\'nski and J. L. Sinclair.
\newblock{Generalized tempered stable processes},
\newblock{\em {Banach Center Publication}}, 90, 153-170, 2010.

\bibitem{Sato:1999}
K. Sato.
\newblock{\em {L\'{e}vy processes and infinitely divisible
distributions}},
\newblock Cambridge University Press, 1999.

\bibitem{Tankov}
P. Tankov.
\newblock {Pricing and hedging in exponential L\'evy models: review of recent results},
\newblock{\em {Paris-Princeton Lecture Notes in Mathematical Finance}}, Springer 2010.


\end{thebibliography}

\end{document}